\documentclass[twocolumn,resetfootnote]{aastex701}

\usepackage{amsmath}

\begin{document}
\title{A radially broad collisional cascade in the debris disk of $\gamma$~Ophiuchi observed by JWST}

\author[orcid=0000-0002-2106-0403,sname=Han]{Yinuo Han}
\affiliation{Division of Geological and Planetary Sciences, California Institute of Technology, 1200 E. California Blvd., Pasadena, CA 91125, USA}
\email[show]{yinuo@caltech.edu}

\author[orcid=0000-0001-9064-5598,sname=Wyatt]{Mark Wyatt}
\affiliation{Institute of Astronomy, University of Cambridge, Madingley Road, Cambridge, CB3 0HA, UK}
\email{wyatt@ast.cam.ac.uk}

\author[orcid=0000-0002-3532-5580,sname=Su]{Kate Y. L. Su}
\affiliation{Space Science Institute, 4750 Walnut Street, Suite 205, Boulder, CO 80301, USA}
\email{ksu@spacescience.org}

\author[orcid=0000-0003-4623-1165,sname=Sefilian]{Antranik A. Sefilian}
\affiliation{Department of Astronomy and Steward Observatory, University of Arizona, Tucson, AZ 85721, USA}
\email{sefilian.antranik@gmail.com}

\author[orcid=0000-0002-4248-5443,sname=Lovell]{Joshua B. Lovell}
\affiliation{Center for Astrophysics, Harvard \& Smithsonian, 60 Garden Street, Cambridge, MA 02138-1516, USA}
\email{joshualovellastro@gmail.com}

\author[orcid=0000-0002-8949-5200,sname=del Burgo]{Carlos del Burgo}
\affiliation{Instituto de Astrof\'\i sica de Canarias, V\'\i a L\'actea S/N, La Laguna, 38200, Tenerife, Spain}
\affiliation{Departamento de Astrof\'\i sica, Universidad de la Laguna, La Laguna, 38200, Tenerife, Spain}
\email{cburgo@ull.edu.es}

\author[orcid=0000-0001-6208-1801,sname=Marshall]{Jonathan P. Marshall}
\affiliation{Institute of Astronomy and Astrophysics, Academia Sinica, 11F of AS/NTU Astronomy-Mathematics Building, No. 1, Sec. 4, Roosevelt Rd, Taipei 106319, Taiwan}
\email{jmarshal@asiaa.sinica.edu.tw}

\author[orcid=0000-0002-5352-2924,sname=Marino]{Sebastian Marino}
\affiliation{Department of Physics and Astronomy, University of Exeter, Stocker Road, Exeter EX4 4QL, UK}
\email{s.marino-estay@exeter.ac.uk}

\author[orcid=0000-0003-1526-7587,sname=Wilner]{David J. Wilner}
\affiliation{Center for Astrophysics, Harvard \& Smithsonian, 60 Garden Street, Cambridge, MA 02138-1516, USA}
\email{dwilner@cfa.harvard.edu}

\author[orcid=0000-0003-3017-9577,sname=Matthews]{Brenda C. Matthews}
\affiliation{Herzberg Astronomy \& Astrophysics Research Centre, National Research Council of
Canada, 5071 West Saanich Road, Victoria, BC, Canada, V9E 2E7}
\affiliation{Department
of Physics \& Astronomy, University of Victoria, 3800 Finnerty Rd,
Victoria, BC, Canada, V8P 5C2}
\email{bcmatthews.herzberg@gmail.com}

\author[orcid=0000-0003-4761-5785,sname=Sommer]{Max Sommer}
\affiliation{Institute of Astronomy, University of Cambridge, Madingley Road, Cambridge, CB3 0HA, UK}
\email{ms3078@cam.ac.uk}

\author[orcid=0000-0002-4803-6200,sname=Hughes]{A. Meredith Hughes}
\affiliation{Department of Astronomy, Van Vleck Observatory, Wesleyan University, 96 Foss Hill Dr., Middletown, CT, 06459, USA}
\email{amhughes@wesleyan.edu}

\author[orcid=0000-0003-2251-0602,sname=Carpenter]{John M. Carpenter}
\affiliation{Joint ALMA Observatory, Avenida Alonso de C\'ordova 3107, Vitacura, Santiago, Chile}
\email{john.carpenter@alma.cl}

\author[orcid=0000-0001-7891-8143,sname=MacGregor]{Meredith A. MacGregor}
\affiliation{Department of Physics and Astronomy, Johns Hopkins University, 3400 N Charles Street, Baltimore, MD 21218, USA}
\email{mmacgregor@jhu.edu}

\author[orcid=0000-0002-9385-9820,sname=Pawellek]{Nicole Pawellek}
\affiliation{Department of Astrophysics, University of Vienna, T\"urkenschanzstra{\ss}e 17, 1180 Vienna, Austria}
\email{nicole.pawellek@univie.ac.at}

\author[orcid=0000-0002-1493-300X,sname=Henning]{Thomas Henning}
\affiliation{Max Planck Institute for Astronomy, K\"onigstuhl 17, 69117 Heidelberg, Germany}
\email{henning@mpia.de}

\begin{abstract}
The A1V star $\gamma$~Oph, at a distance of 29.7\,pc, is known from Spitzer imaging to host a debris disk with a large radial extent and from its spectral energy distribution to host inner warm dust. We imaged $\gamma$~Oph with JWST/MIRI at 15 and 25.5\,$\mu$m, revealing smooth and radially broad emission that extends to a radius of at least 250\,au at 25.5\,$\mu$m. In contrast to JWST findings of an inner small-grain component with distinct ringed structures in Fomalhaut and Vega, the mid-infrared radial profile combined with prior ALMA imaging suggests a radially broad steady-state collisional cascade with the same grain size distribution throughout the disk. This further suggests that the system is populated by a radially broad planetesimal belt from tens of au or less to well over 200\,au, rather than a narrow planetesimal belt from which the observed dust is displaced to appear broad. The disk is also found to be asymmetric, which could be modelled by a stellocentric offset corresponding to a small eccentricity of $\sim$0.03. Such a disk eccentricity could be induced by a mildly eccentric $<$$10\,M_\mathrm{Jup}$ giant planet outside 10\,au, or a more eccentric companion up to stellar mass at a few au, without producing a resolvable radial gap in the disk. 
\end{abstract}

\keywords{\uat{Debris disks}{363} --- \uat{Circumstellar disks}{235} --- \uat{Planetary-disk interactions}{2204} --- \uat{Exoplanet systems}{484} --- \uat{Exoplanet evolution}{491}}


\section{Introduction} \label{sec:intro}
Solid bodies in planetary systems initially assemble from dust embedded in gas-rich protoplanetary disks \citep{Williams2011}. Following the dispersal of the primordial gas disk, planetesimals begin undergoing destructive mutual collisions, re-generating dust that populates a second-generation disk known as a debris disk \citep{Wyatt2008}. These dust belts around main-sequence stars are some of the main observable signatures of extrasolar planetesimals \citep{Hughes2018}. 

Planetesimal belts are expected to inherit the structures of protoplanetary disks to some extent. Observationally, this is partly supported by similarities in the statistical distribution of ring widths in debris disks and protoplanetary disks \citep{Han2025b}. However, ongoing physical processes continue to shape debris disks. These dynamical processes generally occur in three categories. Firstly, solid bodies in debris disks frequently collide, producing successively smaller bodies down to micron-sized dust grains, forming a steady-state collisional cascade in which the size distribution can be characterised by a power law \citep{Dohnanyi1969, Wyatt2007, Kenyon2008, Marshall2025b}. Dust grains smaller than a given threshold (in the micrometre regime) are ejected from the system by radiation pressure \citep{Burns1979, Krivov2010}, thereby setting a minimum grain size. In addition to determining the grain size distribution, these collisions could influence the evolution of the disk structure, as collision rates are generally higher for closer-in orbits, resulting in faster depletion \citep{Kennedy2010}. A collisionally eroded debris disk is expected to exhibit an inner edge surface density proportional to $r^{7/3}$ \citep{Kennedy2010, ImazBlanco2023}, and the outwards propagation of peak emission could contribute to the observed tentative increase in disk radius as a function of age on $\sim$100\,Myr timescales \citep{Eiroa2013, Matra2025, Han2025}. 

Secondly, gravitational interactions with planets could perturb the spatial distribution of the planetesimal belt within which dust is produced. For example, planets could carve sharper disk edges than that expected from collisional evolution alone (e.g., \citealp{Mustill2012, Marino2021, ImazBlanco2023, Pearce2024}) or result in radial gaps \citep[e.g.][]{Yelverton2018, Friebe2022, Sefilian2021, Sefilian2023}, which have been observed in five debris disks based on Atacama Large Millimeter/sub-millimeter Array (ALMA) observations \citep{Marino2018, MacGregor2019, Daley2019, Marino2019, Marino2020, Nederlander2021, Han2025b}. 
Planets and/or stellar companions may also induce a non-zero disk eccentricity \citep{Wyatt1999, Pearce2014, Farhat2023}, which can be further modulated by the disk's own gravity \citep{Sefilian2024}.
Planets that have migrated could further scatter planetesimals into high-eccentricity orbits, resulting in a scattered disk component, which has been suggested to be the case in the Solar System's Kuiper Belt \citep{Malhotra1993, Morbidelli2004}, HR\,8799 \citep{Geiler2019}, $\beta$~Pic \citep{Matra2019} and q$^1$~Eri \citep{Lovell2021}. 

Finally, stellar radiation displaces the spatial distribution of dust from their site of production \citep{Lee2016}. These effects include radiation pressure, which forces smaller grains produced from collisions onto higher eccentricity orbits until micron-sized grains are ejected from the system \citep{Strubbe2006}; and Poynting-Robertson drag (PR drag, e.g., \citealp{Wyatt2005} and references therein), which causes dust grains to spiral inwards towards the star, causing observable signatures such as the Zodiacal dust in the Solar System. In young late-type stars, stellar wind drag could also become significant and result in the inward migration of dust grains, similar to the effect of PR drag \citep{Plavchan2005, Wolff2025}. Additionally, interactions with the interstellar medium (ISM) have also been proposed to shape certain asymmetries in the distribution of small grains \citep{Hines2007, Gaspar2008, Schneider2014}. 

Resolved imaging of debris disks across a range of wavelengths has provided observational constraints on the different dynamical forces in action. At mm wavelengths, observations are sensitive to thermal emission predominantly from mm-sized grains ($\sim$100s of $\mu$m to cm-sized grains) which are minimally impacted by radiative forces. These observations are thought to closely trace the distribution of the planetesimals from which dust is produced, thereby mapping the parent belts that feed the collisional cascade \citep{Hughes2018}. ALMA observations have imaged dozens of these cold outer belts at tens of au and beyond \citep{Matra2025}, of which two dozen have been mapped at high resolution by the ALMA survey to Resolve exoKuiper belt Substructures \citep[ARKS,][]{Marino2025}, revealing substructures such as radial gaps \citep{Han2025}, various vertical distributions (e.g., non-Gaussian or double-Gaussian vertical profiles, \citealp{Matra2019, Zawadzki2025}) and azimuthal asymmetries \citep{Lovell2025}. 

At optical and near-infrared wavelengths, observations are sensitive to micron-sized dust grains that scatter stellar radiation. These observations have revealed evidence of the radiatively displaced small grains in the form of halos, sometimes with outwardly displaced peak surface densities relative to the mm grains \citep{Milli2025}, possibly affected by drag forces if gas were to be present in the system \citep{Jankovic2025, Olofsson2025}. 

More recently, JWST has offered significantly improved sensitivity and resolution for debris disk imaging in the mid-infrared over prior instruments. These wavelengths are sensitive to thermal emission from small grains in the inner tens of au of debris disks, revealing the structures of disks in these inner regions where previous instruments were not able to. 
Previously, \textit{Spitzer} observations identified the presence of inner dust components in systems such as Vega \citep{Stapelfeldt2004}, Fomalhaut \citep{Su2005}, and $\gamma$~Oph \citep{Su2008}. Around the A stars Fomalhaut and Vega, the \textit{James Webb Space Telescope} (JWST) has recently imaged the distribution of this dust at an order of magnitude higher resolution \citep{Gaspar2023, Su2024}. Importantly, modelling of the distribution of the warm dust component interior to the outer ALMA ring in each system has suggested that the warm dust originates from small grains migrating inwards from the parent belt via PR drag \citep{Su2024, Sommer2025}. In the debris disk of the K star $\epsilon$~Eri, imaging by JWST's Mid-InfraRed Instrument (MIRI, \citealp{Wright2023}) identified inner dust consistent with stellar wind drag \citep{Wolff2025}. While weaker effects from PR drag is expected from later spectral types, this leaves the question of whether all debris disks found around early-type stars could likewise be affected by PR drag. 

In light of these findings, the debris disk of $\gamma$~Oph (HD\,161868) offers an important point of comparison to Vega and Fomalhaut. At a distance of 29.7\,pc and an age of $300 \pm 200$\,Myr (\citealp{Vican2012, Gaspar2016, Trevor2015}), $\gamma$~Oph is of a similar spectral type (A1V) and age to Vega (A0V, $460 \pm 10$\,Myr, \citealp{Yoon2010}) and Fomalhaut (A4V, $440 \pm 40$\,Myr, \citealp{Mamajek2012}). \textit{Spitzer} observations at 24\,$\mu$m show the disk to be greatly extended, reaching 265\,au (adjusted for the updated distance used here) radially (or 9$^{\prime\prime}$, compared to a PSF FWHM of 6$^{\prime\prime}$) or beyond \citep{Su2008}. \textit{Herschel} far-infrared observations at 70 to 160~$\mu$m spatially resolve the disc's outer belt with inner and outer edges between 50 to 300~au \citep{Pawellek2014, Moor2015}; a Gaussian belt model for the disc architecture obtains a radius of 95~au and a FWHM of 70~au \citep{Marshall2021}. 

ALMA observations revealed a broad disk extending to a similar outer radius as found by \textit{Spitzer}, with a resolved but shallow central cavity (i.e., with non-zero emission within the central cavity, \citealp{Marino2025, Han2025b}). Proper motion anomalies in the system suggest the possibility of undetected planets \citep{Kervella2022, Milli2025}, but no signatures of planet-disk interactions (such as radial gaps or asymmetries) have yet been detected in the disk. 

We recently imaged the debris disk of $\gamma$~Oph with JWST/MIRI to test whether the PR drag scenario in Vega and Fomalhaut applies to $\gamma$~Oph's radially broad belt. These images resolve the disk in the mid-infrared at more than 7 times higher resolution compared to prior imaging and form the basis of this study. We describe the observations in Section~\ref{sec:obs} and the data reduction in Section~\ref{sec:data}. We model the mid-infrared disk structure in Section~\ref{sec:results}, which we compare to ALMA observations in Section~\ref{sec:discussion}. Our findings are summarised in Section~\ref{sec:conclusions}.

\section{Observations} \label{sec:obs}
We observed $\gamma$~Oph with the JWST MIRI Imager on 13 Aug 2024 UT (programme ID 5709, \citealp{Han2024_JWST_GamOph}). Full aperture direct images were taken with the F1500W and F2550W broadband filters. Smaller MIRI detector subarrays have higher saturation limits at the expense of a more limited field of view. To prevent saturation within each filter while still maximising the field of view to the extent possible, we employed the largest subarray that offers sufficiently fast readout speed with the FASTR1 readout pattern that avoids saturating the bright stellar core. 
The expected emission levels for the star and disk were determined from the spectral energy distribution (SED) of $\gamma$~Oph from which the optimum subarrays were identified (the SED will be discussed in Section~\ref{sec:discussion} and is shown in Fig.~\ref{fig:sed}).
For the F1500W and F2550W filters, this corresponded to the SUB256 and SUB128 detector subarray, respectively. 

For each filter, one exposure was taken at each of 4 dither positions under the default 4-Point-Sets dithering pattern and optimised for extended sources, starting at set number 1 and proceeding for 1 set. 
The F1500W observations were taken with 5 groups per integration and 310 integrations per exposure, totalling 885\,s of exposure time. The nominal RMS noise achieved for regions with significant disk emission as estimated by the JWST pipeline is 0.1\,$\mu$Jy per pixel, with a plate scale of 0.11\,arcsec/pixel (or 0.03\,mJy/arcsec$^2$). The F2550W observations were taken with 10 groups per integration and 45 integrations per exposure, which amounted to 592\,s of exposure time. The RMS noise is 0.3\,$\mu$Jy per pixel as estimated by the JWST pipeline, with the same plate scale as for F1500W ($\equiv$ 0.01\,mJy/arcsec$^2$). 

To characterise the point-spread function (PSF), subtract stellar emission and model disk emission, we also observed $\zeta$~Ser as the PSF reference star. $\zeta$~Ser is of nearly identical mid-infrared flux density to the photosphere of $\gamma$~Oph (within $\sim$1\%), thereby mitigating differences between the science target and reference star PSF sizes that could arise due to mismatches in the flux density \citep{Argyriou2023}. At mid-infrared wavelengths, the impact of spectral mismatch on PSF subtraction is low. Any potential impact is further mitigated by the fact that $\zeta$~Ser (F2V) is of a sufficiently similar spectral type to $\gamma$~Oph, such that no loss in sensitivity is expected even at near-infrared wavelengths, where effects of spectral type mismatches are more pronounced\footnote{JWST User Documentation}. $\zeta$~Ser has also been vetted as a single star (using the SearchCal database, \citealp{SearchCal}, without known infrared excess). 

Given the faint emission of the disk around $\gamma$~Oph, we also performed background observations to accurately subtract off any detector artefacts. The background region observed was chosen to be centred at equatorial coordinates (RA, Dec) at the epoch of J2000 of (17h 47m 58.6080s, +02d 43m 24.32s) and was selected for its proximity to $\gamma$~Oph and relative sparseness of background stars as suggested by \textit{Spitzer} observations at comparable wavelengths. All PSF and background observations were carried out with identical exposure parameters as for $\gamma$~Oph within the corresponding filters.

\section{Data reduction} \label{sec:data}

\begin{figure*}
    \centering
    \includegraphics[width=0.8\linewidth]{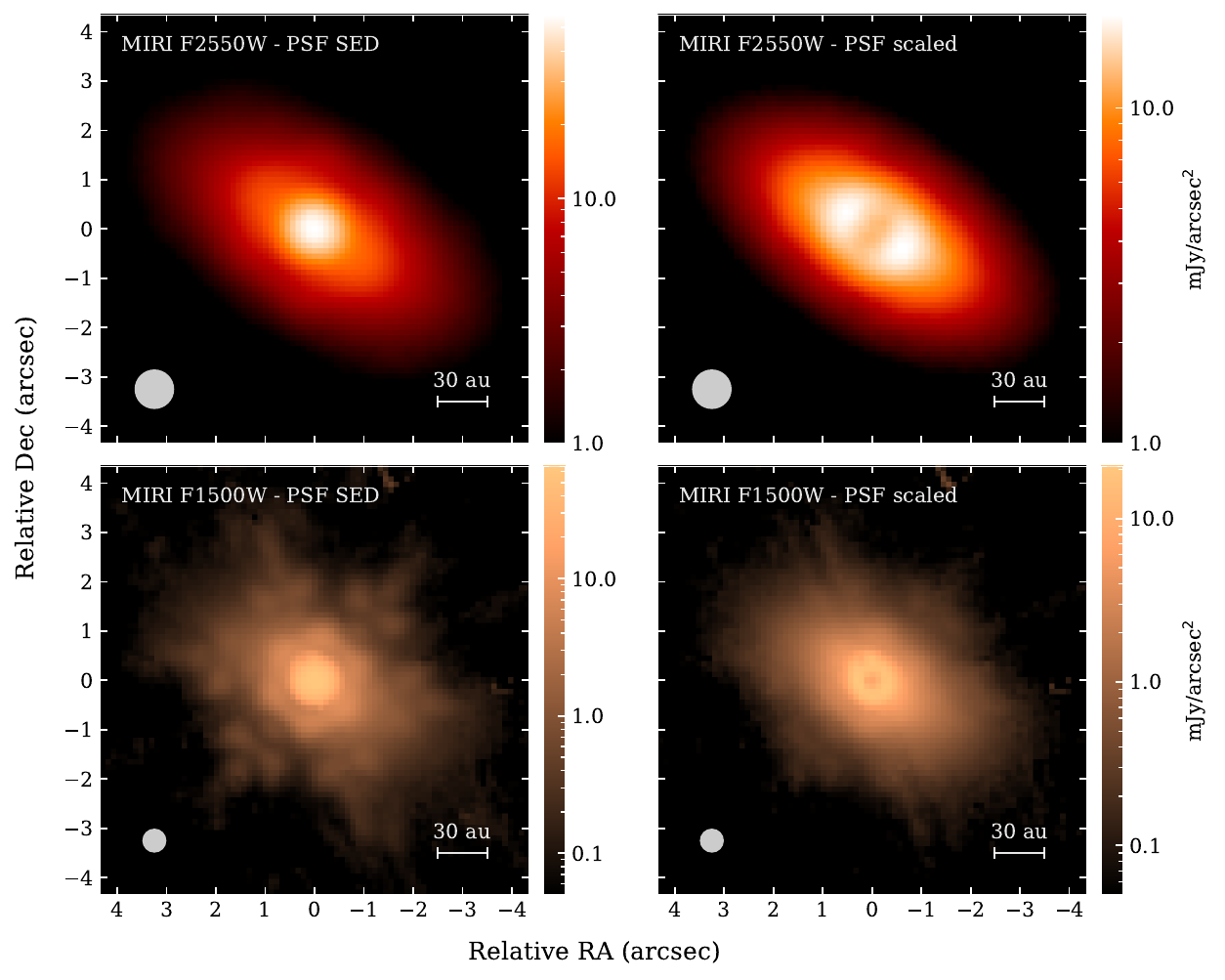}
    \caption{PSF-subtracted MIRI F2550W (top row) and F1500W (bottom row) images. The left column subtracts off a point source centred on the star scaled to the stellar flux density inferred from the SED (205\,mJy for F2550W and 591\,mJy for F1500W; see Section~\ref{sec:almased} and Fig.~\ref{fig:sed}), in which the central component remains bright. The right column subtracts off a higher PSF flux density (246\,mJy for F2550W and 615\,mJy for F1500W) to further emphasise the structure of the resolved disk component. PSF subtraction was performed dither by dither with stage 2 products before combining the subtracted dithers with the stage 3 pipeline to produce the images displayed here. The images are displayed on a logarithmic scale. The PSF FWHM is denoted by the shaded ellipse in the bottom left corner of each panel.
    The orientation of the images is North up, East left.}
    \label{fig:obs}
\end{figure*}

We processed the MIRI observations with the JWST pipeline version 1.18.0 using the corresponding JWST Calibration Reference Data System context \texttt{jwst\_1364.pmap}. No pixels were flagged as saturated across any observations of $\gamma$~Oph and $\zeta$~Ser, as expected from the observational setup (Section~\ref{sec:obs}). 

We first ran the default pipeline to obtain Stage 3 data products (i.e., science-ready mosaics) for the science target ($\gamma$~Oph), PSF star ($\zeta$~Ser) and background observations across both filters, before performing background and PSF subtraction. However, we noticed that such a reduction results in subtle high-contrast artefacts in the PSF-subtracted image, which are likely introduced in the resampling step of the Stage 3 Imaging pipeline, when individual dithers containing both the bright stellar core and the faint disk emission are distortion corrected and aligned \citep{Gaspar2023, Su2024}. 

To mitigate against these artefacts, we instead performed dither-by-dither background and PSF subtraction based on the Stage 2 pipeline products of (calibrated slope) images taken at each dither position, before combining the subtracted disk image at each dither using the Stage 3 pipeline \citep{Su2024}. Specifically, at each dither position, we subtracted the background image from the science target and reference star images taken with the corresponding filters based on the Stage 2 products. To account for slight pointing offsets between the science target and reference star at each dither, we aligned the reference star to the science target by fitting 2D Gaussians to the bright stellar cores of both, noting that the disk emission is significantly fainter than stellar emission at both wavelengths. We then subtracted a scaled version of the PSF observations from the science target to obtain the PSF-subtracted image for each dither. 

We produced two versions of the PSF-subtracted imaged. The first applies PSF subtraction by scaling the reference star to the stellar component flux of the science target inferred from their SEDs, which estimates the stellar component of $\gamma$~Oph to be 205\,mJy at 25.5\,$\mu$m and 591\,mJy at 15\,$\mu$m, and $\zeta$~Ser to be 207\,mJy at 25.5\,$\mu$m and 597\,mJy at 15\,$\mu$m (Fig.~\ref{fig:sed} discussed in Section~\ref{sec:discussion}). The second version increases this scaling to further suppress any unresolved central emission and emphasise the structure of the resolved disk component without producing a negative PSF core. For this version, we increased the PSF scaling by a factor of 1.2 relative to the SED flux of $\gamma$~Oph's stellar component at 25.5\,$\mu$m (equivalent to 246\,mJy) and a factor of 1.04 at 15\,$\mu$m (equivalent to 615\,mJy).

Although the background observations were taken with the same exposure parameters as the science target and reference star, a slight residual background remains in the F2550W observations of the science target, resulting in its background regions showing negative emission on average. We therefore corrected for the background over-subtraction by further subtracting the (negative) median background level, which is measured using the background region beyond an $11^{\prime\prime}$ radius from the star within the field of view. The reduced F2550W and F1500W images are displayed in Fig.~\ref{fig:obs}. 

The per-pixel RMS noise measured in the PSF-subtracted image is 0.07\,mJy/arcsec$^2$ for F2550W and 0.02\,mJy/arcsec$^2$ for F1500W. These noise levels are higher than those estimated by the JWST pipeline for observations of the science target alone, in part due to the fact that they carry additional uncertainties from the background and PSF observations. Within a circular aperture of 10$^{\prime\prime}$ (300\,au), the in-band integrated flux densities measured from the Stage 3 pipeline images (with background subtraction but without dither-by-dither PSF subtraction) are $429.55 \pm 0.06$\,mJy at F2550W and $602.64 \pm 0.03$\,mJy at F1500W for $\gamma$~Oph, and $208.41 \pm 0.06$\,mJy at F2550W and $550.67 \pm 0.03$\,mJy at F1500W for $\zeta$~Ser. 

\section{Results} \label{sec:results}

\begin{figure*}
    \centering
    \includegraphics[width=1.0\linewidth]{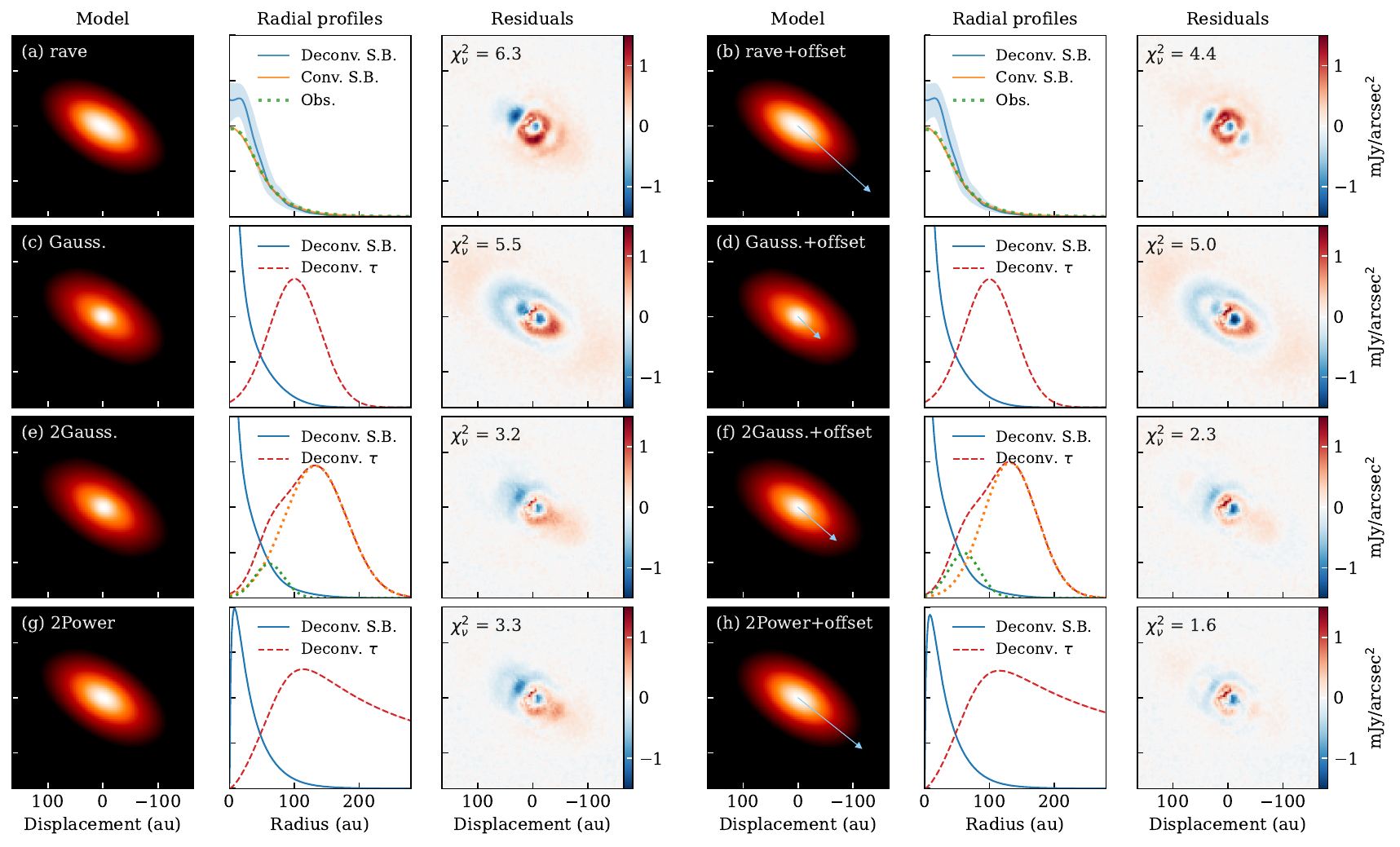}
    \caption{Gallery of nonparametric and parametric models fitted to the MIRI F2550W PSF-subtracted (assuming stellar SED flux) image of $\gamma$~Oph. Each group of 3 panels displays the PSF-convolved disk-only model image (i.e., stellar component not included, displayed on a logarithmic colour scale, to be compared with the top-left panel in Fig.~\ref{fig:obs}), disk-only deconvolved and deprojected radial surface brightness (S.B.) and optical depth ($\tau$) profiles and the residual image (data $-$ model). ``Offset'' indicates that the geometric centre of the disk is offset relative to the star in the model, and is included as (two) free parameters in the model (along two spatial axes). The displacement vector of the disk centre relative to the star is indicated with blue arrows. All radial profile panels are plotted with the same vertical scaling, which span from 0 to 1.8\,mJy/arcsec$^2$. For the \texttt{rave} radial profile panel (panel a), the azimuthally averaged profile of the PSF-subtracted (assuming the stellar flux of the best-fit grid point in the 4D \texttt{rave} grid) observations and the PSF-convolved best-fit disk model are overplotted, as this is the 1D quantity that \texttt{rave} directly fits to. Where multiple Gaussian components are invoked (panels e and f), the constituent Gaussian components are individually overplotted. Uncertainties are indicated with shaded regions for the main deconvolved profiles, but these are generally too narrow to be visible for the parametric models. The per-pixel RMS noise measured from the background is 0.07\,mJy/arcsec$^2$. North is oriented upwards and east to the left. 
    }
    \label{fig:models25}
\end{figure*}

The primary aim of this section is to characterise the spatial structure of the disk seen at mid-infrared wavelengths by experimenting with a range of models. We begin with a nonparametric approach to gauge the shape of the radial profile of the disk at 25\,$\mu$m, before testing a range of parametric models in search for a satisfactory fit. We then consider the 15\,$\mu$m image in light of the 25\,$\mu$m structure that we find and attempt to recover any structural information about the disk from its significantly more centrally peaked emission.

\subsection{Nonparametric modelling at 25\,$\mu$m}
We nonparametrically fitted the PSF-subtracted (assuming SED stellar flux) F2500W image with the \texttt{rave} package \citep{Han2022}. This approach determines the deconvolved and deprojected radial surface brightness profile by fitting concentric annuli to the disk image, assuming that the disk is azimuthally symmetric and optically thin. The placement of annuli boundaries are randomised to marginalise over their effect on the fitted profile, and the fit is repeated 100 times. We applied the non-edge-on optimisation of the code \citep{Han2025}, assuming that the vertical density distribution of the disk is Gaussian, with $\rho(z) \propto \exp{\left[ - z^2/(2h^2) \right]}$, and that the vertical aspect ratio, $h = H(r) / r$, is constant across the disk. Note that $H(r)$ here is the standard deviation (rather than FWHM) of the vertical profile. 

The application of \texttt{rave} requires an assumption on the position angle ($\theta$) and, for the non-edge-on version of the code, the inclination ($i$) and vertical aspect ratio ($h$). Moreover, while the PSF has been subtracted assuming the SED flux of the stellar component, there could be (positive or negative) residual stellar flux and unresolved central dust emission, both of which we absorb into a central point source flux parameter ($F_*$) that also includes the 205\,mJy already subtracted during dither-by-dither PSF subtraction. 
We therefore applied \texttt{rave} over a four-dimensional grid of ($\theta$, $i$, $F_*$, $h$), repeating the radial profile fitting procedure with 12 annuli\footnote{Chosen based on the largest number of annuli without producing oscillatory artefacts due to noise \citep{Han2022}. The surface brightness within each annulus is modelled as being constant.} between 0 and 245\,au from the star for each combination of the four parameters. The PSF observations were used to convolve the model. We computed the squared residuals (the sum over the squared residual image) for each median \texttt{rave} model, masking the region within 1.5$^{\prime\prime}$ (45\,au) from the star, which contains significant PSF subtraction artefacts. The 4D grid was defined by the closed intervals [$57.5^\circ$, $59.5^\circ$] with $0.5^\circ$ spacing for $\theta$, [61.0, 66.5] with $0.5^\circ$ spacing for $i$ and [230\,mJy, 250\,mJy] with 5\,mJy spacing for $F_*$ and the approximately logarithmically spaced sample points 0, 0.01, 0.02, 0.05, 0.1 and 0.2 for $h$. 

We found that across the 4D grid, the best combination of parameters are $\theta = 59.0^\circ$, $I = 64.5^\circ$, $F_* = 240$\,mJy and $h = 0.1$, as defined by the \texttt{rave} model with the lowest squared residuals. 
When examining each parameter individually and summing over all other axes of the grid, the four parameters each individually minimise the residuals at identical values to the best-fitting combination stated above. 
Note that we do not aim to estimate the uncertainties on these parameters in this section and leave this for parametric modelling in subsequent sections, which is expected to better characterise covariances between these parameters with more continuous sampling rather than a grid-based approach. 
Note also that any marginally resolved dust as part of the disk is likely degenerate with the central point source, preventing a robust separation of stellar and disk flux at small radii. This is further investigated with parametric models in Section~\ref{sec:parametric} and compared with the stellar SED in Section~\ref{sec:discussion}. We attempt to further investigate the inner disk structure with parametric modelling in Section~\ref{sec:parametric}. 

We used the \texttt{rave} profile fitted at the best-fitting grid point as the final nonparametric model. The fitted radial profile and residuals are displayed in Fig.~\ref{fig:models25}a and key model assumptions are listed in Table~\ref{tab:params25}. 

We find that the 1D azimuthally averaged profile of the PSF-subtracted observations is well-fit by the median \texttt{rave} model as plotted in Fig.~\ref{fig:models25}. However, significant residuals structures are found in the 2D residual image. While residual structures within $1^{\prime\prime}$ from the star likely correspond to PSF subtraction artefacts, which we discuss in more detail in the context of parametric modelling, extended asymmetries are seen in the resolved disk at a level of approximately 0.5\,mJy/arcsec$^2$ relative to the symmetric \texttt{rave} model (i.e., twice of this value when comparing the two sides of the disk relative to each other). Assuming an RMS noise per pixel of 0.07\,mJy/arcsec$^2$, the S/N per pixel of this asymmetry is approximately 7$\sigma$. For comparison, the disk emission at this region is approximately 9\,mJy/arcsec$^2$, plausibly suggesting the presence of asymmetric residual emission at the level of $\sim$5\%. At 25\,$\mu$m, the star has a comparable flux to that of the total flux of the spatially extended disk, so the core of the image is thus dominated by the star and the stellar position is well-constrained by the data. This disk asymmetry about the star is thus unlikely to be due to a wrong stellar location being assumed. Removing this asymmetry requires a $\sim$1\,au (or 0.3 MIRI pixels) shift in the stellar location assumed, which would result in the stellar PSF appearing significantly off-centre. 

To further explore the suggestion of an asymmetric disk, we repeated the \texttt{rave} fitting procedure over the 4D grid, but based on the disk image shifted by an amount that minimises its self-subtracted (i.e., disk image $-$ disk image rotated by 180$^\circ$ about the star) squared residuals after PSF subtraction by $F_*$. We find that the optimal ($\theta$, $i$, $F_*$, $h$) grid point and the values of each of the four parameters that individually minimize the squared residuals are the same as before (i.e., the case without a disk offset). The \texttt{rave} model with a stellocentric offset fitted assuming the optimal grid point is shown in Fig.~\ref{fig:models25}b. The central region near the star remains the region with the most significant residuals, in part because the sharpness of features in non-parametric fitting with \texttt{rave} is partly limited by the resolution of the observations \citep{Han2022}. Sharper features may be produced in the model by parametrising the radial profile with specific functional forms, which we explore in Section~\ref{sec:parametric}.

Fig.~\ref{fig:models25} and Table~\ref{tab:params25} indicate the direction and magnitude of the offset relative to the star. The magnitude of this offset is small (less than 1\,au, but magnified by a factor of 100 in Fig.~\ref{fig:models25} for display), and its effect on the fitted radial profile is minor, however, the offset disk model provides improved residuals over the star-centred model, with symmetric residual structures and lower squared residuals summed over the image. We further quantify the offset with parametric modelling in Section~\ref{sec:parametric}.

\subsection{Parametric modelling at 25\,$\mu$m} \label{sec:parametric}
Parametric modelling can yield further insights by restricting structural recovery to a few parameters of interest if a reasonably well-fitting radial profile functional form were to be assumed. Here, we begin with relatively simple functions to model the radial profile before increasing model complexity, primarily aiming to better understand the structure (e.g., radius, slope and any gaps) of the inner and outer edge of the disk. 

Throughout this section, we parametrise and fit the optical depth profile ($\tau$) rather than the surface brightness profile (S.B.), so we begin by briefly describing the conversion between the two before delving into the fitting. Converting from the optical depth to surface brightness requires assumptions on the optical properties of the emitting dust. We modelled the optical properties of dust grains using the \texttt{astrodust\_optprops} code \citep{Sommer2025}, which calculates the absorption efficiency, $Q_\mathrm{abs}(a_g, \lambda)$ (where $\lambda$ is the wavelength and $a_g$ is the grain size), the temperature profile, $T(r, a_g)$, and the ratio between radiation pressure and gravity, $\beta$, using either Mie theory, Rayleigh-Gans theory or geometric optics, depending on the grain size relative to the wavelength. The stellar luminosity was assumed to be 24.4\,$L_\odot$, as inferred from its SED (Fig.~\ref{fig:sed}), and the mass was assumed to be 2.11\,$M_\odot$ \citep{Marino2025}.

Each dust grain was assumed to consist of a silicate core and an organic refractory mantle, with ice and voids embedded in pores \citep{Li1997}. The optical constants of the aggregate were combined from its constituents using Maxwell-Garnett effective medium theory \citep{Bohren1983}. The grain composition is parametrised by a silicate fraction, $q_\mathrm{sil}$, ice fraction, $q_\mathrm{ice}$, and porosity, $p$, such that as a fraction of the total volume, $(1-p) \, q_\mathrm{sil}$ is silicates, $(1-p) \, (1-q_\mathrm{sil})$ is organics, $p \, q_\mathrm{ice}$ is ice and $p \, (1-q_\mathrm{ice})$ is voids. 

Using the best-fit grain parameters fitted to MIRI observations of Fomalhaut, we assumed that the grains are described by $q_\mathrm{sil} = 0.4$, $q_\mathrm{ice} = 1.0$ and $p = 0.7$, which is comparable to Solar System values \citep{Sommer2025}. 
We further assumed that the grain size distribution follows a power law, i.e., $dN/da \propto a^{-\gamma}$, where we set $\gamma = 3.5$ expected for a steady-state collisional cascade \citep{Dohnanyi1969, Wyatt2008, Hughes2018}. 
The minimum grain size (i.e., diameter) was assumed to be $s_\mathrm{min} = 15\,\mu$m, which is just below the blowout size limit in the disk for this dust composition estimated to be at 17\,$\mu$m (below which $\beta > 0.5$, causing fragments released from circular-orbit progenitors to be ejected from the system). The maximum ``grain'' size was set to be 1\,m, although larger grains do not contribute significantly to the emission provided that they are much larger than the wavelength considered, which is satisfied by this choice of maximum size. 

The grain size distribution assumptions are consistent with previous modelling of $\gamma$~Oph suggesting $\gamma = 3.6^{+0.2}_{-0.3}$ \citep{Marshall2025b}, as well as with typical observationally inferred values of $\gamma$ \citep{Hughes2018, Marshall2025b} and the expectation that $s_\mathrm{min}$ is generally found to be similar to or smaller than the blowout size for A-type stars \citep{Pawellek2015, Marshall2025b}. Theoretically, it is possible for the grain composition to change across the disk, as the water ice line is expected at $\sim1^{\prime\prime}$ from the star. Observationally, however, we do not see strong evidence for abrupt changes in emission at this region $\gamma$~Oph or in Fomalhaut (which is closer to us, thus the ice line region is better resolved), so we do not model spatially varying grain compositions here.
Note that while we have not explored the effect of grain composition assumptions, the fitting is not expected to sensitively depend on $q_\mathrm{sil}$, however $p$ and $q_\mathrm{ice}$ could have a stronger effect \citep{Sommer2025}. 

\subsubsection{Gaussian} \label{sec:gaussian}
We fitted a disk model assuming that the disk's face-on optical depth is radially Gaussian, given by:
\begin{equation}
\label{eq:1gauss}
    \tau(r) = A_1 \exp{\left[ - \frac{(r - \mu_{r1})^2}{2 \sigma_{r1}^2} \right]}.
\end{equation}
Similar to the \texttt{rave} fitting performed, the disk was assumed to be vertically Gaussian with a vertical height aspect ratio $h$. In addition to $A_1$, $\mu_{r1}$ and $\sigma_{r1}$, we included $\theta$, $i$, $F_*$ and $h$ as free parameters which we used to simulate a PSF-convolved model image to compare with the observations. Note that the observed PSF ($\zeta$~Ser) rather than a simulated PSF was used. We sampled the parameter space with a Markov chain Monte Carlo (MCMC) approach implemented with the \texttt{emcee} package \citep{emcee}, using 64 walkers and running the chain for at least 5,000 steps after burn-in. The noise in each pixel was assumed to be independent and Gaussian with a standard deviation equal to the rms noise described in Section~\ref{sec:data}, such that the log-likelihood function being sampled was computed as the sum of that for each pixel. Although in reality noise is expected to scale with flux density, constant noise is a reasonable approximation here given that disk emission is significantly fainter than the thermal background. Using the 16th, 50th and 84th percentiles of the marginalised posterior distribution of each parameter, we find that the model converges on the parameters displayed in Table~\ref{tab:params25}. 

The median model image of the disk, the fitted radial profile and the residual image are shown in Fig.~\ref{fig:models25}c. Compared to the \texttt{rave} model, we find that the Gaussian model lacks the wider ``wing'' on the outer edge that drops off more slowly towards larger radii. 

To account for the misaligned geometric disk centre and stellar PSF centre, we also fitted a second version of the Gaussian model, which includes two additional free parameters that describe the spatial offset of the disk along the RA and Dec directions. Note that the stellar location is fixed in the model and aligns with the PSF centre in the data, and that only the disk component is offset relative to the star. The results of this model are shown in Fig.~\ref{fig:models25}d. While residuals due to the choice of functional form of the radial profile are retained in this model as expected, the residuals are more symmetric for the version of the model with an offset, and the best-fit stellocentric offset values indicate that the offset or asymmetry of the disk is statistically significant under this choice of parametrisation. 

To aid with model comparison, we computed the reduced $\chi^2$ for each model and display these on the residual images in Fig.~\ref{fig:models25}, which suggest a slightly improved model by incorporating an offset. The Akaike's and Bayesian information criteria (AIC and BIC) are often used to evaluate the balance between the goodness of fit and the number of free parameters. As we have assumed each of the $\sim10^4$ pixels to be an independent data point, even small improvements in the reduced $\chi^2$  (e.g., by 0.1) causes the log-likelihood term in the AIC or BIC to dominate over the penalty term which only varies by a few free parameters. Indeed within the scope of this study, the AIC and BIC preference is always consistent with the model with the lower reduced $\chi^2$, hence for simplicity throughout this section we base the comparison directly on the reduced $\chi^2$.

\begin{sidewaystable}
    \centering
    \caption{Assumed parameters for \texttt{rave} and fitted values for parametric models at 25\,$\mu$m. }
    \label{tab:params25}
\begin{tabular}{lllllllll}
\hline \hline
 & \texttt{rave} &  & Gaussian &  & Two Gaussians &  & Two power laws &  \\
Parameter & No offset & Offset & No offset & Offset & No offset & Offset & No offset & Offset \\
\hline
$F_*$ (mJy) & 240 & 240 & $220.68^{+0.05}_{-0.05}$ & $220.98^{+0.05}_{-0.05}$ & $226.5^{+0.2}_{-0.2}$ & $229.30^{+0.08}_{-0.08}$ & $235.78^{+0.04}_{-0.04}$ & $236.32^{+0.04}_{-0.04}$ \\
$A_1$ &  &  & $1.034^{+0.001}_{-0.001} \times 10^{-4}$ & $1.031^{+0.001}_{-0.001} \times 10^{-4}$ & $1.063^{+0.003}_{-0.003} \times 10^{-4}$ & $1.089^{+0.003}_{-0.003} \times 10^{-4}$ & $1.337^{+0.003}_{-0.003} \times 10^{-4}$ & $1.300^{+0.003}_{-0.003} \times 10^{-4}$ \\
$\mu_{r1}$ (au) &  &  & $100.48^{+0.10}_{-0.09}$ & $99.99^{+0.09}_{-0.10}$ & $132.7^{+0.6}_{-0.6}$ & $132.0^{+0.4}_{-0.4}$ &  &  \\
$\sigma_{r1}$ (au) &  &  & $39.53^{+0.04}_{-0.04}$ & $39.27^{+0.04}_{-0.04}$ & $48.0^{+0.3}_{-0.3}$ & $41.8^{+0.2}_{-0.2}$ &  &  \\
$A_2$ &  &  &  &  & $2.75^{+0.10}_{-0.09} \times 10^{-5}$ & $3.61^{+0.05}_{-0.05} \times 10^{-5}$ &  &  \\
$\mu_{r2}$ (au) &  &  &  &  & $61.3^{+0.4}_{-0.4}$ & $60.8^{+0.3}_{-0.3}$ &  &  \\
$\sigma_{r2}$ (au) &  &  &  &  & $22.8^{+0.4}_{-0.4}$ & $24.4^{+0.1}_{-0.1}$ &  &  \\
$R_c$ (au) &  &  &  &  &  &  & $102.4^{+0.3}_{-0.3}$ & $98.0^{+0.3}_{-0.3}$ \\
$\alpha_\mathrm{in}$ &  &  &  &  &  &  & $1.443^{+0.002}_{-0.002}$ & $1.472^{+0.002}_{-0.002}$ \\
$\alpha_\mathrm{out}$ &  &  &  &  &  &  & $-0.89^{+0.02}_{-0.02}$ & $-0.72^{+0.02}_{-0.02}$ \\
$i$ (deg) & 64.5 & 64.5 & $62.94^{+0.02}_{-0.02}$ & $62.93^{+0.02}_{-0.02}$ & $63.30^{+0.02}_{-0.02}$ & $63.28^{+0.02}_{-0.02}$ & $63.30^{+0.02}_{-0.02}$ & $63.29^{+0.02}_{-0.02}$ \\
$\theta$ (deg) & 59.0 & 59.0 & $58.81^{+0.02}_{-0.02}$ & $58.81^{+0.02}_{-0.02}$ & $58.82^{+0.02}_{-0.02}$ & $58.84^{+0.02}_{-0.02}$ & $58.82^{+0.02}_{-0.02}$ & $58.83^{+0.02}_{-0.02}$ \\
$h$ & 0.1 & 0.1 & $0.1045^{+0.0009}_{-0.0009}$ & $0.1064^{+0.0009}_{-0.0009}$ & $0.0223^{+0.0007}_{-0.0006}$ & $0.0226^{+0.0008}_{-0.0008}$ & $0.0222^{+0.0006}_{-0.0006}$ & $0.0227^{+0.0006}_{-0.0006}$ \\
$\Delta_\mathrm{RA}$ (au) &  &  &  & $-0.325^{+0.006}_{-0.006}$ &  & $-0.614^{+0.009}_{-0.009}$ &  & $-1.07^{+0.01}_{-0.01}$ \\
$\Delta_\mathrm{Dec}$ (au) &  &  &  & $-0.322^{+0.005}_{-0.005}$ &  & $-0.535^{+0.007}_{-0.007}$ &  & $-0.855^{+0.008}_{-0.008}$ \\
\hline
\end{tabular}
\end{sidewaystable}

\subsubsection{Two Gaussians}
\label{sec:2gauss}
To account for the broad ``wings'' on the outer edge that the Gaussian model residuals suggest, we fitted a more complex model with two Gaussian radial components, with the optical depth given by:
\begin{equation}
    \tau(r) = A_1 \exp{\left[ - \frac{(r - \mu_{r1})^2}{2 \sigma_{r1}^2} \right]} + A_2 \exp{\left[ - \frac{(r - \mu_{r2})^2}{2 \sigma_{r2}^2} \right]}.
\end{equation}

The best-fit double-Gaussian model is shown in Figs.~\ref{fig:models25}e and f, with the corresponding fitted parameters displayed in Table~\ref{tab:params25}. This model shows less prominent residual disk structures than the ring-like residuals seen in the single-Gaussian models. The residual features are more similar to the star-centred \texttt{rave} model, although at a lower amplitude and with a lower $\chi_\nu^2$. 

To reduce the asymmetry of the residuals, we also tested a double-Gaussian model with a stellocentric offset, which indeed further improves the fit. The best-fit star--disk offset is oriented along the same direction as the single-Gaussian model, with a slightly larger amplitude. 

\subsubsection{Power law edges}
To test a simple model capable of producing inner and outer edges with different slopes, we parameterised the optical depth of the edges as power laws with different exponents. We used the double-power-law parametrisation from \citet{Han2025}, given by:
\begin{equation}
    \tau(r) = \frac{A_1}{\sqrt{\left( r / R_c \right)^{-2 \alpha_\mathrm{in}} + \left( r / R_c \right)^{-2 \alpha_\mathrm{out}}}},
\end{equation}
which introduces a smooth transition where the two edges meet near the peak. We fitted a model both with and without a stellocentric offset. The results are shown in Table~\ref{tab:params25} and visualised in Figs.~\ref{fig:models25}g and h. 

Similar to the previous models, introducing a star--disk offset noticeably improves the symmetry of the residuals and produced the lowest reduced $\chi^2$ among all models that we fitted to the F2550W image. The residual features near the star show an oscillatory pattern between narrow positive and negative annuli, which is likely due to imperfections in PSF subtraction despite the relatively high PSF stability of JWST, rather than real dust features. Note that while the reduced $\chi^2$ of 1.6 is still greater than 1, this includes the effects of imperfect PSF subtraction near the star. We adopt this power-law model with a disk offset as the best-fitting parametric model at 25\,$\mu$m.

\subsection{Parametric modelling at 15\,$\mu$m}

\begin{figure*}
    \centering
    \includegraphics[width=1.0\linewidth]{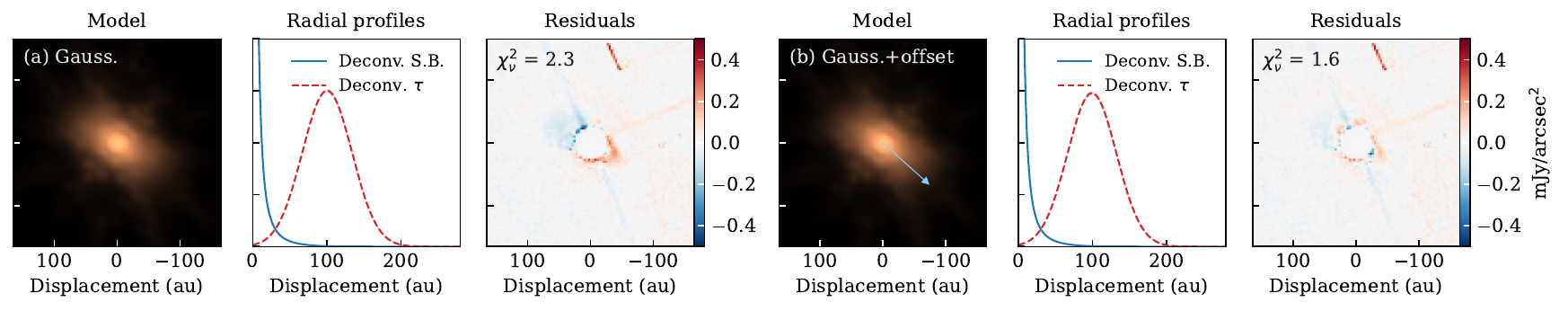}
    \caption{Gaussian model fitted to the MIRI F1500W image of $\gamma$~Oph. The panels are the same as those described in Fig.~\ref{fig:models25} for the F2550W modelling and the model image should be compared with the bottom-right panel of Fig.~\ref{fig:obs}. The central region has been masked as it is dominated by PSF subtraction artefacts and was not used for model fitting. The per-pixel RMS noise measured from the image background is 0.02\,mJy/arcsec$^2$. }
    \label{fig:models15}
\end{figure*}

\begin{table}
    \centering
    \caption{Assumed (*) and fitted values for parametric models at 15\,$\mu$m. }
    \label{tab:params15}
\begin{tabular}{lll}
\hline \hline
 & Gaussian &  \\
Parameter & No offset & Offset \\
\hline
$F_*$ (mJy) & $614.1^{+0.2}_{-0.2}$ & $615.2^{+0.2}_{-0.2}$ \\
$A_1$ & $2.19^{+0.02}_{-0.02} \times 10^{-4}$ & $2.16^{+0.02}_{-0.02} \times 10^{-4}$ \\
$\mu_{r1}$ (au) & $100.4^{+0.5}_{-0.5}$ & $99.4^{+0.5}_{-0.5}$ \\
$\sigma_{r1}$ (au) & $33.4^{+0.2}_{-0.2}$ & $33.1^{+0.2}_{-0.2}$ \\
$i$ (deg) & 63 & 63 \\
$\theta$ (deg) & 59 & 59 \\
$h$ & 0.11 & 0.11 \\
$\Delta_\mathrm{RA}$ (au) &  & $-0.66^{+0.01}_{-0.01}$ \\
$\Delta_\mathrm{Dec}$ (au) &  & $-0.59^{+0.01}_{-0.01}$ \\
\hline
\end{tabular}
\end{table}

The significantly more compact emission and higher star--disk contrast at 15\,$\mu$m compared to 25\,$\mu$m makes it difficult to fit a meaningful nonparametric model at 15\,$\mu$m that attempts to deconvolve marginally resolved disk emission from the bright PSF core. Even given the PSF stability of JWST, PSF uncertainties produce PSF subtraction artefacts that make even parametric fitting challenging. Nonetheless, we attempted to fit a simple model with a restricted set of free parameters to extract basic structural information about the disk at this wavelength. 

We attempted to fit a Gaussian model analogous to that fitted to the F2550W image as described in Section~\ref{sec:gaussian}, however the model did not converge to a single solution. The disk emission is compact and faint emission relative to imaging artefacts, and parameters such as $i$, $\theta$ and $h$ become degenerate with the radial profile.
We therefore restricted the model by fixing the $i$, $\theta$ and $h$ parameters based on the best-fit values of the 25\,$\mu$m Gaussian disk model, fitting only to $F_*$, $A_1$, $\mu_{r1}$ and $\sigma_{r1}$ (Eq.~\eqref{eq:1gauss}). We also masked out the central 33\,au (i.e., 10 MIRI pixels) where PSF subtraction artefacts are most prominent. 
The best-fit parameters of this model are shown in Table~\ref{tab:params15}, and the fitted model image, radial profile and residuals are shown in Fig.~\ref{fig:models15}a. The central radius of the Gaussian models are similar at 25 and 15\,$\mu$m, although at 15\,$\mu$m the best-fit Gaussian is slightly narrower. Note that the small uncertainties reported in the table do not fully account for systematic uncertainties associated with PSF subtraction and are conditioned on the assumptions of the viewing geometry based on the 25\,$\mu$m imaging.

This restricted Gaussian model is broadly able to account for the disk emission at 15\,$\mu$m, though asymmetric residual structures exist. These residuals are significantly reduced with the introduction of a star--disk offset as shown in Fig.~\ref{fig:models15}b, with the best-fit offset comparable to those found at 25\,$\mu$m. While many of the residual features are imaging artefacts, there appear to be subtle residual emission that align with the orientation of the disk.

To obtain a rough estimate of the blackbody radius of any unresolved inner dust, we estimated the unresolved excess flux ratio to be $\sim$0.77 between 15 and 25.5\,$\mu$m based on the best-fit model at each wavelength (Gaussian model with an offset at 15\,$\mu$m and power law model with an offset at 25.5\,$\mu$m). The corresponding colour temperature assuming a blackbody is $\sim$220\,K, for which $B_\nu$ peaks close to 23\,$\mu$m. The corresponding radius with such a blackbody equilibrium temperature is $\sim$8\,au. This is smaller than the PSF FWHM at either wavelength (14\,au at 15\,$\mu$m and 24\,au at 25.5\,$\mu$m).

\section{Discussion} \label{sec:discussion}

\begin{figure}
    \centering
    \includegraphics[width=1.0\linewidth]{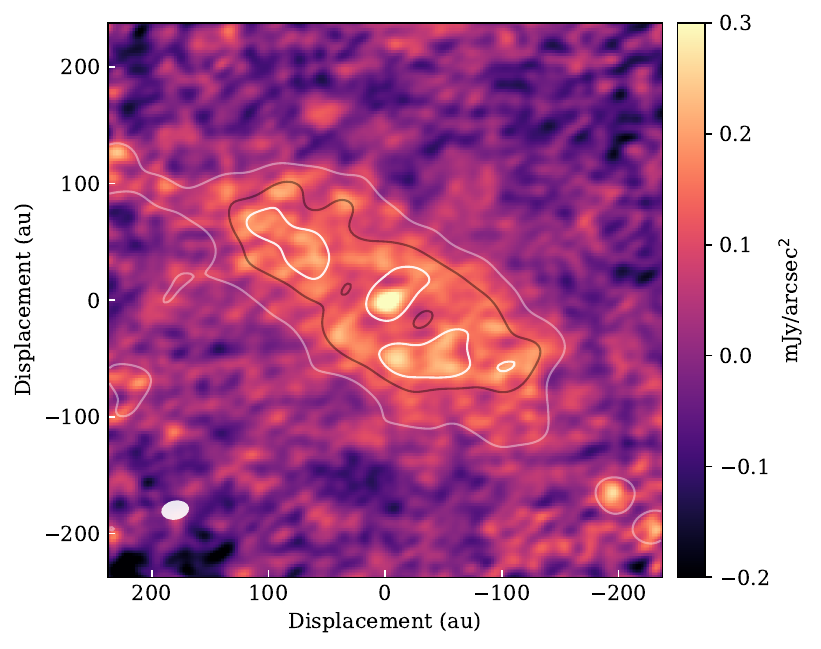}
    \caption{ALMA Band 7 observations presented in \citet{Marino2025}, imaged with \texttt{clean} using a robust parameter of 2.0 and with primary beam correction applied. The beam FWHM is indicated with a white ellipse in the bottom-left corner. Stellar emission has not been subtracted from the image. Contours are drawn at 0.06, 0.12 and 0.15\,mJy\,arcsec$^{-2}$ based on the image smoothed with a 1\,arcsec UV taper, in which the noise level is 0.03\,mJy\,arcsec$^{-2}$.}
    \label{fig:alma}
\end{figure}

\begin{figure*} 
    \centering
    \includegraphics[width=0.8\linewidth]{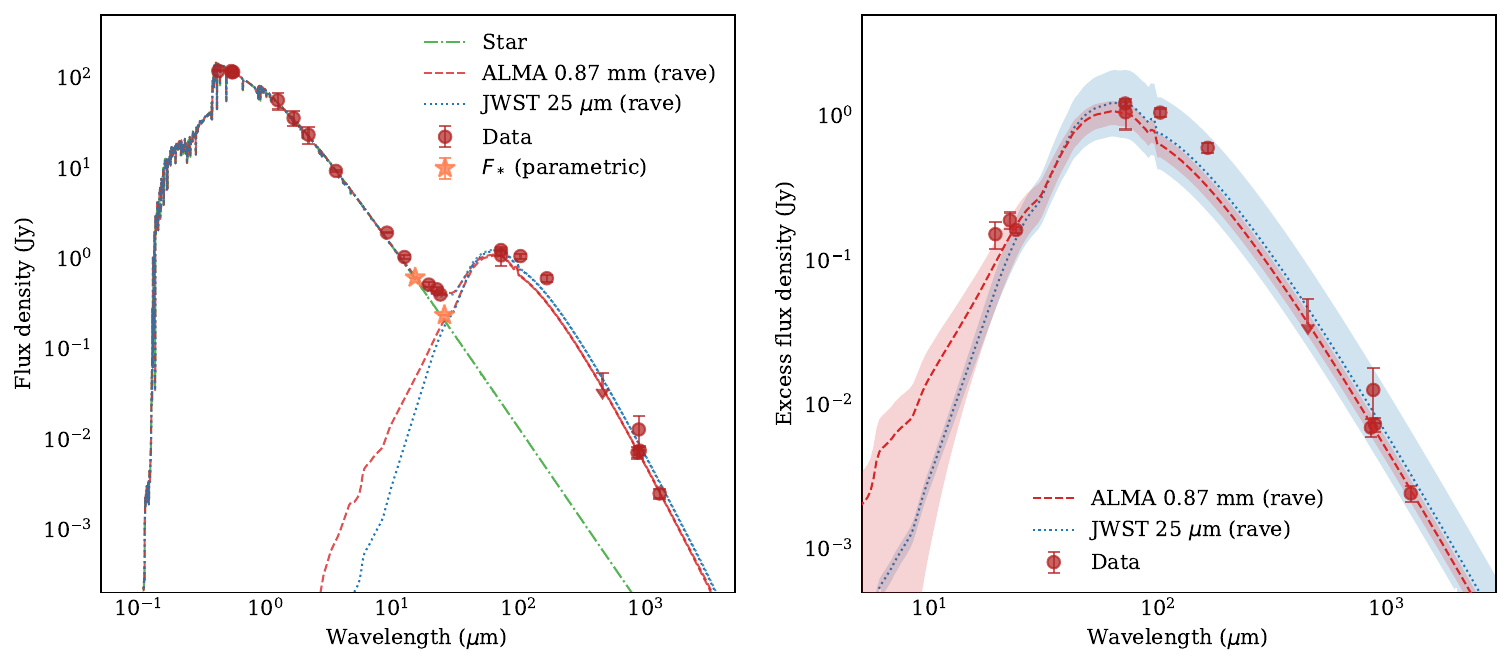}
    \caption{The SED of $\gamma$~Oph. The photometric data points were collected from \textit{Hipparcos} \citep{Hipparcos1997, Hog2000}, the \citet{Mermilliod2006} UBV catalogue, 2MASS \citep{Cutri2003}, \textit{Gaia} \citep{Gaia2018}, AKARI \citep{Ishihara2010}, WISE \citep{Wright2010}, \textit{Spitzer} \citep{Su2006, Chen2014, SEIP2020}, \textit{Herschel} \citep{Pilbratt2010, Pawellek2014, Moor2015}, JCMT \citep{Holland2017}, APEX \citep{Nilsson2010} and ALMA \citep{Matra2025, Marino2025}. The $F_*$ points represent the fitted central component flux under then double-power-law-with-offset model at 25\,$\mu$m and the Gaussian-with-offset model at 15\,$\mu$m. The stellar spectrum corresponds to a Phoenix stellar model \citep{Husser2013} with $T_\mathrm{eff} = 9050$\,K and $\log(g) = 3.9$. The SED models of the disk are overplotted, including those derived using \texttt{rave} on the ALMA 0.87\,mm and JWST 25\,$\mu$m images, both with the same grain composition and size distribution as described in Section~\ref{sec:parametric}. }
    \label{fig:sed}
\end{figure*}

We compare $\gamma$~Oph observations across wavelength in this section and begin by pointing out two pieces of information referenced throughout the discussion. Firstly, ALMA observations of $\gamma$~Oph have been taken as part of the REASONS sample \citep{Matra2025}, and more recently as part of the ARKS programme \citep{Marino2025}. Here we focus on the latter, which were carried out in Band 7 (0.87\,~mm) at higher resolution (15\,au) than the REASONS observations. A \texttt{clean} image of the ALMA observations is shown in Fig.~\ref{fig:alma}. Secondly, we assembled the SED of $\gamma$~Oph, which is shown in Fig.~\ref{fig:sed}. Specific data points and stellar models used are described in the legend of the SED.

\subsection{Mid-infrared disk structure}
\label{sec:discussion_structure}

The modelling approaches in Section~\ref{sec:results} together support four main conclusions on the disk morphology in the mid-infrared. 

Firstly, the outer edge of the disk exhibits a smooth drop-off with no evidence for radial gaps or additional rings based on the MIRI dataset. 
The optical depth (or surface density) profile appears to be shallower at the outer edge than the inner edge, as suggested by the better fit of the power-law-edges model compared to the radially symmetric Gaussian model. 
The deconvolved radial profile obtained with \texttt{rave} drops to 0 upon reaching 250\,au, which is consistent with the furthest emission detected in the azimuthally averaged radial profile measured directly from the (PSF-convolved) observations (see Appendix~\ref{sec:appendix}). 
This MIRI 25.5\,$\mu$m disk extent is therefore similar to the 260\,au lower limit of the disk radius inferred from \textit{Spitzer} observations at 24\,$\mu$m \citep{Su2008}. Note that the \textit{Spitzer} observations reached a per-pixel RMS noise of 0.9\,$\mu$Jy/arcsec$^2$ for its 2\farcs55 MIPS24 pixels. The MIRI F2550W RMS noise achieved here corresponds to 3\,$\mu$Jy/arcsec$^2$ per \textit{Spitzer} MIPS24 pixel (binning MIRI pixels and adding noise in quadrature), which is comparable but slightly noisier than the \textit{Spitzer} observations accounting for differences in the central wavelength. At 15\,$\mu$m, the outermost dust detected is at approximately 150\,au from the star based on the azimuthally averaged radial profile of the (convolved) observations (Fig.~\ref{fig:halo}). 

Secondly, the MIRI images do not directly resolve an inner disk cavity in surface brightness, though the presence of an inner edge in the underlying optical depth/surface density is inferred from the surface brightness profiles derived. 
The disk and central component flux (i.e., stellar flux + any unresolved inner disk emission) from nonparametric modelling are plotted on the SED of the system shown in Fig.~\ref{fig:sed}. 
The relative consistency between the predicted stellar spectrum and the fitted central point source flux suggests that despite an inner edge not being resolved, the level of emission contributed by any unresolved inner dust is low compared to the stellar flux. 

Thirdly, the disk appears to be asymmetric, and can be modelled by an axisymmetric disk with a geometric centre offset from the star. It is possible that the asymmetry could alternatively be modelled by a geometrically axisymmetric disk centred on the star but with azimuthally asymmetric emission, or a combination of the two scenarios in a more physically realistic model such as due to a nonzero eccentricity \citep{Wyatt2005b, Lovell2021, Lynch2022, Lovell2023}, but these models are not tested in this study. Taking the stellocentric offset values from the power-law-edges model, which offered the best fit, the projected offset from the star is $1.370 \pm 0.009$\,au along $231.4 \pm 0.4^\circ$ counterclockwise from north (for comparison, the major axis position angle is $58.7^\circ$), corresponding to a $1.41 \pm 0.01$\,au offset along $16.2 \pm 0.8^\circ$ clockwise from the ascending (western) node when deprojected given the fitted inclination and position angle of the disk. Assuming a semimajor axis of 50\,au where the asymmetry in the axisymmetric model residuals peak (see Fig.~\ref{fig:models25}), such an offset corresponds to an eccentricity of $\sim$0.03. Note that the semimajor axis assumed is a factor of 2 smaller than the peak radius in optical depth, and is used only as an approximate estimate for the eccentricity here. ALMA observations of $\gamma$~Oph did not identify any asymmetries in the disk \citep{Marino2025, Lovell2025}. However, an asymmetry at the same level as that seen by JWST would not have been detectable in the ALMA images, as the disk at peak emission is only detected at 6$\sigma$ in these observations. Future deeper ALMA observations may be able to search for evidence of the subtle asymmetry that we find here. 

Finally, based on the models with an axisymmetric disk with a stellocentric offset, the vertical aspect ratio seen by MIRI appears to be large ($\sim$0.1) based on nonparametric modelling, however the values of $h$ returned by parametric modelling ranges widely between 0.02 and 0.11 depending on the choice of parametrisation, suggesting that the aspect ratio is in practice not tightly constrained. 
For comparison, the best-fit aspect ratios measured with ALMA generally range between 0.13 and 0.16 (\citealp{Terrill2023, Matra2025, Han2025, Zawadzki2025}), and this range expands to be 0.08 to 0.18 when considering the uncertainties of individual measurements. An exception to this general range is an upper limit of 0.08 placed using the \texttt{frank} code \citep{Jennings2020, Terrill2023} based on the ARKS dataset \citep{Marino2025}.

If the mid-infrared aspect ratio were to be 0.13, we estimate that at a radius of 100\,au where the disk's optical depth peaks (see Section~\ref{sec:almased} and Fig.~\ref{fig:optical_depth}), this corresponds to a deprojected vertical standard deviation of 0.4$^{\prime\prime}$, or a deprojected vertical FWHM approximately 30\% larger than the PSF FWHM. An aspect ratio of 0.13 or larger would then be measurable in theory given the high sensitivity of the MIRI observations, even when taking into account projection effects given the $\sim$60$^\circ$ inclination of the disk. The fact that all MIRI models prefer an aspect ratio below this value could suggest that the MIRI aspect ratio is lower than, or near the lower end of, the likely scale height range measured from ALMA. However, the MIRI models prefer a lower (more face-on) inclination than the ALMA models, which could bias the MIRI models towards a lower scale height, so without further joint modelling, it is difficult to draw a robust comparison. A MIRI scale height smaller than that found with ALMA would be consistent with that suggested in AU~Mic by comparing ALMA observations at different bands \citep{Vizgan2022}, which is consistent with predictions of smaller inclination dispersions among smaller grains from collisional damping \citep{Pan2012}. However, studies have also suggested that the efficiency of collisional damping and its effect on the scale height could be limited \citep{Marshall2023, Jankovic2024, Marshall2025}.

\subsection{Planet constraints}
\label{sec:planets}

\begin{figure}
    \centering
    \includegraphics[width=1.0\linewidth]{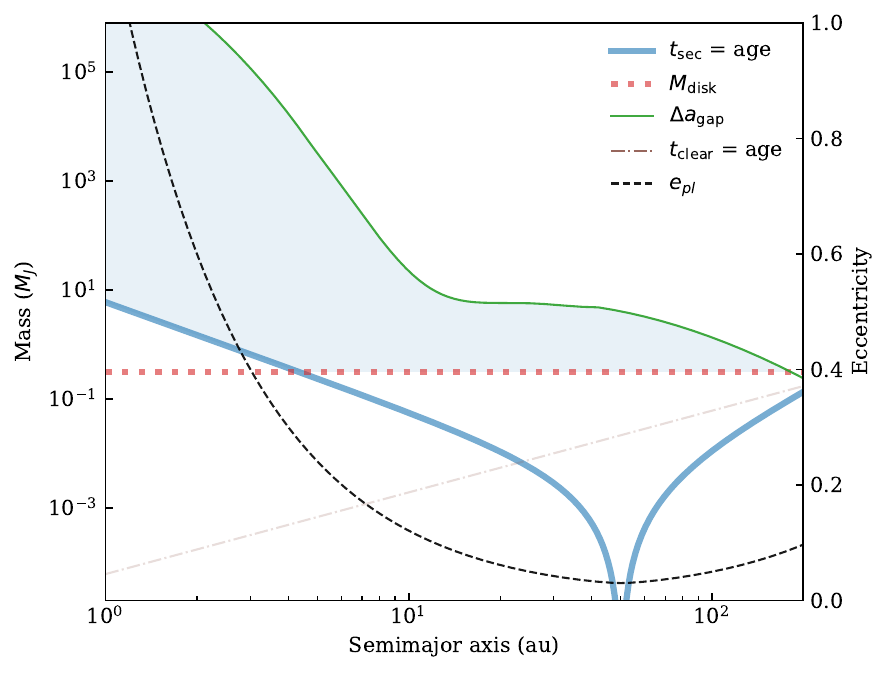}
    \caption{Constraints on the mass (left axis), eccentricity (right axis) and semimajor axis of a hypothetical planet required to explain the stellocentric offset of the disk. Lower bounds in the mass--semimajor axis parameter space are plotted with thick lines and upper bounds with thin lines. Origins of lower bounds include a sufficiently fast secular perturbation timescale such that the disk becomes eccentric within the age of the system ($t_\mathrm{sec} = \mathrm{age}$), and a sufficiently massive planet to overcome the disk's gravity ($M_\mathrm{disk}$). Upper bounds are set by the lack of a radial gap detected in MIRI observations, either due to the gap being too narrow ($\Delta a_\mathrm{gap}$), or due to a gap having not yet been carved by the planet within the age of the system ($t_\mathrm{clear} = \mathrm{age}$). The shaded region simultaneously satisfies all such constraints described.
    The eccentricity required of the planet to excite a forced eccentricity of 0.03 in the disk at 50\,au is overplotted on a separate vertical axis with a black dashed line ($e$). }
    \label{fig:planets}
\end{figure}

Since the presence of sufficiently massive planets in the disk is expected to carve resolvable gaps within the age of the system, the lack of gaps resolved by MIRI places upper limits on the mass of any planets that could reside within the disk. 

A planet could hide within the disk if any radial gap that it carves is too narrow to be detected. For a planet on a circular orbit, we estimated the width of gaps carved by the chaotic overlap of first order mean-motion resonances as:
\begin{equation}
\label{eq:chaoticzone}
\Delta a \approx 3 a_\mathrm{pl} \left( \frac{M_\mathrm{pl}} {M_\mathrm{*}} \right)^{2/7} ,
\end{equation}
i.e., the chaotic zone combined from both sides of the planet \citep{Wisdom1980, Morrison2015}, where $a_\mathrm{pl}$ is the semimajor axis of the planet, $M_\mathrm{pl}$ is the mass of the planet and $M_\mathrm{*}$ is the mass of the star, which is inferred to be 2.11\,$M_\odot$ for $\gamma$~Oph \citep{Marino2025}. 

To test the widest gap that could hide in the disk while remaining undetected, we simulated observations using the \texttt{rave} radial profile derived in Section~\ref{sec:results} and added a Gaussian gap that is fully cleared at the local minimum within the gap. 
We determined the widest gap such that the median profile returned by \texttt{rave} detects a change in concavity and that the range of possible models rejects the possibility of a smooth radial profile without abrupt slope changes at the location of the injected gap.
The maximum undetectable gap width depends on the radius at which the gap is centred, which we find ranges from a quarter of a PSF FWHM for a gap at 10\,au, to half at 25\,au, to a full PSF above 75\,au (due to the emission being too faint) or less than 5\,au (due to the PSF core). 
For reference, the PSF FWHM is 24\,au. We interpolated between sample points of the maximum undetectable gap width as a function of semimajor axis and plot the smoothed constraints from Eq.~\eqref{eq:chaoticzone} in Fig.~\ref{fig:planets}. The deviation of the curve from a straight line reflects the higher sensitivity to fully cleared gaps at intermediate semimajor axes. 

A second way in which a planet in the disk could avoid leaving a detectable gap is if the gap-clearing timescale is longer than the age of the system. Following \cite{Shannon2016}, the timescale required to clear the gap can be estimated as: 
\begin{equation}
t_\mathrm{clear} \approx 4 \, \mathrm{Myr} \, (M_\mathrm{pl}/M_\oplus)^{-1} \, (a_\mathrm{pl}/\mathrm{au})^{3/2} \,  (M_\mathrm{*} / M_\odot)^{1/2} .
\end{equation} 
This is plotted in Fig.~\ref{fig:planets} assuming an age of 300\,Myr, which shows that any planet capable of eventually carving a gap wide enough for detection can do so within a short enough timescale, such that the effective upper limit is set by the $\Delta a$ rather than $t_\mathrm{clear}$ constraint. 

If the stellocentric offset of the disk is due to a forced eccentricity of 0.03 (see Section~\ref{sec:discussion_structure}) that is imposed by an eccentric planet, the secular timescale is required to be shorter than the age of the system, placing a lower limit on the perturbing planet's mass. We estimated the secular timescale \citep{Murray1999, Wyatt2005, Sefilian2021} via
\begin{equation}
    t_\mathrm{sec} = 2 \pi \left[\frac{1}{4} n \frac{M_\mathrm{pl}}{M_*} \alpha \bar{\alpha} b_{3/2}^{1} (\alpha) \right]^{-1},
\end{equation}
where $\alpha = \min(a_\mathrm{disk}, a_\mathrm{pl}) / \max(a_\mathrm{disk}, a_\mathrm{pl})$, $n = \sqrt{G (M_* + M_\mathrm{pl})/a_\mathrm{disk}^3} $, $\bar{\alpha} = \min(1, a_\mathrm{disk}/a_\mathrm{pl})$, $b_s^{m}(\alpha)$ are the standard Laplace coefficients, and we set $a_\mathrm{disk} = 50$\,au where the surface brightness asymmetry peaks. 
This condition is plotted in Fig.~\ref{fig:planets}. 

Furthermore, the planet would also need to be larger than the disk mass to dominate over the disk's self-gravity \citep{Sefilian2024, Sefilian2025}, otherwise secular resonances would have led to gapped structures which we don't find \citep{Sefilian2021, Sefilian2023}\footnote{The threshold planet mass to overcome the disk's self-gravity in principle depends on the planet's semimajor axis and the disk's surface density profile \citep{Sefilian2024}, but we ignore this dependence here and only show an indicative threshold given that the disk mass itself is highly uncertain.}. There is considerable uncertainty surrounding the total mass of debris disks, although extrapolations of the collisional cascade from dust to planetesimals suggest values above tens of $M_\oplus$ \citep{Krivov2021}. In Fig.~\ref{fig:planets}, we plot an indicative disk mass of 100\,$M_\oplus$ based on the lower end of disk masses interpolated from a collisional cascade that is also within the estimated gravitational stability limit during planet formation \citep{Krivov2021}.

In addition to the mass, we can also constrain the eccentricity required of a perturbing planet. Assuming a disk forced eccentricity of $e_\mathrm{f} = 0.03$ as before at a peak surface brightness asymmetry of $a_\mathrm{disk} = 50$\,au, we estimated the required perturber's eccentricity via
\begin{equation}
    \label{eq:e}
    e_\mathrm{pl} = \frac{ b_{3/2}^{1} (\alpha) }{ b_{3/2}^{2} (\alpha) } \, e_\mathrm{f}
\end{equation}
\citep{Wyatt1999, Sefilian2021}. This eccentricity condition is plotted in Fig.~\ref{fig:planets} on a separate vertical axis from the mass axis. Note that Eq.~\eqref{eq:e} is valid for massless disks; the inclusion of disk gravity modifies the forced eccentricity, both in terms of its amplitude and radial dependence \citep{Sefilian2024}. 

The intersection of the MIRI mass--semimajor axis constraints is shown with the shaded region in Fig.~\ref{fig:planets}. The strongest constraints are primarily set by the maximum hidden gap width, which permit giant planets up to 10\,$M_\mathrm{Jup}$ to reside outside 10\,au, with an eccentricity required to be a few percent to induce the disk asymmetry due to secular perturbation. Such a scenario is possible even if the disk is as massive as assumed. 
If the perturber were to reside within 10\,au, the mass threshold becomes significantly larger and a brown dwarf or close binary companion could remain hidden while perturbing the disk. 
Such a perturber is compatible with proper motion anomalies of $\gamma$~Oph that have led to a prediction of a 9--100\,$M_\mathrm{Jup}$ companion at 3--25\,au \citep{Kervella2022, Milli2025}, which overlaps with the allowable mass range based on MIRI for semimajor axes under 10\,au. 

\subsection{Disk structure across wavelength} \label{sec:almased}

\begin{figure*}
    \centering
    \includegraphics[width=1.0\linewidth]{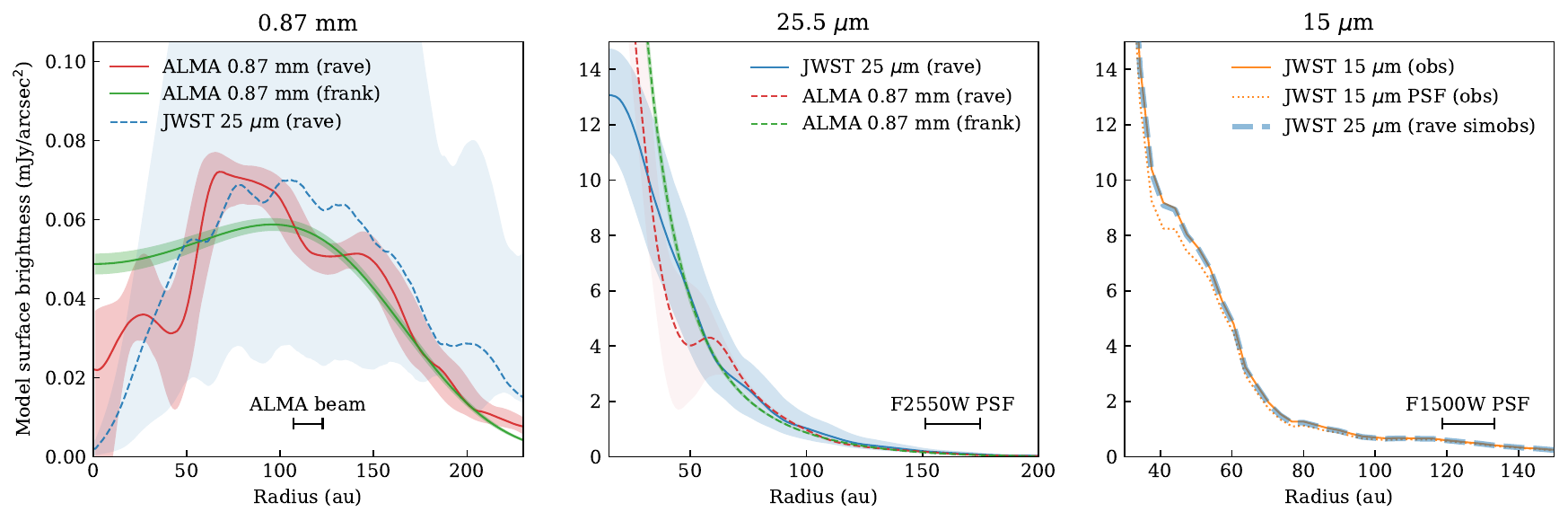}
    \caption{Radial surface brightness profiles at the ALMA Band 7 (left), MIRI F2550W (middle) and MIRI F1500W (right) wavelengths. Solid lines indicate deconvolved and deprojected disk-only (i.e., star-subtracted) profiles fitted directly to observations at the corresponding wavelength, with an exception being the F1500W profile, which is the PSF-convolved profile measured directly by azimuthally averaging the (star-included) observations given the difficulty in robustly deconvolving the compact disk emission. Dashed profiles indicate those converted from observations at other wavelengths assuming a spatially uniform grain size distribution and grain composition as described in Section~\ref{sec:parametric}. In the right panel which displays 15\,$\mu$m profiles, the dotted line indicates the azimuthally averaged profile of the PSF, which is scaled to a flux density of 0.64\,Jy. The dash-dotted line is the sum of the azimuthally-averaged PSF-convolved disk model and the aforementioned dotted line corresponding to the PSF. }
    \label{fig:radial_profiles}
\end{figure*}

The aim of this section is to compare the radial structure of the disk across wavelengths, particularly between JWST and ALMA. 
Specifically, we compare the MIRI radial profile fitted nonparametrically with \texttt{rave} with those fitted using the same method to the ARKS ALMA observations (Fig.~\ref{fig:alma}). We also include the ALMA nonparametric profile fitted with the \texttt{frank} method \citep{Jennings2020, Terrill2023} in this comparison, which fits directly to ALMA visibilities (rather than the \texttt{CLEAN} image as in the case of \texttt{rave}). The ALMA profiles are fitted to star-subtracted observations assuming a stellar flux of 0.161\,mJy. Details of the ALMA modelling are described in \citet{Han2025b}.

\subsubsection{Collisional cascade}
Fig.~\ref{fig:radial_profiles} displays the surface brightness profiles at 25\,$\mu$m and 0.87\,mm. 
The ALMA observations resolve an inner edge and central cavity, although the exact distribution at the inner edge is somewhat uncertain, as demonstrated by the different radial profile shapes recovered by \texttt{rave} and \texttt{frank}. It should be pointed out that the ``wiggles'' in the ALMA \texttt{rave} profile likely reflect noise in the image, as the ALMA observations reach a lower S/N than the MIRI observations. Note that the ALMA beam size is slightly smaller than the MIRI F2550W PSF (by $\sim1/3$), as indicated in Fig.~\ref{fig:models25}. The ALMA position angle is measured to be $57.6 \pm 1.6^\circ$ and the inclination $66.1 \pm 1.5^\circ$ by fitting a Gaussian ring model \citep{Marino2025}. This is consistent with the MIRI position angle of $58.83 \pm 0.02^\circ$ and slightly larger than the inclination of $63.29 \pm 0.02^\circ$ determined from the power-law-edges-with-offset model.  

To compare these observations, we convert the surface brightness profiles between wavelength assuming the same optical properties as those described in Section~\ref{sec:parametric}. We find that a steady-state collisional cascade with $\gamma = 3.5$ \citep{Dohnanyi1969, Wyatt2008, Hughes2018} and a minimum grain size of 15\,$\mu$m is consistent with the 25\,$\mu$m and 0.87\,mm surface brightness profiles that we observe. The surface brightness profile converted between wavelength assuming a common underlying surface density profile are overplotted in Fig.~\ref{fig:radial_profiles}. No vertical scaling is applied between profiles within each panel. While deviations exist at $<$50\,au in the 25\,$\mu$m panel, this is likely a reflection of degeneracies between disk emission at small radii and any unresolved central component, since even a small nonzero flux in the ALMA radial profile becomes significantly magnified when converted to 25\,$\mu$m. Similarly, the large uncertainties of the MIRI 25\,$\mu$m profile in the 0.87\,mm panel reflects the large fractional uncertainties in the region with low-amplitude emission in the MIRI observations. 

While the deconvolved radial profile is difficult to determine at 15\,$\mu$m due to the compact and faint emission, we converted the 25\,$\mu$m radial profile to 15\,$\mu$m assuming the same optical properties as described above. We convolved the resulting model image with the F1500W PSF to derive the azimuthally averaged surface brightness profile, which we find is consistent with the same quantity measured from the observations. 

Furthermore, we modelled the SED of the disk assuming either the 25\,$\mu$m or 0.87\,mm \texttt{rave} profiles, as shown in Fig.~\ref{fig:sed}, which is broadly consistent with available infrared to mm photometry. The multi-wavelength analysis therefore suggests that despite the large radial extent of $\gamma$~Oph, the disk can be broadly described by a common grain size distribution throughout the disk when observed at mid-infrared and mm wavelengths. Such a single power law size distribution is expected in an idealised disk of planetesimals as part of a steady-state collisional cascade, which appears to have offered a reasonable approximation in this disk. In theory, a more realistic size distribution could depart from a power law by exhibiting a wavy pattern, especially affecting intermediate grain sizes that are not probed by our observations \citep{Thebault2007}. Furthermore, small grains produced in the inner disk could also populate the outer disk. Radially extended halos have been linked to small grains displaced from the parent planetesimal belt due to processes such as radiation pressure in debris disks systems such as Vega \citep{Su2005}, $\beta$~Pic \citep{Ballering2016} and HD\,32297 \citep{Olofsson2022b}. 

To further visualise the consistency of the observations with a uniform size distribution, we plot in Fig.~\ref{fig:optical_depth} the geometric optical depth profiles derived from both JWST and ALMA under the steady-state grain size distribution with $\gamma=3.5$. These nonparametric profiles show a high degree of consistency without the need to introduce any vertical scaling. 

\subsubsection{Inner and outer edges}
The most simple parametric model that we fitted to the 25.5\,$\mu$m image was one with a Gaussian radial profile, which naturally assumes inner and outer edges that are symmetric about the radial peak. In the version with a stellocentric offset, we find that the optical depth peaks at 100\,au, with a standard deviation of 39\,au (Table~\ref{tab:params25}). In comparison, in ALMA observations, the best-fit Gaussian is centred at $121 \pm 4$\,au, with a standard deviation of $57 \pm 3$\,au \citep{Marino2025}. The mid-IR optical depth under this parametrisation therefore appears to be slightly interior and narrower than that found from ALMA.

We also explored models in which the inner and outer edges are asymmetric. Based on the double-power-law-with-offset model, in which the inner and outer edges are parametrised by power laws, the steepness of the inner edge is fitted by a power law index of 1.5, and the outer edge a power law index of -0.7 (Table~\ref{tab:params25}). In comparison, the ALMA inner edge under the same parametrisation finds the power law index to be $1.0 \pm 0.1$ for the inner edge and $-3.7 \pm 0.8$ for the outer edge \citep{Han2025b}. The inner edge inferred from the mid-IR observations under this power-law parametrisation is therefore slightly steeper than that inferred from mm-wavelength observations, whereas the outer edge is significantly shallower, overall exhibiting a stronger degree of radial asymmetry.

For a disk evolving under steady-state collisions without truncating planets, the inner edge is expected to scale as $r^{7/3}$ in optical depth \citep{Kennedy2010}. Such a profile is consistent with the inner edge of JWST and ALMA \texttt{rave} profiles within uncertainties, as shown in Fig.~\ref{fig:optical_depth}. The power-law inner edge fitted is shallower than this steepness. This difference could in part be due to the double-power-law parametrisation attempting to recover a wide peak that is favoured by the nonparametric model, but which this parametrisation cannot naturally produce.

Assuming that the initial surface density profile of planetesimals formed from the protoplanetary disk scales with the minimum mass solar nebula \citep{Hayashi1981}, the age of the system and the size of the largest bodies within the disk determine the radius up to which the disk has been collisionally eroded into the $r^{7/3}$ slope \citep{ImazBlanco2023}. The \texttt{rave} profile exhibits a turnover radius at 100\,au from the star, beyond which it deviates from the $r^{7/3}$ slope. Assuming an age of 300\,Myr and emission originating from dust grains with a diameter of 15\,$\mu$m, Eq.~(21) in the pre-stirred model of \citet{ImazBlanco2023} predicts the largest bodies in the disk to be approximately 150\,m in diameter, in order for the radial profile within 100\,au to have eroded from MMSN-like to $r^{7/3}$-like.

In comparison, this is significantly smaller than the 2000-km stirring bodies which are inferred to be present in the disk based on the mm vertical height \citep{Zawadzki2025}, assuming stirring up to heights set by the stirrers' escape velocities while also taking into account collisional damping and dynamical friction \citep{Pan2012}. However, the stirring bodies do not have to be part of the cascade, so they can be larger than the 150\,m inferred from the \citet{ImazBlanco2023} calculation. Constraints on those stirrers should also take into account the total mass in that population and the time over which stirring has been ongoing \citep{Ida1993}, so stirrers larger than 2000\,km are also possible for a sufficient total mass in the stirrer population (Jankovic et al. in prep), as is the possibility of the disk containing no bodies larger than 150\,m, such as if its scale height was somehow imprinted during the formation process.

While a shallower mid-IR outer edge than ALMA is hinted at by the double-power-law parametrisations, the nonparametric modelling which do not explicitly enforce noise and functional form assumption return large uncertainties. Although we have shown that there exists a steady-state grain size distribution that consistently explains the observations from the nonparametric radial profiles, given the steep emission dropoff as function of radius at MIRI wavelengths, large fractional uncertainties associated with the faint outer disk in the translate to uncertain optical depth profiles in the outer edge, and we are unable to place more stringent constraints on the level of any halo emission in Fig.~\ref{fig:halo} based on the nonparametric \texttt{rave} profile. 

Note also that the minimum grain size of 15\,$\mu$m assumed in our model is slightly smaller than the 17\,$\mu$m radiation pressure blowout size estimated from the optical properties assumed. 
The minimum grain size in debris disks has been found to be on average similar to the expected blowout size around A stars \citep{Pawellek2014}, however more recent analyses using a larger sample with ALMA-measured radii have found that for early A stars, the minimum grain size is on average smaller than the blowout size \citep{Marshall2025b}, as unbound sub-blowout grains could still contribute to dust emission \citep{Thebault2019}. 
Furthermore, small deviations are expected as our models of the optical properties of dust are imperfect, and the truncated power-law grain size distribution is an approximation of the true grain size distribution \citep{Thebault2007}.

\subsubsection{Warm inner dust}
The presence of warm dust interior to ALMA-imaged belts has been a notable finding from debris disk imaging with MIRI so far \citep{Gaspar2023, Su2024}. While $\gamma$~Oph adds to the growing list of debris disks showing well-resolved inner dust emission absent from ALMA observations, the interpretation for $\gamma$~Oph based on our findings appears to differ somewhat from Fomalhaut \citep{Gaspar2023} and Vega \citep{Su2024}. 

In the case of Fomalhaut and Vega, the ALMA observations suggest a relatively sharp inner edge \citep{MacGregor2017Fom, Matra2020}, while the inner emission observed by MIRI suggests an overabundance of small grains in the inner region that cannot be explained by the same size distribution as in the outer belt seen in ALMA observations. The inner dust has instead been modelled by dust dragged inwards from the outer belt by Poynting-Robertson drag in Fomalhaut \citep{Sommer2025} and Vega \citep{Su2024}, which very closely reproduces the observations. 

$\gamma$~Oph differs from Fomalhaut and Vega in that the mm-wavelength belt is broad with a shallow inner edge, and that the ALMA and MIRI radial profiles are in fact consistent with a single grain size distribution, without the need to invoke PR drag. A steady-state collisional model consistent between the 15\,$\mu$m, 25\,$\mu$m and 0.87\,mm is notable as a single power-law grain size model has not been found for other spatially resolved, radially broad disks such as HR\,8799 \citep{Su2009} and $\beta$~Pic \citep{Ballering2016}. $\gamma$~Oph could therefore be an archetype for a radially broad steady-state collisional cascade. 

\begin{figure}
    \centering
    \includegraphics[width=1.0\linewidth]{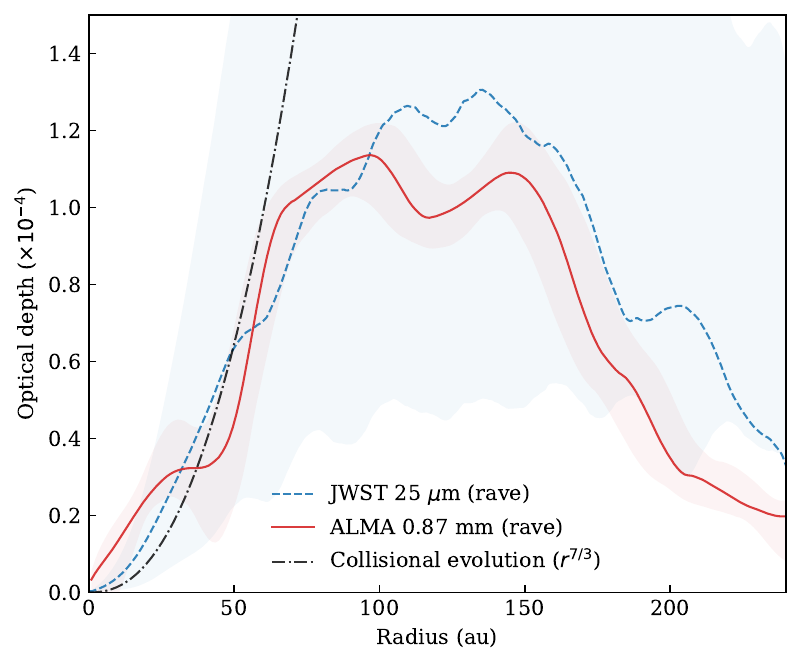}
    \caption{The geometric optical depth profile of the debris disk of $\gamma$~Oph inferred from surface brightness profiles shown in Fig.~\ref{fig:radial_profiles}. These profiles assume a steady-state collisional cascade with a size distribution power-law index of -3.5 between 15\,$\mu$m and 1\,m and a grain composition as described in Section~\ref{sec:parametric}. The dash-dotted black line corresponds to an inner edge optical depth profile proportional to $r^{7/3}$, as expected of a collisionally evolved broad disk \citep{Kennedy2010}. }
    \label{fig:optical_depth}
\end{figure}

While the radial profiles suggest that the smaller (micron-sized) grains are no more abundant in the inner disk (within 50\,au) than expected from a steady-state collisional cascade from larger (mm-sized) grains (Figs.~\ref{fig:radial_profiles} and \ref{fig:optical_depth}), uncertainties in this region are large. Noting that Vega and Fomalhaut are close (the 3rd and 4th nearest A stars, both at 7.7\,pc), whereas $\gamma$~Oph is significantly further (the 59th closest A star at 29.7\,pc), it is possible that proximity may have more clearly revealed any inner disk formed from PR drag in Vega and Fomalhaut but not in $\gamma$~Oph. 

It is possible that some level of PR-drag dust component is hiding as unresolved dust emission. In $\gamma$~Oph at 25.5\,$\mu$m, the fitted central point source flux of between 220 and 240\,mJy (depending on choice of model) is slightly greater than the stellar component flux of 205\,mJy inferred from the SED, suggesting that any unresolved dust is less than approximately 30\,mJy. 
This level of dust is comparable to the flux density of the inner warm dust component in Fomalhaut and Vega when adjusted for their 4 times closer distance. 

Alternatively, the balance between the PR drag timescale and the collisional timescale in $\gamma$~Oph could have caused PR drag to exert a less prominent effect on the dust distribution. The PR drag timescale in $\gamma$~Oph, Fomalhaut and Vega are expected to be similar given their comparable stellar masses and luminosities, however the collisional timescale is likely significantly shorter in $\gamma$~Oph. The collisional timescale can be estimated as $t_\mathrm{col} = 1/(n \sigma v) = 1/(\tau \sigma \Omega_K)$, where $n$ is the dust number density, $\sigma$ is the cross sectional area, $v$ is the relative speed and $\Omega_K$ is the Keplerian angular speed. Given comparable stellar luminosities, a direct comparison of the surface brightness of the the main/outer-belt at $\sim$150\,au in all three disks shows $\gamma$~Oph to be over an order of magnitude brighter when adjusted for distance, and thus likely of a significantly higher optical depth. This is supported by optical depth modelling, which infers a six times lower optical depth in the broad outer belt of Vega \citep{Su2024} compared to the broad belt of $\gamma$~Oph. The outer belt of Fomalhaut is narrow, so dust grains are expected to spend more time in the broad inner dust component immediately interior to the main belt, which has a similar optical depth to Vega at 100\,au \citep{Sommer2025}. As a result, for dust grains of a comparable size and orbit, the effective collisional timescale in $\gamma$~Oph is half an order of magnitude shorter than in Vega and Fomalhaut, which we hypothesise without further modelling could be sufficient to weaken the degree of inward dust transport observed by MIRI \citep{Wyatt2005}.

\section{Conclusions} \label{sec:conclusions}
We imaged the debris disk of $\gamma$~Oph with JWST/MIRI at 15 and 25.5 $\mu$m, finding a smooth and radially broad disk with inner emission extending to the star and outer emission detected out to approximately 250\,au at 25.5\,$\mu$m and 150\,au at 15\,$\mu$m. The radial structure of the disk inferred from the MIRI observations, available ALMA imaging \citep{Marino2025} and the SED of the system are consistent with a steady-state collisional cascade characterised by a single power-law grain size distribution throughout the disk, with a minimum grain size within $\sim$10\% (just below) the radiation pressure blowout size. This broad consistency suggests that the resolved disk is likely populated by a radially broad planetesimal belt with the observed dust population predominantly formed in situ, rather than displaced by radiative forces. 

The disk appears to be azimuthally asymmetric, which can be modelled by a stellocentric offset corresponding to a mild eccentricity of $\sim$0.03. It is plausible for an eccentric giant planet, or a stellar-mass companion within 10\,au, to induce this disk eccentricity within the 300\,Myr age of the system without clearing an observable radial gap. 

Overall, the observations seem to indicate a broad collisional disk rather than one seen to be broad due to PR drag or radiation pressure, making $\gamma$~Oph a thus far rare example of a smooth and broad debris disk without significant evidence for radiative transport of dust grains in the mid-infrared. 
Such a result contrasts with other A stars including Vega \citep{Su2024} and Fomalhaut \citep{Gaspar2023}, for which MIRI imaging revealed an inner disk component caused by PR drag \citep{Sommer2025}. The dense and radially broad nature of the disk in $\gamma$~Oph could contribute to shorter collisional timescales and thus less prominent effects from PR drag. 
These results encourage further observational exploration of the diversity of mid-infrared debris disk structures and the dynamical processes to which they are sensitive.

\begin{acknowledgments}
YH is grateful to Andr\'as G\'asp\'ar for checking our observing sequence.
This work is based on observations made with the NASA/ESA/CSA \textit{James Webb Space Telescope}. The data were obtained from the Mikulski Archive for Space Telescopes at the Space Telescope Science Institute, which is operated by the Association of Universities for Research in Astronomy, Inc., under NASA contract NAS 5-03127 for JWST. 
The data described here may be obtained from the MAST archive at \dataset[https://doi.org/10.17909/kzyx-pn27]{https://doi.org/10.17909/kzyx-pn27}. 
Support for program \#5709 was provided by NASA through a grant from the Space Telescope Science Institute, which is operated by the Association of Universities for Research in Astronomy, Inc., under NASA contract NAS 5–26555.
This paper makes use of the following ALMA data: ADS/JAO.ALMA\# 2022.1.00338.L. 
ALMA is a partnership of ESO (representing its member states), NSF (USA) and NINS (Japan), together with NRC (Canada), MOST and ASIAA (Taiwan), and KASI (Republic of Korea), in cooperation with the Republic of Chile. The Joint ALMA Observatory is operated by ESO, AUI/NRAO and NAOJ. The National Radio Astronomy Observatory is a facility of the National Science Foundation operated under cooperative agreement by Associated Universities, Inc. 
This research used the Canadian Advanced Network For Astronomy Research (CANFAR) operated in partnership by the Canadian Astronomy Data Centre and The Digital Research Alliance of Canada with support from the National Research Council of Canada the Canadian Space Agency, CANARIE and the Canadian Foundation for Innovation.
YH is supported by a Caltech Barr Fellowship. 
This work was supported by the Science and Technology Facilities Council grant UKRI~1198.
AAS is supported by the Heising-Simons Foundation through a 51 Pegasi b Fellowship.
JBL acknowledges the Smithsonian Institute for funding via a Submillimeter Array (SMA) Fellowship.
CdB acknowledges support from the Spanish Ministerio de Ciencia, Innovaci\'on y Universidades (MICIU) and the European Regional Development Fund (ERDF) under reference PID2023-153342NB-I00/10.13039/501100011033, from the Beatriz Galindo Senior Fellowship BG22/00166 funded by the MICIU, and the support from the Universidad de La Laguna (ULL) and the Consejer\'ia de Econom\'ia, Conocimiento y Empleo of the Gobierno de Canarias.
JPM acknowledges research support by the National Science and Technology Council of Taiwan under grant NSTC 112-2112-M-001-032-MY3.
SM acknowledges funding by the Royal Society through a Royal Society University Research Fellowship (URF-R1-221669) and the European Union through the FEED ERC project (grant number 101162711). Views and opinions expressed are however those of the author(s) only and do not necessarily reflect those of the European Union or the European Research Council Executive Agency. Neither the European Union nor the granting authority can be held responsible for them.
AMH gratefully acknowledges support from the National Science Foundation under Grant No. AST-2307920. 
\end{acknowledgments}

\begin{contribution}
YH planned the observations and wrote the manuscript. All authors contributed to proposal preparation and manuscript editing. 

\end{contribution}

\facilities{JWST(MIRI), ALMA}

\software{\texttt{NumPy} \citep{numpy}, 
          \texttt{SciPy} \citep{scipy},
          \texttt{Matplotlib} \citep{matplotlib},
          \texttt{Astropy} \citep{astropy:2022},
          JWST Science Calibration Pipeline \citep{jwst_pipeline}
          }


\appendix

\section{Azimuthally avevraged profiles}
\label{sec:appendix}
\begin{figure*}
    \centering
    \includegraphics[width=1.0\linewidth]{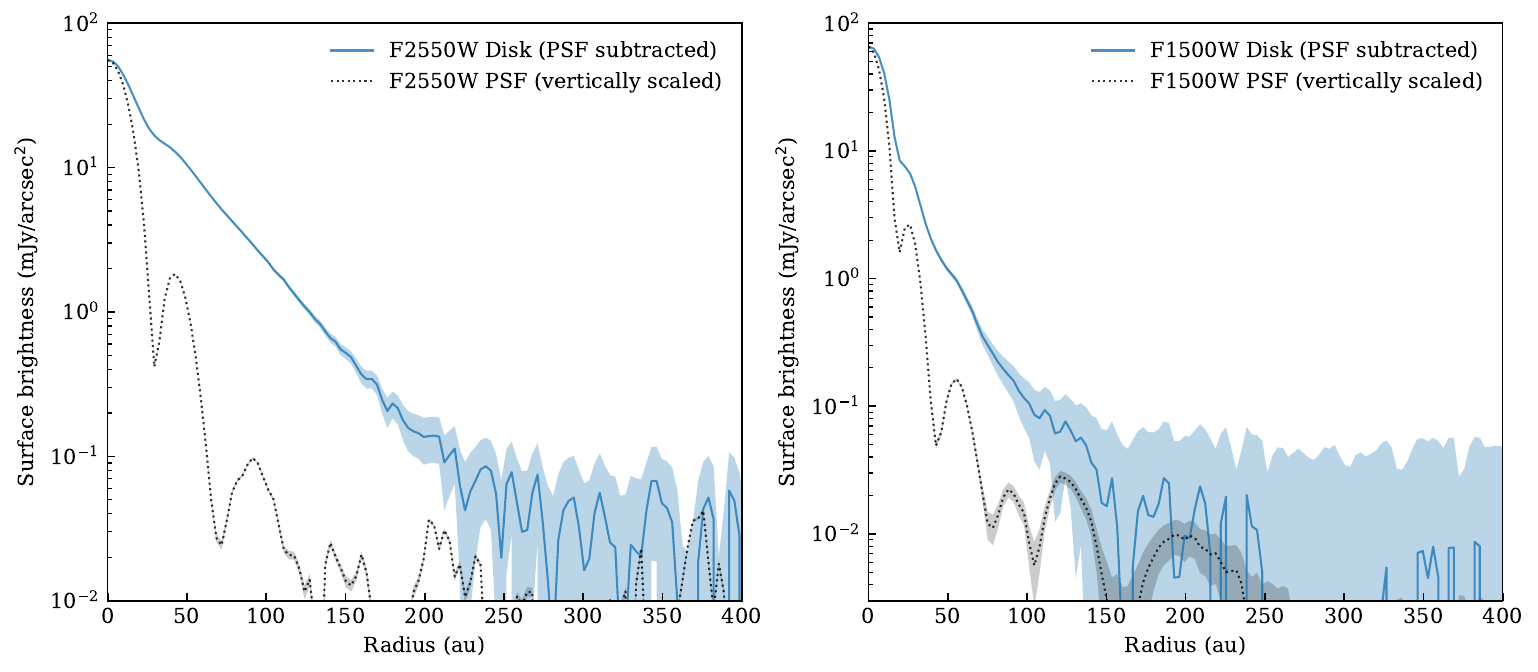}
    \caption{The azimuthally averaged F2550W (left) and F1500W (right) radial profiles measured from the PSF-subtracted (assuming SED flux) image, which are displayed in solid blue lines. The black dotted lines correspond to the PSF observations scaled to match the radial profile of the disk at its centre. }
    \label{fig:halo}
\end{figure*}

The azimuthally averaged radial profiles measured from the images are shown in Fig.~\ref{fig:halo}. 

\bibliography{references}{}

@ARTICLE{Marshall2025b,
       author = {{Marshall}, J.~P. and {Hengst}, S. and {Young}, R. and {Kemper}, F. and {Matr{\`a}}, L. and {Pawellek}, N. and {Kobayashi}, H. and {Scicluna}, P. and {Zeegers}, S.~T.},
        title = "{Systematic determination of dust properties for a sample of 133 spatially resolved debris discs}",
      journal = {\mnras},
         year = 2025,
        month = dec,
          doi = {10.1093/mnras/staf2221},
archivePrefix = {arXiv},
       eprint = {2512.07573},
 primaryClass = {astro-ph.EP},
       adsurl = {https://ui.adsabs.harvard.edu/abs/2025MNRAS.tmp.2118M}
}

@ARTICLE{Eiroa2013,
       author = {{Eiroa}, C. and {Marshall}, J.~P. and {Mora}, A. and {Montesinos}, B. and {Absil}, O. and {Augereau}, J. Ch. and {Bayo}, A. and {Bryden}, G. and {Danchi}, W. and {del Burgo}, C. and {Ertel}, S. and {Fridlund}, M. and {Heras}, A.~M. and {Krivov}, A.~V. and {Launhardt}, R. and {Liseau}, R. and {L{\"o}hne}, T. and {Maldonado}, J. and {Pilbratt}, G.~L. and {Roberge}, A. and {Rodmann}, J. and {Sanz-Forcada}, J. and {Solano}, E. and {Stapelfeldt}, K. and {Th{\'e}bault}, P. and {Wolf}, S. and {Ardila}, D. and {Ar{\'e}valo}, M. and {Beichmann}, C. and {Faramaz}, V. and {Gonz{\'a}lez-Garc{\'\i}a}, B.~M. and {Guti{\'e}rrez}, R. and {Lebreton}, J. and {Mart{\'\i}nez-Arn{\'a}iz}, R. and {Meeus}, G. and {Montes}, D. and {Olofsson}, G. and {Su}, K.~Y.~L. and {White}, G.~J. and {Barrado}, D. and {Fukagawa}, M. and {Gr{\"u}n}, E. and {Kamp}, I. and {Lorente}, R. and {Morbidelli}, A. and {M{\"u}ller}, S. and {Mutschke}, H. and {Nakagawa}, T. and {Ribas}, I. and {Walker}, H.},
        title = "{DUst around NEarby Stars. The survey observational results}",
      journal = {\aap},
         year = 2013,
        month = jul,
       volume = {555},
          eid = {A11},
        pages = {A11},
          doi = {10.1051/0004-6361/201321050},
archivePrefix = {arXiv},
       eprint = {1305.0155},
 primaryClass = {astro-ph.SR},
       adsurl = {https://ui.adsabs.harvard.edu/abs/2013A&A...555A..11E}
}

@ARTICLE{Yelverton2018,
       author = {{Yelverton}, Ben and {Kennedy}, Grant M.},
        title = "{Empty gaps? Depleting annular regions in debris discs by secular resonance with a two-planet system}",
      journal = {\mnras},
         year = 2018,
        month = sep,
       volume = {479},
       number = {2},
        pages = {2673-2691},
          doi = {10.1093/mnras/sty1678},
archivePrefix = {arXiv},
       eprint = {1806.08802},
 primaryClass = {astro-ph.EP},
       adsurl = {https://ui.adsabs.harvard.edu/abs/2018MNRAS.479.2673Y}
}

@ARTICLE{Sefilian2021,
       author = {{Sefilian}, Antranik A. and {Rafikov}, Roman R. and {Wyatt}, Mark C.},
        title = "{Formation of Gaps in Self-gravitating Debris Disks by Secular Resonance in a Single-planet System. I. A Simplified Model}",
      journal = {\apj},
         year = 2021,
        month = mar,
       volume = {910},
       number = {1},
          eid = {13},
        pages = {13},
          doi = {10.3847/1538-4357/abda46},
archivePrefix = {arXiv},
       eprint = {2010.15617},
 primaryClass = {astro-ph.EP},
       adsurl = {https://ui.adsabs.harvard.edu/abs/2021ApJ...910...13S}
}

@ARTICLE{Sefilian2023,
       author = {{Sefilian}, Antranik A. and {Rafikov}, Roman R. and {Wyatt}, Mark C.},
        title = "{Formation of Gaps in Self-gravitating Debris Disks by Secular Resonance in a Single-planet System. II. Toward a Self-consistent Model}",
      journal = {\apj},
         year = 2023,
        month = sep,
       volume = {954},
       number = {1},
          eid = {100},
        pages = {100},
          doi = {10.3847/1538-4357/ace68e},
archivePrefix = {arXiv},
       eprint = {2305.00951},
 primaryClass = {astro-ph.EP},
       adsurl = {https://ui.adsabs.harvard.edu/abs/2023ApJ...954..100S}
}

@ARTICLE{Sefilian2024,
       author = {{Sefilian}, Antranik A.},
        title = "{Massive Debris Disks May Hinder Secular Stirring by Planetary Companions: An Analytic Proof of Concept}",
      journal = {\apj},
         year = 2024,
        month = may,
       volume = {966},
       number = {1},
          eid = {140},
        pages = {140},
          doi = {10.3847/1538-4357/ad32d1},
archivePrefix = {arXiv},
       eprint = {2401.18020},
 primaryClass = {astro-ph.EP},
       adsurl = {https://ui.adsabs.harvard.edu/abs/2024ApJ...966..140S}
}

@ARTICLE{Pearce2024,
       author = {{Pearce}, Tim D. and {Krivov}, Alexander V. and {Sefilian}, Antranik A. and {Jankovic}, Marija R. and {L{\"o}hne}, Torsten and {Morgner}, Tobias and {Wyatt}, Mark C. and {Booth}, Mark and {Marino}, Sebastian},
        title = "{The effect of sculpting planets on the steepness of debris-disc inner edges}",
      journal = {\mnras},
         year = 2024,
        month = jan,
       volume = {527},
       number = {2},
        pages = {3876-3899},
          doi = {10.1093/mnras/stad3462},
archivePrefix = {arXiv},
       eprint = {2311.04265},
 primaryClass = {astro-ph.EP},
       adsurl = {https://ui.adsabs.harvard.edu/abs/2024MNRAS.527.3876P}
}

@article{Milli2025, author = { {Milli}, J. and {Olofsson}, J and {Bonduelle}, M. and others}, title = "{The ALMA survey to Resolve exoKuiper belt Substructures (ARKS) V: Comparison between scattered light and thermal emission}", journal = {\aap}, year = 2026, doi = {10.1051/0004-6361/202556523}, volume = {705}, pages={A199} }

@article{Zawadzki2025, author = { {Zawadzki}, B. and {Fehr}, A. and {Hughes}, A.~M. and others}, title = "{The ALMA survey to Resolve exoKuiper belt Substructures (ARKS) III: The Vertical Structure of Debris Discs}", journal = {\aap}, year = 2026, doi = {10.1051/0004-6361/202556505}, volume = {705}, pages={A197} }

@article{Jankovic2025, author = { {Jankovic}, M.~R. and {Pawellek}, N. and {Zander}, J. and others}, title = "{The ALMA survey to Resolve exoKuiper belt Substructures (ARKS) X: Interpreting the peculiar dust rings around HD 131835}", journal = {\aap}, year = 2026, doi = {10.1051/0004-6361/202556637}, volume = {705}, pages={A204} }

@article{Olofsson2025,
    author = {{Olofsson et al.}, Johan},
    title = "{Scattered light vs ALMA radial offset}",
    journal = {\aap},
    year = {submitted}
}

@ARTICLE{Trevor2015,
       author = {{David}, Trevor J. and {Hillenbrand}, Lynne A.},
        title = "{The Ages of Early-type Stars: Str{\"o}mgren Photometric Methods Calibrated, Validated, Tested, and Applied to Hosts and Prospective Hosts of Directly Imaged Exoplanets}",
      journal = {\apj},
         year = 2015,
        month = may,
       volume = {804},
       number = {2},
          eid = {146},
        pages = {146},
          doi = {10.1088/0004-637X/804/2/146},
archivePrefix = {arXiv},
       eprint = {1501.03154},
 primaryClass = {astro-ph.SR},
       adsurl = {https://ui.adsabs.harvard.edu/abs/2015ApJ...804..146D}
}

@ARTICLE{MacGregor2019,
       author = {{MacGregor}, Meredith A. and {Weinberger}, Alycia J. and {Nesvold}, Erika R. and {Hughes}, A. Meredith and {Wilner}, D.~J. and {Currie}, Thayne and {Debes}, John H. and {Donaldson}, Jessica K. and {Redfield}, Seth and {Roberge}, Aki and {Schneider}, Glenn},
        title = "{Multiple Rings of Millimeter Dust Emission in the HD 15115 Debris Disk}",
      journal = {\apjl},
         year = 2019,
        month = jun,
       volume = {877},
       number = {2},
          eid = {L32},
        pages = {L32},
          doi = {10.3847/2041-8213/ab21c2},
archivePrefix = {arXiv},
       eprint = {1905.08258},
 primaryClass = {astro-ph.EP},
       adsurl = {https://ui.adsabs.harvard.edu/abs/2019ApJ...877L..32M}
}

@ARTICLE{Matra2019,
       author = {{Matr{\`a}}, L. and {Wyatt}, M.~C. and {Wilner}, D.~J. and {Dent}, W.~R.~F. and {Marino}, S. and {Kennedy}, G.~M. and {Milli}, J.},
        title = "{Kuiper Belt-like Hot and Cold Populations of Planetesimal Inclinations in the {\ensuremath{\beta}} Pictoris Belt Revealed by ALMA}",
      journal = {\aj},
         year = 2019,
        month = apr,
       volume = {157},
       number = {4},
          eid = {135},
        pages = {135},
          doi = {10.3847/1538-3881/ab06c0},
archivePrefix = {arXiv},
       eprint = {1902.04081},
 primaryClass = {astro-ph.EP},
       adsurl = {https://ui.adsabs.harvard.edu/abs/2019AJ....157..135M}
}

@ARTICLE{Matra2020,
       author = {{Matr{\`a}}, Luca and {Dent}, William R.~F. and {Wilner}, David J. and {Marino}, Sebasti{\'a}n and {Wyatt}, Mark C. and {Marshall}, Jonathan P. and {Su}, Kate Y.~L. and {Chavez}, Miguel and {Hales}, Antonio and {Hughes}, A. Meredith and {Greaves}, Jane S. and {Corder}, Stuartt A.},
        title = "{Dust Populations in the Iconic Vega Planetary System Resolved by ALMA}",
      journal = {\apj},
         year = 2020,
        month = aug,
       volume = {898},
       number = {2},
          eid = {146},
        pages = {146},
          doi = {10.3847/1538-4357/aba0a4},
archivePrefix = {arXiv},
       eprint = {2006.16257},
 primaryClass = {astro-ph.EP},
       adsurl = {https://ui.adsabs.harvard.edu/abs/2020ApJ...898..146M}
}

@ARTICLE{Daley2019,
       author = {{Daley}, Cail and {Hughes}, A. Meredith and {Carter}, Evan S. and {Flaherty}, Kevin and {Lambros}, Zachary and {Pan}, Margaret and {Schlichting}, Hilke and {Chiang}, Eugene and {Wyatt}, Mark and {Wilner}, David and {Andrews}, Sean and {Carpenter}, John},
        title = "{The Mass of Stirring Bodies in the AU Mic Debris Disk Inferred from Resolved Vertical Structure}",
      journal = {\apj},
         year = 2019,
        month = apr,
       volume = {875},
       number = {2},
          eid = {87},
        pages = {87},
          doi = {10.3847/1538-4357/ab1074},
archivePrefix = {arXiv},
       eprint = {1904.00027},
 primaryClass = {astro-ph.EP},
       adsurl = {https://ui.adsabs.harvard.edu/abs/2019ApJ...875...87D}
}

@ARTICLE{Hines2007,
       author = {{Hines}, Dean C. and {Schneider}, Glenn and {Hollenbach}, David and {Mamajek}, Eric E. and {Hillenbrand}, Lynne A. and {Metchev}, Stanimir A. and {Meyer}, Michael R. and {Carpenter}, John M. and {Moro-Mart{\'\i}n}, Amaya and {Silverstone}, Murray D. and {Kim}, Jinyoung Serena and {Henning}, Thomas and {Bouwman}, Jeroen and {Wolf}, Sebastian},
        title = "{The Moth: An Unusual Circumstellar Structure Associated with HD 61005}",
      journal = {\apjl},
         year = 2007,
        month = dec,
       volume = {671},
       number = {2},
        pages = {L165-L168},
          doi = {10.1086/525016},
       adsurl = {https://ui.adsabs.harvard.edu/abs/2007ApJ...671L.165H}
}

@ARTICLE{Wyatt2008,
       author = {{Wyatt}, M.~C.},
        title = "{Evolution of debris disks.}",
      journal = {\araa},
         year = 2008,
        month = sep,
       volume = {46},
        pages = {339-383},
          doi = {10.1146/annurev.astro.45.051806.110525},
       adsurl = {https://ui.adsabs.harvard.edu/abs/2008ARA&A..46..339W}
}

@ARTICLE{Krivov2010,
       author = {{Krivov}, Alexander V.},
        title = "{Debris disks: seeing dust, thinking of planetesimals and planets}",
      journal = {Research in Astronomy and Astrophysics},
         year = 2010,
        month = may,
       volume = {10},
       number = {5},
        pages = {383-414},
          doi = {10.1088/1674-4527/10/5/001},
archivePrefix = {arXiv},
       eprint = {1003.5229},
 primaryClass = {astro-ph.EP},
       adsurl = {https://ui.adsabs.harvard.edu/abs/2010RAA....10..383K}
}

@ARTICLE{Geiler2019,
       author = {{Geiler}, Fabian and {Krivov}, Alexander V. and {Booth}, Mark and {L{\"o}hne}, Torsten},
        title = "{The scattered disc of HR 8799}",
      journal = {\mnras},
         year = 2019,
        month = feb,
       volume = {483},
       number = {1},
        pages = {332-341},
          doi = {10.1093/mnras/sty3160},
archivePrefix = {arXiv},
       eprint = {1811.09470},
 primaryClass = {astro-ph.EP},
       adsurl = {https://ui.adsabs.harvard.edu/abs/2019MNRAS.483..332G}
}

@ARTICLE{Marino2020,
       author = {{Marino}, S. and {Zurlo}, A. and {Faramaz}, V. and {Milli}, J. and {Henning}, Th and {Kennedy}, G.~M. and {Matr{\`a}}, L. and {P{\'e}rez}, S. and {Delorme}, P. and {Cieza}, L.~A. and {Hughes}, A.~M.},
        title = "{Insights into the planetary dynamics of HD 206893 with ALMA}",
      journal = {\mnras},
         year = 2020,
        month = oct,
       volume = {498},
       number = {1},
        pages = {1319-1334},
          doi = {10.1093/mnras/staa2386},
archivePrefix = {arXiv},
       eprint = {2010.12582},
 primaryClass = {astro-ph.EP},
       adsurl = {https://ui.adsabs.harvard.edu/abs/2020MNRAS.498.1319M}
}

@ARTICLE{Hughes2018,
       author = {{Hughes}, A. Meredith and {Duch{\^e}ne}, Gaspard and {Matthews}, Brenda C.},
        title = "{Debris Disks: Structure, Composition, and Variability}",
      journal = {ARAA},
         year = 2018,
        month = sep,
       volume = {56},
        pages = {541-591},
          doi = {10.1146/annurev-astro-081817-052035},
archivePrefix = {arXiv},
       eprint = {1802.04313},
 primaryClass = {astro-ph.EP},
       adsurl = {https://ui.adsabs.harvard.edu/abs/2018ARA&A..56..541H}
}

@ARTICLE{Pan2012,
       author = {{Pan}, Margaret and {Schlichting}, Hilke E.},
        title = "{Self-consistent Size and Velocity Distributions of Collisional Cascades}",
      journal = {\apj},
         year = 2012,
        month = mar,
       volume = {747},
       number = {2},
          eid = {113},
        pages = {113},
          doi = {10.1088/0004-637X/747/2/113},
archivePrefix = {arXiv},
       eprint = {1111.0667},
 primaryClass = {astro-ph.EP},
       adsurl = {https://ui.adsabs.harvard.edu/abs/2012ApJ...747..113P}
}

@ARTICLE{Gaia2018,
       author = {{Gaia Collaboration} and {Brown}, A.~G.~A. and {Vallenari}, A. and {Prusti}, T. and {de Bruijne}, J.~H.~J. and {Babusiaux}, C. and {Bailer-Jones}, C.~A.~L. and {Biermann}, M. and {Evans}, D.~W. and {Eyer}, L. and {Jansen}, F. and {Jordi}, C. and {Klioner}, S.~A. and {Lammers}, U. and {Lindegren}, L. and {Luri}, X. and {Mignard}, F. and {Panem}, C. and {Pourbaix}, D. and {Randich}, S. and {Sartoretti}, P. and {Siddiqui}, H.~I. and {Soubiran}, C. and {van Leeuwen}, F. and {Walton}, N.~A. and {Arenou}, F. and {Bastian}, U. and {Cropper}, M. and {Drimmel}, R. and {Katz}, D. and {Lattanzi}, M.~G. and {Bakker}, J. and {Cacciari}, C. and {Casta{\~n}eda}, J. and {Chaoul}, L. and {Cheek}, N. and {De Angeli}, F. and {Fabricius}, C. and {Guerra}, R. and {Holl}, B. and {Masana}, E. and {Messineo}, R. and {Mowlavi}, N. and {Nienartowicz}, K. and {Panuzzo}, P. and {Portell}, J. and {Riello}, M. and {Seabroke}, G.~M. and {Tanga}, P. and {Th{\'e}venin}, F. and {Gracia-Abril}, G. and {Comoretto}, G. and {Garcia-Reinaldos}, M. and {Teyssier}, D. and {Altmann}, M. and {Andrae}, R. and {Audard}, M. and {Bellas-Velidis}, I. and {Benson}, K. and {Berthier}, J. and {Blomme}, R. and {Burgess}, P. and {Busso}, G. and {Carry}, B. and {Cellino}, A. and {Clementini}, G. and {Clotet}, M. and {Creevey}, O. and {Davidson}, M. and {De Ridder}, J. and {Delchambre}, L. and {Dell'Oro}, A. and {Ducourant}, C. and {Fern{\'a}ndez-Hern{\'a}ndez}, J. and {Fouesneau}, M. and {Fr{\'e}mat}, Y. and {Galluccio}, L. and {Garc{\'\i}a-Torres}, M. and {Gonz{\'a}lez-N{\'u}{\~n}ez}, J. and {Gonz{\'a}lez-Vidal}, J.~J. and {Gosset}, E. and {Guy}, L.~P. and {Halbwachs}, J. -L. and {Hambly}, N.~C. and {Harrison}, D.~L. and {Hern{\'a}ndez}, J. and {Hestroffer}, D. and {Hodgkin}, S.~T. and {Hutton}, A. and {Jasniewicz}, G. and {Jean-Antoine-Piccolo}, A. and {Jordan}, S. and {Korn}, A.~J. and {Krone-Martins}, A. and {Lanzafame}, A.~C. and {Lebzelter}, T. and {L{\"o}ffler}, W. and {Manteiga}, M. and {Marrese}, P.~M. and {Mart{\'\i}n-Fleitas}, J.~M. and {Moitinho}, A. and {Mora}, A. and {Muinonen}, K. and {Osinde}, J. and {Pancino}, E. and {Pauwels}, T. and {Petit}, J. -M. and {Recio-Blanco}, A. and {Richards}, P.~J. and {Rimoldini}, L. and {Robin}, A.~C. and {Sarro}, L.~M. and {Siopis}, C. and {Smith}, M. and {Sozzetti}, A. and {S{\"u}veges}, M. and {Torra}, J. and {van Reeven}, W. and {Abbas}, U. and {Abreu Aramburu}, A. and {Accart}, S. and {Aerts}, C. and {Altavilla}, G. and {{\'A}lvarez}, M.~A. and {Alvarez}, R. and {Alves}, J. and {Anderson}, R.~I. and {Andrei}, A.~H. and {Anglada Varela}, E. and {Antiche}, E. and {Antoja}, T. and {Arcay}, B. and {Astraatmadja}, T.~L. and {Bach}, N. and {Baker}, S.~G. and {Balaguer-N{\'u}{\~n}ez}, L. and {Balm}, P. and {Barache}, C. and {Barata}, C. and {Barbato}, D. and {Barblan}, F. and {Barklem}, P.~S. and {Barrado}, D. and {Barros}, M. and {Barstow}, M.~A. and {Bartholom{\'e} Mu{\~n}oz}, S. and {Bassilana}, J. -L. and {Becciani}, U. and {Bellazzini}, M. and {Berihuete}, A. and {Bertone}, S. and {Bianchi}, L. and {Bienaym{\'e}}, O. and {Blanco-Cuaresma}, S. and {Boch}, T. and {Boeche}, C. and {Bombrun}, A. and {Borrachero}, R. and {Bossini}, D. and {Bouquillon}, S. and {Bourda}, G. and {Bragaglia}, A. and {Bramante}, L. and {Breddels}, M.~A. and {Bressan}, A. and {Brouillet}, N. and {Br{\"u}semeister}, T. and {Brugaletta}, E. and {Bucciarelli}, B. and {Burlacu}, A. and {Busonero}, D. and {Butkevich}, A.~G. and {Buzzi}, R. and {Caffau}, E. and {Cancelliere}, R. and {Cannizzaro}, G. and {Cantat-Gaudin}, T. and {Carballo}, R. and {Carlucci}, T. and {Carrasco}, J.~M. and {Casamiquela}, L. and {Castellani}, M. and {Castro-Ginard}, A. and {Charlot}, P. and {Chemin}, L. and {Chiavassa}, A. and {Cocozza}, G. and {Costigan}, G. and {Cowell}, S. and {Crifo}, F. and {Crosta}, M. and {Crowley}, C. and {Cuypers}, J. and {Dafonte}, C. and {Damerdji}, Y. and {Dapergolas}, A. and {David}, P. and {David}, M. and {de Laverny}, P. and {De Luise}, F. and {De March}, R. and {de Martino}, D. and {de Souza}, R. and {de Torres}, A. and {Debosscher}, J. and {del Pozo}, E. and {Delbo}, M. and {Delgado}, A. and {Delgado}, H.~E. and {Di Matteo}, P. and {Diakite}, S. and {Diener}, C. and {Distefano}, E. and {Dolding}, C. and {Drazinos}, P. and {Dur{\'a}n}, J. and {Edvardsson}, B. and {Enke}, H. and {Eriksson}, K. and {Esquej}, P. and {Eynard Bontemps}, G. and {Fabre}, C. and {Fabrizio}, M. and {Faigler}, S. and {Falc{\~a}o}, A.~J. and {Farr{\`a}s Casas}, M. and {Federici}, L. and {Fedorets}, G. and {Fernique}, P. and {Figueras}, F. and {Filippi}, F. and {Findeisen}, K. and {Fonti}, A. and {Fraile}, E. and {Fraser}, M. and {Fr{\'e}zouls}, B. and {Gai}, M. and {Galleti}, S. and {Garabato}, D. and {Garc{\'\i}a-Sedano}, F. and {Garofalo}, A. and {Garralda}, N. and {Gavel}, A. and {Gavras}, P. and {Gerssen}, J. and {Geyer}, R. and {Giacobbe}, P. and {Gilmore}, G. and {Girona}, S. and {Giuffrida}, G. and {Glass}, F. and {Gomes}, M. and {Granvik}, M. and {Gueguen}, A. and {Guerrier}, A. and {Guiraud}, J. and {Guti{\'e}rrez-S{\'a}nchez}, R. and {Haigron}, R. and {Hatzidimitriou}, D. and {Hauser}, M. and {Haywood}, M. and {Heiter}, U. and {Helmi}, A. and {Heu}, J. and {Hilger}, T. and {Hobbs}, D. and {Hofmann}, W. and {Holland}, G. and {Huckle}, H.~E. and {Hypki}, A. and {Icardi}, V. and {Jan{\ss}en}, K. and {Jevardat de Fombelle}, G. and {Jonker}, P.~G. and {Juh{\'a}sz}, {\'A}. L. and {Julbe}, F. and {Karampelas}, A. and {Kewley}, A. and {Klar}, J. and {Kochoska}, A. and {Kohley}, R. and {Kolenberg}, K. and {Kontizas}, M. and {Kontizas}, E. and {Koposov}, S.~E. and {Kordopatis}, G. and {Kostrzewa-Rutkowska}, Z. and {Koubsky}, P. and {Lambert}, S. and {Lanza}, A.~F. and {Lasne}, Y. and {Lavigne}, J. -B. and {Le Fustec}, Y. and {Le Poncin-Lafitte}, C. and {Lebreton}, Y. and {Leccia}, S. and {Leclerc}, N. and {Lecoeur-Taibi}, I. and {Lenhardt}, H. and {Leroux}, F. and {Liao}, S. and {Licata}, E. and {Lindstr{\o}m}, H.~E.~P. and {Lister}, T.~A. and {Livanou}, E. and {Lobel}, A. and {L{\'o}pez}, M. and {Managau}, S. and {Mann}, R.~G. and {Mantelet}, G. and {Marchal}, O. and {Marchant}, J.~M. and {Marconi}, M. and {Marinoni}, S. and {Marschalk{\'o}}, G. and {Marshall}, D.~J. and {Martino}, M. and {Marton}, G. and {Mary}, N. and {Massari}, D. and {Matijevi{\v{c}}}, G. and {Mazeh}, T. and {McMillan}, P.~J. and {Messina}, S. and {Michalik}, D. and {Millar}, N.~R. and {Molina}, D. and {Molinaro}, R. and {Moln{\'a}r}, L. and {Montegriffo}, P. and {Mor}, R. and {Morbidelli}, R. and {Morel}, T. and {Morris}, D. and {Mulone}, A.~F. and {Muraveva}, T. and {Musella}, I. and {Nelemans}, G. and {Nicastro}, L. and {Noval}, L. and {O'Mullane}, W. and {Ord{\'e}novic}, C. and {Ord{\'o}{\~n}ez-Blanco}, D. and {Osborne}, P. and {Pagani}, C. and {Pagano}, I. and {Pailler}, F. and {Palacin}, H. and {Palaversa}, L. and {Panahi}, A. and {Pawlak}, M. and {Piersimoni}, A.~M. and {Pineau}, F. -X. and {Plachy}, E. and {Plum}, G. and {Poggio}, E. and {Poujoulet}, E. and {Pr{\v{s}}a}, A. and {Pulone}, L. and {Racero}, E. and {Ragaini}, S. and {Rambaux}, N. and {Ramos-Lerate}, M. and {Regibo}, S. and {Reyl{\'e}}, C. and {Riclet}, F. and {Ripepi}, V. and {Riva}, A. and {Rivard}, A. and {Rixon}, G. and {Roegiers}, T. and {Roelens}, M. and {Romero-G{\'o}mez}, M. and {Rowell}, N. and {Royer}, F. and {Ruiz-Dern}, L. and {Sadowski}, G. and {Sagrist{\`a} Sell{\'e}s}, T. and {Sahlmann}, J. and {Salgado}, J. and {Salguero}, E. and {Sanna}, N. and {Santana-Ros}, T. and {Sarasso}, M. and {Savietto}, H. and {Schultheis}, M. and {Sciacca}, E. and {Segol}, M. and {Segovia}, J.~C. and {S{\'e}gransan}, D. and {Shih}, I. -C. and {Siltala}, L. and {Silva}, A.~F. and {Smart}, R.~L. and {Smith}, K.~W. and {Solano}, E. and {Solitro}, F. and {Sordo}, R. and {Soria Nieto}, S. and {Souchay}, J. and {Spagna}, A. and {Spoto}, F. and {Stampa}, U. and {Steele}, I.~A. and {Steidelm{\"u}ller}, H. and {Stephenson}, C.~A. and {Stoev}, H. and {Suess}, F.~F. and {Surdej}, J. and {Szabados}, L. and {Szegedi-Elek}, E. and {Tapiador}, D. and {Taris}, F. and {Tauran}, G. and {Taylor}, M.~B. and {Teixeira}, R. and {Terrett}, D. and {Teyssandier}, P. and {Thuillot}, W. and {Titarenko}, A. and {Torra Clotet}, F. and {Turon}, C. and {Ulla}, A. and {Utrilla}, E. and {Uzzi}, S. and {Vaillant}, M. and {Valentini}, G. and {Valette}, V. and {van Elteren}, A. and {Van Hemelryck}, E. and {van Leeuwen}, M. and {Vaschetto}, M. and {Vecchiato}, A. and {Veljanoski}, J. and {Viala}, Y. and {Vicente}, D. and {Vogt}, S. and {von Essen}, C. and {Voss}, H. and {Votruba}, V. and {Voutsinas}, S. and {Walmsley}, G. and {Weiler}, M. and {Wertz}, O. and {Wevers}, T. and {Wyrzykowski}, {\L}. and {Yoldas}, A. and {{\v{Z}}erjal}, M. and {Ziaeepour}, H. and {Zorec}, J. and {Zschocke}, S. and {Zucker}, S. and {Zurbach}, C. and {Zwitter}, T.},
        title = "{Gaia Data Release 2. Summary of the contents and survey properties}",
      journal = {\aap},
         year = 2018,
        month = aug,
       volume = {616},
          eid = {A1},
        pages = {A1},
          doi = {10.1051/0004-6361/201833051},
archivePrefix = {arXiv},
       eprint = {1804.09365},
 primaryClass = {astro-ph.GA},
       adsurl = {https://ui.adsabs.harvard.edu/abs/2018A&A...616A...1G}
}

@ARTICLE{Marino2021,
       author = {{Marino}, Sebastian},
        title = "{Constraining planetesimal stirring: how sharp are debris disc edges?}",
      journal = {\mnras},
         year = 2021,
        month = may,
       volume = {503},
       number = {4},
        pages = {5100-5114},
          doi = {10.1093/mnras/stab771},
archivePrefix = {arXiv},
       eprint = {2104.02072},
 primaryClass = {astro-ph.EP},
       adsurl = {https://ui.adsabs.harvard.edu/abs/2021MNRAS.503.5100M}
}

@ARTICLE{Schneider2014,
       author = {{Schneider}, Glenn and {Grady}, Carol A. and {Hines}, Dean C. and {Stark}, Christopher C. and {Debes}, John H. and {Carson}, Joe and {Kuchner}, Marc J. and {Perrin}, Marshall D. and {Weinberger}, Alycia J. and {Wisniewski}, John P. and {Silverstone}, Murray D. and {Jang-Condell}, Hannah and {Henning}, Thomas and {Woodgate}, Bruce E. and {Serabyn}, Eugene and {Moro-Martin}, Amaya and {Tamura}, Motohide and {Hinz}, Phillip M. and {Rodigas}, Timothy J.},
        title = "{Probing for Exoplanets Hiding in Dusty Debris Disks: Disk Imaging, Characterization, and Exploration with HST/STIS Multi-roll Coronagraphy}",
      journal = {\aj},
         year = 2014,
        month = oct,
       volume = {148},
       number = {4},
          eid = {59},
        pages = {59},
          doi = {10.1088/0004-6256/148/4/59},
archivePrefix = {arXiv},
       eprint = {1406.7303},
 primaryClass = {astro-ph.IM},
       adsurl = {https://ui.adsabs.harvard.edu/abs/2014AJ....148...59S}
}

@ARTICLE{emcee,
       author = {{Foreman-Mackey}, Daniel and {Hogg}, David W. and {Lang}, Dustin and {Goodman}, Jonathan},
        title = "{emcee: The MCMC Hammer}",
      journal = {\pasp},
         year = 2013,
        month = mar,
       volume = {125},
       number = {925},
        pages = {306},
          doi = {10.1086/670067},
archivePrefix = {arXiv},
       eprint = {1202.3665},
 primaryClass = {astro-ph.IM},
       adsurl = {https://ui.adsabs.harvard.edu/abs/2013PASP..125..306F}
}

@Article{matplotlib, Author = {Hunter, J. D.}, Title = {Matplotlib: A 2D graphics environment}, Journal = {Computing In Science \& Engineering}, Volume = {9}, Number = {3}, Pages = {90--95}, publisher = {IEEE COMPUTER SOC}, year = 2007 }

@ARTICLE{astropy:2022,
       author = {{Astropy Collaboration} and {Price-Whelan}, Adrian M. and {Lim}, Pey Lian and {Earl}, Nicholas and {Starkman}, Nathaniel and {Bradley}, Larry and {Shupe}, David L. and {Patil}, Aarya A. and {Corrales}, Lia and {Brasseur}, C.~E. and {N{"o}the}, Maximilian and {Donath}, Axel and {Tollerud}, Erik and {Morris}, Brett M. and {Ginsburg}, Adam and {Vaher}, Eero and {Weaver}, Benjamin A. and {Tocknell}, James and {Jamieson}, William and {van Kerkwijk}, Marten H. and {Robitaille}, Thomas P. and {Merry}, Bruce and {Bachetti}, Matteo and {G{"u}nther}, H. Moritz and {Aldcroft}, Thomas L. and {Alvarado-Montes}, Jaime A. and {Archibald}, Anne M. and {B{'o}di}, Attila and {Bapat}, Shreyas and {Barentsen}, Geert and {Baz{'a}n}, Juanjo and {Biswas}, Manish and {Boquien}, M{'e}d{'e}ric and {Burke}, D.~J. and {Cara}, Daria and {Cara}, Mihai and {Conroy}, Kyle E. and {Conseil}, Simon and {Craig}, Matthew W. and {Cross}, Robert M. and {Cruz}, Kelle L. and {D'Eugenio}, Francesco and {Dencheva}, Nadia and {Devillepoix}, Hadrien A.~R. and {Dietrich}, J{"o}rg P. and {Eigenbrot}, Arthur Davis and {Erben}, Thomas and {Ferreira}, Leonardo and {Foreman-Mackey}, Daniel and {Fox}, Ryan and {Freij}, Nabil and {Garg}, Suyog and {Geda}, Robel and {Glattly}, Lauren and {Gondhalekar}, Yash and {Gordon}, Karl D. and {Grant}, David and {Greenfield}, Perry and {Groener}, Austen M. and {Guest}, Steve and {Gurovich}, Sebastian and {Handberg}, Rasmus and {Hart}, Akeem and {Hatfield-Dodds}, Zac and {Homeier}, Derek and {Hosseinzadeh}, Griffin and {Jenness}, Tim and {Jones}, Craig K. and {Joseph}, Prajwel and {Kalmbach}, J. Bryce and {Karamehmetoglu}, Emir and {Ka{l}uszy{'n}ski}, Miko{l}aj and {Kelley}, Michael S.~P. and {Kern}, Nicholas and {Kerzendorf}, Wolfgang E. and {Koch}, Eric W. and {Kulumani}, Shankar and {Lee}, Antony and {Ly}, Chun and {Ma}, Zhiyuan and {MacBride}, Conor and {Maljaars}, Jakob M. and {Muna}, Demitri and {Murphy}, N.~A. and {Norman}, Henrik and {O'Steen}, Richard and {Oman}, Kyle A. and {Pacifici}, Camilla and {Pascual}, Sergio and {Pascual-Granado}, J. and {Patil}, Rohit R. and {Perren}, Gabriel I. and {Pickering}, Timothy E. and {Rastogi}, Tanuj and {Roulston}, Benjamin R. and {Ryan}, Daniel F. and {Rykoff}, Eli S. and {Sabater}, Jose and {Sakurikar}, Parikshit and {Salgado}, Jes{'u}s and {Sanghi}, Aniket and {Saunders}, Nicholas and {Savchenko}, Volodymyr and {Schwardt}, Ludwig and {Seifert-Eckert}, Michael and {Shih}, Albert Y. and {Jain}, Anany Shrey and {Shukla}, Gyanendra and {Sick}, Jonathan and {Simpson}, Chris and {Singanamalla}, Sudheesh and {Singer}, Leo P. and {Singhal}, Jaladh and {Sinha}, Manodeep and {Sip{H{o}}cz}, Brigitta M. and {Spitler}, Lee R. and {Stansby}, David and {Streicher}, Ole and {{{S}}umak}, Jani and {Swinbank}, John D. and {Taranu}, Dan S. and {Tewary}, Nikita and {Tremblay}, Grant R. and {Val-Borro}, Miguel de and {Van Kooten}, Samuel J. and {Vasovi{'c}}, Zlatan and {Verma}, Shresth and {de Miranda Cardoso}, Jos{'e} Vin{'i}cius and {Williams}, Peter K.~G. and {Wilson}, Tom J. and {Winkel}, Benjamin and {Wood-Vasey}, W.~M. and {Xue}, Rui and {Yoachim}, Peter and {Zhang}, Chen and {Zonca}, Andrea and {Astropy Project Contributors}},
        title = "{The Astropy Project: Sustaining and Growing a Community-oriented Open-source Project and the Latest Major Release (v5.0) of the Core Package}",
      journal = {\apj},
         year = 2022,
        month = aug,
       volume = {935},
       number = {2},
          eid = {167},
        pages = {167},
          doi = {10.3847/1538-4357/ac7c74},
archivePrefix = {arXiv},
       eprint = {2206.14220},
 primaryClass = {astro-ph.IM},
       adsurl = {https://ui.adsabs.harvard.edu/abs/2022ApJ...935..167A}
}

@ARTICLE{scipy,
  author  = {Virtanen, Pauli and Gommers, Ralf and Oliphant, Travis E. and
            Haberland, Matt and Reddy, Tyler and Cournapeau, David and
            Burovski, Evgeni and Peterson, Pearu and Weckesser, Warren and
            Bright, Jonathan and {van der Walt}, St{\'e}fan J. and
            Brett, Matthew and Wilson, Joshua and Millman, K. Jarrod and
            Mayorov, Nikolay and Nelson, Andrew R. J. and Jones, Eric and
            Kern, Robert and Larson, Eric and Carey, C J and
            Polat, {\.I}lhan and Feng, Yu and Moore, Eric W. and
            {VanderPlas}, Jake and Laxalde, Denis and Perktold, Josef and
            Cimrman, Robert and Henriksen, Ian and Quintero, E. A. and
            Harris, Charles R. and Archibald, Anne M. and
            Ribeiro, Ant{\^o}nio H. and Pedregosa, Fabian and
            {van Mulbregt}, Paul and {SciPy 1.0 Contributors}},
  title   = {{{SciPy} 1.0: Fundamental Algorithms for Scientific
            Computing in Python}},
  journal = {Nature Methods},
  year    = {2020},
  volume  = {17},
  pages   = {261--272},
  adsurl  = {https://rdcu.be/b08Wh},
  doi     = {10.1038/s41592-019-0686-2},
}

@ARTICLE{numpy,
  author  = {Harris, Charles R. and Millman, K. Jarrod and
            van der Walt, Stéfan J and Gommers, Ralf and
            Virtanen, Pauli and Cournapeau, David and
            Wieser, Eric and Taylor, Julian and Berg, Sebastian and
            Smith, Nathaniel J. and Kern, Robert and Picus, Matti and
            Hoyer, Stephan and van Kerkwijk, Marten H. and
            Brett, Matthew and Haldane, Allan and
            Fernández del Río, Jaime and Wiebe, Mark and
            Peterson, Pearu and Gérard-Marchant, Pierre and
            Sheppard, Kevin and Reddy, Tyler and Weckesser, Warren and
            Abbasi, Hameer and Gohlke, Christoph and
            Oliphant, Travis E.},
  title   = {Array programming with {NumPy}},
  journal = {Nature},
  year    = {2020},
  volume  = {585},
  pages   = {357–362},
  doi     = {10.1038/s41586-020-2649-2}
}

@ARTICLE{MacGregor2017Fom,
       author = {{MacGregor}, Meredith A. and {Matr{\`a}}, Luca and {Kalas}, Paul and {Wilner}, David J. and {Pan}, Margaret and {Kennedy}, Grant M. and {Wyatt}, Mark C. and {Duchene}, Gaspard and {Hughes}, A. Meredith and {Rieke}, George H. and {Clampin}, Mark and {Fitzgerald}, Michael P. and {Graham}, James R. and {Holland}, Wayne S. and {Pani{\'c}}, Olja and {Shannon}, Andrew and {Su}, Kate},
        title = "{A Complete ALMA Map of the Fomalhaut Debris Disk}",
      journal = {\apj},
         year = 2017,
        month = jun,
       volume = {842},
       number = {1},
          eid = {8},
        pages = {8},
          doi = {10.3847/1538-4357/aa71ae},
archivePrefix = {arXiv},
       eprint = {1705.05867},
 primaryClass = {astro-ph.EP},
       adsurl = {https://ui.adsabs.harvard.edu/abs/2017ApJ...842....8M}
}

@ARTICLE{Wyatt1999,
       author = {{Wyatt}, M.~C. and {Dermott}, S.~F. and {Telesco}, C.~M. and {Fisher}, R.~S. and {Grogan}, K. and {Holmes}, E.~K. and {Pi{\~n}a}, R.~K.},
        title = "{How Observations of Circumstellar Disk Asymmetries Can Reveal Hidden Planets: Pericenter Glow and Its Application to the HR 4796 Disk}",
      journal = {\apj},
         year = 1999,
        month = dec,
       volume = {527},
       number = {2},
        pages = {918-944},
          doi = {10.1086/308093},
archivePrefix = {arXiv},
       eprint = {astro-ph/9908267},
 primaryClass = {astro-ph},
       adsurl = {https://ui.adsabs.harvard.edu/abs/1999ApJ...527..918W}
}

@ARTICLE{Wyatt2005,
       author = {{Wyatt}, M.~C.},
        title = "{The insignificance of P-R drag in detectable extrasolar planetesimal belts}",
      journal = {\aap},
         year = 2005,
        month = apr,
       volume = {433},
       number = {3},
        pages = {1007-1012},
          doi = {10.1051/0004-6361:20042073},
archivePrefix = {arXiv},
       eprint = {astro-ph/0501038},
 primaryClass = {astro-ph},
       adsurl = {https://ui.adsabs.harvard.edu/abs/2005A&A...433.1007W}
}

@ARTICLE{wyatt2005b,
       author = {{Wyatt}, M.~C.},
        title = "{Spiral structure when setting up pericentre glow: possible giant planets at hundreds of AU in the HD 141569 disk}",
      journal = {\aap},
         year = 2005,
        month = sep,
       volume = {440},
       number = {3},
        pages = {937-948},
          doi = {10.1051/0004-6361:20053391},
archivePrefix = {arXiv},
       eprint = {astro-ph/0506208},
 primaryClass = {astro-ph},
       adsurl = {https://ui.adsabs.harvard.edu/abs/2005A&A...440..937W}
}

@ARTICLE{Krivov2021,
       author = {{Krivov}, Alexander V. and {Wyatt}, Mark C.},
        title = "{Solution to the debris disc mass problem: planetesimals are born small?}",
      journal = {\mnras},
         year = 2021,
        month = jan,
       volume = {500},
       number = {1},
        pages = {718-735},
          doi = {10.1093/mnras/staa2385},
archivePrefix = {arXiv},
       eprint = {2008.07406},
 primaryClass = {astro-ph.EP},
       adsurl = {https://ui.adsabs.harvard.edu/abs/2021MNRAS.500..718K}
}

@ARTICLE{Hayashi1981,
       author = {{Hayashi}, C.},
        title = "{Structure of the Solar Nebula, Growth and Decay of Magnetic Fields and Effects of Magnetic and Turbulent Viscosities on the Nebula}",
      journal = {Progress of Theoretical Physics Supplement},
         year = 1981,
        month = jan,
       volume = {70},
        pages = {35-53},
          doi = {10.1143/PTPS.70.35},
       adsurl = {https://ui.adsabs.harvard.edu/abs/1981PThPS..70...35H}
}

@ARTICLE{Marino2019,
       author = {{Marino}, S. and {Yelverton}, B. and {Booth}, M. and {Faramaz}, V. and {Kennedy}, G.~M. and {Matr{\`a}}, L. and {Wyatt}, M.~C.},
        title = "{A gap in HD 92945's broad planetesimal disc revealed by ALMA}",
      journal = {\mnras},
         year = 2019,
        month = mar,
       volume = {484},
       number = {1},
        pages = {1257-1269},
          doi = {10.1093/mnras/stz049},
archivePrefix = {arXiv},
       eprint = {1901.01406},
 primaryClass = {astro-ph.EP},
       adsurl = {https://ui.adsabs.harvard.edu/abs/2019MNRAS.484.1257M}
}

@ARTICLE{Marino2018,
       author = {{Marino}, S. and {Carpenter}, J. and {Wyatt}, M.~C. and {Booth}, M. and {Casassus}, S. and {Faramaz}, V. and {Guzman}, V. and {Hughes}, A.~M. and {Isella}, A. and {Kennedy}, G.~M. and {Matr{\`a}}, L. and {Ricci}, L. and {Corder}, S.},
        title = "{A gap in the planetesimal disc around HD 107146 and asymmetric warm dust emission revealed by ALMA}",
      journal = {\mnras},
         year = 2018,
        month = oct,
       volume = {479},
       number = {4},
        pages = {5423-5439},
          doi = {10.1093/mnras/sty1790},
archivePrefix = {arXiv},
       eprint = {1805.01915},
 primaryClass = {astro-ph.EP},
       adsurl = {https://ui.adsabs.harvard.edu/abs/2018MNRAS.479.5423M}
}

@ARTICLE{Vizgan2022,
       author = {{Vizgan}, David and {Hughes}, A. Meredith and {Carter}, Evan S. and {Flaherty}, Kevin M. and {Pan}, Margaret and {Chiang}, Eugene and {Schlichting}, Hilke and {Wilner}, David J. and {Andrews}, Sean M. and {Carpenter}, John M. and {Mo{\'o}r}, Attila and {MacGregor}, Meredith A.},
        title = "{Multiwavelength Vertical Structure in the AU Mic Debris Disk: Characterizing the Collisional Cascade}",
      journal = {\apj},
         year = 2022,
        month = aug,
       volume = {935},
       number = {2},
          eid = {131},
        pages = {131},
          doi = {10.3847/1538-4357/ac80b8},
archivePrefix = {arXiv},
       eprint = {2207.05277},
 primaryClass = {astro-ph.EP},
       adsurl = {https://ui.adsabs.harvard.edu/abs/2022ApJ...935..131V}
}

@ARTICLE{Terrill2023,
       author = {{Terrill}, James and {Marino}, Sebastian and {Booth}, Richard A. and {Han}, Yinuo and {Jennings}, Jeff and {Wyatt}, Mark C.},
        title = "{Deprojecting and constraining the vertical thickness of exoKuiper belts}",
      journal = {\mnras},
         year = 2023,
        month = sep,
       volume = {524},
       number = {1},
        pages = {1229-1245},
          doi = {10.1093/mnras/stad1847},
archivePrefix = {arXiv},
       eprint = {2306.09715},
 primaryClass = {astro-ph.EP},
       adsurl = {https://ui.adsabs.harvard.edu/abs/2023MNRAS.524.1229T}
}

@ARTICLE{Li1997,
       author = {{Li}, A. and {Greenberg}, J.~M.},
        title = "{A unified model of interstellar dust.}",
      journal = {\aap},
         year = 1997,
        month = jul,
       volume = {323},
        pages = {566-584},
       adsurl = {https://ui.adsabs.harvard.edu/abs/1997A&A...323..566L}
}

@BOOK{Bohren1983,
       author = {{Bohren}, Craig F. and {Huffman}, Donald R.},
        title = "{Absorption and scattering of light by small particles}",
         year = 1983,
       adsurl = {https://ui.adsabs.harvard.edu/abs/1983asls.book.....B},
    publisher = {Wiley}
}

@ARTICLE{Pawellek2015,
       author = {{Pawellek}, Nicole and {Krivov}, Alexander V.},
        title = "{The dust grain size-stellar luminosity trend in debris discs}",
      journal = {\mnras},
         year = 2015,
        month = dec,
       volume = {454},
       number = {3},
        pages = {3207-3221},
          doi = {10.1093/mnras/stv2142},
archivePrefix = {arXiv},
       eprint = {1509.04032},
 primaryClass = {astro-ph.SR},
       adsurl = {https://ui.adsabs.harvard.edu/abs/2015MNRAS.454.3207P}
}

@ARTICLE{Han2022,
       author = {{Han}, Yinuo and {Wyatt}, Mark C. and {Matr{\`a}}, Luca},
        title = "{RAVE: a non-parametric method for recovering the surface brightness and height profiles of edge-on debris discs}",
      journal = {\mnras},
         year = 2022,
        month = apr,
       volume = {511},
       number = {4},
        pages = {4921-4936},
          doi = {10.1093/mnras/stac373},
archivePrefix = {arXiv},
       eprint = {2202.04475},
 primaryClass = {astro-ph.EP},
       adsurl = {https://ui.adsabs.harvard.edu/abs/2022MNRAS.511.4921H}
}

@ARTICLE{Williams2011,
       author = {{Williams}, Jonathan P. and {Cieza}, Lucas A.},
        title = "{Protoplanetary Disks and Their Evolution}",
      journal = {\araa},
         year = 2011,
        month = sep,
       volume = {49},
       number = {1},
        pages = {67-117},
          doi = {10.1146/annurev-astro-081710-102548},
archivePrefix = {arXiv},
       eprint = {1103.0556},
 primaryClass = {astro-ph.GA},
       adsurl = {https://ui.adsabs.harvard.edu/abs/2011ARA&A..49...67W}
}

@ARTICLE{Jennings2020,
       author = {{Jennings}, Jeff and {Booth}, Richard A. and {Tazzari}, Marco and {Rosotti}, Giovanni P. and {Clarke}, Cathie J.},
        title = "{frankenstein: protoplanetary disc brightness profile reconstruction at sub-beam resolution with a rapid Gaussian process}",
      journal = {\mnras},
         year = 2020,
        month = jul,
       volume = {495},
       number = {3},
        pages = {3209-3232},
          doi = {10.1093/mnras/staa1365},
archivePrefix = {arXiv},
       eprint = {2005.07709},
 primaryClass = {astro-ph.EP},
       adsurl = {https://ui.adsabs.harvard.edu/abs/2020MNRAS.495.3209J}
}

@ARTICLE{Pawellek2014,
       author = {{Pawellek}, Nicole and {Krivov}, Alexander V. and {Marshall}, Jonathan P. and {Montesinos}, Benjamin and {{\'A}brah{\'a}m}, P{\'e}ter and {Mo{\'o}r}, Attila and {Bryden}, Geoffrey and {Eiroa}, Carlos},
        title = "{Disk Radii and Grain Sizes in Herschel-resolved Debris Disks}",
      journal = {\apj},
         year = 2014,
        month = sep,
       volume = {792},
       number = {1},
          eid = {65},
        pages = {65},
          doi = {10.1088/0004-637X/792/1/65},
archivePrefix = {arXiv},
       eprint = {1407.4579},
 primaryClass = {astro-ph.SR},
       adsurl = {https://ui.adsabs.harvard.edu/abs/2014ApJ...792...65P}
}

@ARTICLE{Holland2017,
       author = {{Holland}, Wayne S. and {Matthews}, Brenda C. and {Kennedy}, Grant M. and {Greaves}, Jane S. and {Wyatt}, Mark C. and {Booth}, Mark and {Bastien}, Pierre and {Bryden}, Geoff and {Butner}, Harold and {Chen}, Christine H. and {Chrysostomou}, Antonio and {Davies}, Claire L. and {Dent}, William R.~F. and {Di Francesco}, James and {Duch{\^e}ne}, Gaspard and {Gibb}, Andy G. and {Friberg}, Per and {Ivison}, Rob J. and {Jenness}, Tim and {Kavelaars}, JJ and {Lawler}, Samantha and {Lestrade}, Jean-Fran{\c{c}}ois and {Marshall}, Jonathan P. and {Moro-Martin}, Amaya and {Pani{\'c}}, Olja and {Phillips}, Neil and {Serjeant}, Stephen and {Schieven}, Gerald H. and {Sibthorpe}, Bruce and {Vican}, Laura and {Ward-Thompson}, Derek and {van der Werf}, Paul and {White}, Glenn J. and {Wilner}, David and {Zuckerman}, Ben},
        title = "{SONS: The JCMT legacy survey of debris discs in the submillimetre}",
      journal = {\mnras},
         year = 2017,
        month = sep,
       volume = {470},
       number = {3},
        pages = {3606-3663},
          doi = {10.1093/mnras/stx1378},
archivePrefix = {arXiv},
       eprint = {1706.01218},
 primaryClass = {astro-ph.SR},
       adsurl = {https://ui.adsabs.harvard.edu/abs/2017MNRAS.470.3606H}
}

@ARTICLE{Kennedy2010,
       author = {{Kennedy}, G.~M. and {Wyatt}, M.~C.},
        title = "{Are debris discs self-stirred?}",
      journal = {\mnras},
         year = 2010,
        month = jun,
       volume = {405},
       number = {2},
        pages = {1253-1270},
          doi = {10.1111/j.1365-2966.2010.16528.x},
archivePrefix = {arXiv},
       eprint = {1002.3469},
 primaryClass = {astro-ph.EP},
       adsurl = {https://ui.adsabs.harvard.edu/abs/2010MNRAS.405.1253K}
}

@ARTICLE{ImazBlanco2023,
       author = {{Imaz Blanco}, Amaia and {Marino}, Sebastian and {Matr{\`a}}, Luca and {Booth}, Mark and {Carpenter}, John and {Faramaz}, Virginie and {Henning}, Thomas and {Hughes}, A. Meredith and {Kennedy}, Grant M. and {P{\'e}rez}, Sebasti{\'a}n and {Ricci}, Luca and {Wyatt}, Mark C.},
        title = "{Inner edges of planetesimal belts: collisionally eroded or truncated?}",
      journal = {\mnras},
         year = 2023,
        month = jul,
       volume = {522},
       number = {4},
        pages = {6150-6169},
          doi = {10.1093/mnras/stad1221},
archivePrefix = {arXiv},
       eprint = {2304.12337},
 primaryClass = {astro-ph.EP},
       adsurl = {https://ui.adsabs.harvard.edu/abs/2023MNRAS.522.6150I}
}

@ARTICLE{Morbidelli2004,
       author = {{Morbidelli}, A. and {Emel'yanenko}, V.~V. and {Levison}, H.~F.},
        title = "{Origin and orbital distribution of the trans-Neptunian scattered disc}",
      journal = {\mnras},
         year = 2004,
        month = dec,
       volume = {355},
       number = {3},
        pages = {935-940},
          doi = {10.1111/j.1365-2966.2004.08372.x},
       adsurl = {https://ui.adsabs.harvard.edu/abs/2004MNRAS.355..935M}
}

@ARTICLE{Wyatt2007,
       author = {{Wyatt}, M.~C. and {Smith}, R. and {Su}, K.~Y.~L. and {Rieke}, G.~H. and {Greaves}, J.~S. and {Beichman}, C.~A. and {Bryden}, G.},
        title = "{Steady State Evolution of Debris Disks around A Stars}",
      journal = {\apj},
         year = 2007,
        month = jul,
       volume = {663},
       number = {1},
        pages = {365-382},
          doi = {10.1086/518404},
archivePrefix = {arXiv},
       eprint = {astro-ph/0703608},
 primaryClass = {astro-ph},
       adsurl = {https://ui.adsabs.harvard.edu/abs/2007ApJ...663..365W}
}

@ARTICLE{Matra2025,
       author = {{Matr{\`a}}, L. and {Marino}, S. and {Wilner}, D.~J. and {Kennedy}, G.~M. and {Booth}, M. and {Krivov}, A.~V. and {Williams}, J.~P. and {Hughes}, A.~M. and {del Burgo}, C. and {Carpenter}, J. and {Davies}, C.~L. and {Ertel}, S. and {Kral}, Q. and {Lestrade}, J. -F. and {Marshall}, J.~P. and {Milli}, J. and {{\"O}berg}, K.~I. and {Pawellek}, N. and {Sepulveda}, A.~G. and {Wyatt}, M.~C. and {Matthews}, B.~C. and {MacGregor}, M.},
        title = "{REsolved ALMA and SMA Observations of Nearby Stars (REASONS): A population of 74 resolved planetesimal belts at millimetre wavelengths}",
      journal = {\aap},
         year = 2025,
        month = jan,
       volume = {693},
          eid = {A151},
        pages = {A151},
          doi = {10.1051/0004-6361/202451397},
archivePrefix = {arXiv},
       eprint = {2501.09058},
 primaryClass = {astro-ph.EP},
       adsurl = {https://ui.adsabs.harvard.edu/abs/2025A&A...693A.151M}
}

@ARTICLE{Nederlander2021,
       author = {{Nederlander}, Ava and {Hughes}, A. Meredith and {Fehr}, Anna J. and {Flaherty}, Kevin M. and {Su}, Kate Y.~L. and {Mo{\'o}r}, Attila and {Chiang}, Eugene and {Andrews}, Sean M. and {Wilner}, David J. and {Marino}, Sebastian},
        title = "{Resolving Structure in the Debris Disk around HD 206893 with ALMA}",
      journal = {\apj},
         year = 2021,
        month = aug,
       volume = {917},
       number = {1},
          eid = {5},
        pages = {5},
          doi = {10.3847/1538-4357/abdd32},
archivePrefix = {arXiv},
       eprint = {2101.08849},
 primaryClass = {astro-ph.EP},
       adsurl = {https://ui.adsabs.harvard.edu/abs/2021ApJ...917....5N}
}

@ARTICLE{Su2008,
       author = {{Su}, K.~Y.~L. and {Rieke}, G.~H. and {Stapelfeldt}, K.~R. and {Smith}, P.~S. and {Bryden}, G. and {Chen}, C.~H. and {Trilling}, D.~E.},
        title = "{The Exceptionally Large Debris Disk around {\ensuremath{\gamma}} Ophiuchi}",
      journal = {\apjl},
         year = 2008,
        month = jun,
       volume = {679},
       number = {2},
        pages = {L125},
          doi = {10.1086/589508},
archivePrefix = {arXiv},
       eprint = {0804.2924},
 primaryClass = {astro-ph},
       adsurl = {https://ui.adsabs.harvard.edu/abs/2008ApJ...679L.125S}
}

@ARTICLE{Moor2015,
       author = {{Mo{\'o}r}, A. and {K{\'o}sp{\'a}l}, {\'A}. and {{\'A}brah{\'a}m}, P. and {Apai}, D. and {Balog}, Z. and {Grady}, C. and {Henning}, Th. and {Juh{\'a}sz}, A. and {Kiss}, Cs. and {Krivov}, A.~V. and {Pawellek}, N. and {Szab{\'o}}, Gy. M.},
        title = "{Stirring in massive, young debris discs from spatially resolved Herschel images}",
      journal = {\mnras},
         year = 2015,
        month = feb,
       volume = {447},
       number = {1},
        pages = {577-597},
          doi = {10.1093/mnras/stu2442},
archivePrefix = {arXiv},
       eprint = {1411.5829},
 primaryClass = {astro-ph.SR},
       adsurl = {https://ui.adsabs.harvard.edu/abs/2015MNRAS.447..577M}
}

@ARTICLE{Han2025,
       author = {{Han}, Yinuo and {Wyatt}, Mark C. and {Marino}, Sebastian},
        title = "{Recovering the structure of debris disks non-parametrically from images}",
      journal = {\mnras},
         year = 2025,
        month = feb,
          doi = {10.1093/mnras/staf282},
archivePrefix = {arXiv},
       eprint = {2502.08584},
 primaryClass = {astro-ph.EP},
       adsurl = {https://ui.adsabs.harvard.edu/abs/2025MNRAS.tmp..268H}
}

@article{Marino2025, author = { {Marino}, S. and {Matr\`a}, L. and {Hughes}, A.~M. and others}, title = "{The ALMA survey to Resolve exoKuiper belt Substructures (ARKS) I: Motivation, Sample, Data Reduction and Results Overview}", journal = {\aap}, year = 2026, doi = {10.1051/0004-6361/202556489}, volume = {705}, pages={A195} }

@article{Han2025b, author = { {Han}, Y. and {Mansell}, E. and {Jennings}, J. and others}, title = "{The ALMA survey to Resolve exoKuiper belt Substructures (ARKS) II: The Radial Structure of Debris Discs}", journal = {\aap}, year = 2026, doi = {10.1051/0004-6361/202556450}, volume = {705}, pages={A196} }

@article{Lovell2025, author = { {Lovell}, J.~B. and {Hales}, A.~S. and {Kennedy}, G.~M. and others}, title = "{The ALMA survey to Resolve exoKuiper belt Substructures (ARKS) VI: Asymmetries and Offsets}", journal = {\aap}, year = 2026, doi = {10.1051/0004-6361/202556568}, volume = {705}, pages={A200} }

@ARTICLE{Farhat2023,
       author = {{Farhat}, Mohammad A. and {Sefilian}, Antranik A. and {Touma}, Jihad R.},
        title = "{The case of HD 106906 debris disc: a binary's revenge}",
      journal = {\mnras},
         year = 2023,
        month = may,
       volume = {521},
       number = {2},
        pages = {2067-2086},
          doi = {10.1093/mnras/stad316},
archivePrefix = {arXiv},
       eprint = {2210.07395},
 primaryClass = {astro-ph.EP},
       adsurl = {https://ui.adsabs.harvard.edu/abs/2023MNRAS.521.2067F}
}

@ARTICLE{Friebe2022,
       author = {{Friebe}, Marc F. and {Pearce}, Tim D. and {L{\"o}hne}, Torsten},
        title = "{Gap carving by a migrating planet embedded in a massive debris disc}",
      journal = {\mnras},
         year = 2022,
        month = may,
       volume = {512},
       number = {3},
        pages = {4441-4454},
          doi = {10.1093/mnras/stac664},
archivePrefix = {arXiv},
       eprint = {2203.03611},
 primaryClass = {astro-ph.EP},
       adsurl = {https://ui.adsabs.harvard.edu/abs/2022MNRAS.512.4441F}
}

@ARTICLE{Morrison2015,
       author = {{Morrison}, Sarah and {Malhotra}, Renu},
        title = "{Planetary Chaotic Zone Clearing: Destinations and Timescales}",
      journal = {\apj},
         year = 2015,
        month = jan,
       volume = {799},
       number = {1},
          eid = {41},
        pages = {41},
          doi = {10.1088/0004-637X/799/1/41},
archivePrefix = {arXiv},
       eprint = {1411.1378},
 primaryClass = {astro-ph.EP},
       adsurl = {https://ui.adsabs.harvard.edu/abs/2015ApJ...799...41M}
}

@ARTICLE{Marshall2021,
       author = {{Marshall}, J.~P. and {Wang}, L. and {Kennedy}, G.~M. and {Zeegers}, S.~T. and {Scicluna}, P.},
        title = "{A search for trends in spatially resolved debris discs at far-infrared wavelengths}",
      journal = {\mnras},
         year = 2021,
        month = mar,
       volume = {501},
       number = {4},
        pages = {6168-6180},
          doi = {10.1093/mnras/staa3917},
       adsurl = {https://ui.adsabs.harvard.edu/abs/2021MNRAS.501.6168M}
}

@ARTICLE{Wisdom1980,
       author = {{Wisdom}, J.},
        title = "{The resonance overlap criterion and the onset of stochastic behavior in the restricted three-body problem}",
      journal = {\aj},
         year = 1980,
        month = aug,
       volume = {85},
        pages = {1122-1133},
          doi = {10.1086/112778},
       adsurl = {https://ui.adsabs.harvard.edu/abs/1980AJ.....85.1122W}
}

@ARTICLE{Kenyon2008,
       author = {{Kenyon}, Scott J. and {Bromley}, Benjamin C.},
        title = "{Variations on Debris Disks: Icy Planet Formation at 30-150 AU for 1-3 M$_{{\ensuremath{\odot}}}$ Main-Sequence Stars}",
      journal = {\apjs},
         year = 2008,
        month = dec,
       volume = {179},
       number = {2},
        pages = {451-483},
          doi = {10.1086/591794},
archivePrefix = {arXiv},
       eprint = {0807.1134},
 primaryClass = {astro-ph},
       adsurl = {https://ui.adsabs.harvard.edu/abs/2008ApJS..179..451K}
}

@ARTICLE{Shannon2016,
       author = {{Shannon}, Andrew and {Bonsor}, Amy and {Kral}, Quentin and {Matthews}, Elisabeth},
        title = "{The unseen planets of double belt debris disc systems}",
      journal = {\mnras},
         year = 2016,
        month = oct,
       volume = {462},
       number = {1},
        pages = {L116-L120},
          doi = {10.1093/mnrasl/slw143},
archivePrefix = {arXiv},
       eprint = {1607.04282},
 primaryClass = {astro-ph.EP},
       adsurl = {https://ui.adsabs.harvard.edu/abs/2016MNRAS.462L.116S}
}

@ARTICLE{Pearce2014,
       author = {{Pearce}, Tim D. and {Wyatt}, Mark C.},
        title = "{Dynamical evolution of an eccentric planet and a less massive debris disc}",
      journal = {\mnras},
         year = 2014,
        month = sep,
       volume = {443},
       number = {3},
        pages = {2541-2560},
          doi = {10.1093/mnras/stu1302},
archivePrefix = {arXiv},
       eprint = {1406.7294},
 primaryClass = {astro-ph.EP},
       adsurl = {https://ui.adsabs.harvard.edu/abs/2014MNRAS.443.2541P}
}

@ARTICLE{Mustill2012,
       author = {{Mustill}, Alexander J. and {Wyatt}, Mark C.},
        title = "{Dependence of a planet's chaotic zone on particle eccentricity: the shape of debris disc inner edges}",
      journal = {\mnras},
         year = 2012,
        month = feb,
       volume = {419},
       number = {4},
        pages = {3074-3080},
          doi = {10.1111/j.1365-2966.2011.19948.x},
archivePrefix = {arXiv},
       eprint = {1110.1282},
 primaryClass = {astro-ph.EP},
       adsurl = {https://ui.adsabs.harvard.edu/abs/2012MNRAS.419.3074M}
}

@ARTICLE{Ida1993,
       author = {{Ida}, Shigeru and {Makino}, Junichiro},
        title = "{Scattering of Planetesimals by a Protoplanet: Slowing Down of Runaway Growth}",
      journal = {\icarus},
         year = 1993,
        month = nov,
       volume = {106},
       number = {1},
        pages = {210-227},
          doi = {10.1006/icar.1993.1167},
       adsurl = {https://ui.adsabs.harvard.edu/abs/1993Icar..106..210I}
}

@software{jwst_pipeline,
       author = {{Bushouse}, Howard and {Eisenhamer}, Jonathan and {Dencheva}, Nadia and {Davies}, James and {Greenfield}, Perry and {Morrison}, Jane and {Hodge}, Phil and {Simon}, Bernie and {Grumm}, David and {Droettboom}, Michael and {Slavich}, Edward and {Sosey}, Megan and {Pauly}, Tyler and {Miller}, Todd and {Jedrzejewski}, Robert and {Hack}, Warren and {Davis}, David and {Crawford}, Steven and {Law}, David and {Gordon}, Karl and {Regan}, Michael and {Cara}, Mihai and {MacDonald}, Ken and {Bradley}, Larry and {Shanahan}, Clare and {Jamieson}, William and {Teodoro}, Mairan and {Williams}, Thomas and {Pena-Guerrero}, Maria and {Graham}, Brett and {Molter}, Edward and {Brandt}, Timothy and {Hayes}, Christian and {Cooper}, Rachel and {Clarke}, Melanie and {Filippazzo}, Joseph},
        title = "{JWST Calibration Pipeline}",
         year = 2025,
        month = apr,
          eid = {10.5281/zenodo.6984365},
          doi = {10.5281/zenodo.6984365},
      version = {1.18.0},
    publisher = {Zenodo},
       adsurl = {https://ui.adsabs.harvard.edu/abs/2023zndo...6984365B}
}

@MISC{Han2024_JWST_GamOph,
       author = {{Han}, Yinuo and {Wyatt}, Mark and {Carpenter}, John and {Henning}, Thomas K. and {Hughes}, Meredith and {Kennedy}, Grant and {Kospal}, Agnes and {Lovell}, Josh Bennett and {MacGregor}, Meredith and {Marino}, Sebastian and {Marshall}, Jonathan Peter and {Matthews}, Brenda and {Pawellek}, Nicole and {Sefilian}, Antranik and {Wilner}, David and {del Burgo}, Carlos},
        title = "{What causes warm dust interior to planetesimal belts?}",
 howpublished = {JWST Proposal. Cycle 3, ID. \#5709},
         year = 2024,
        month = feb,
        pages = {5709},
       adsurl = {https://ui.adsabs.harvard.edu/abs/2024jwst.prop.5709H}
}

@ARTICLE{Lynch2022,
       author = {{Lynch}, Elliot M. and {Lovell}, Joshua B.},
        title = "{Eccentric debris disc morphologies - I. Exploring the origin of apocentre and pericentre glows in face-on debris discs}",
      journal = {\mnras},
         year = 2022,
        month = feb,
       volume = {510},
       number = {2},
        pages = {2538-2551},
          doi = {10.1093/mnras/stab3566},
archivePrefix = {arXiv},
       eprint = {2112.02973},
 primaryClass = {astro-ph.EP},
       adsurl = {https://ui.adsabs.harvard.edu/abs/2022MNRAS.510.2538L}
}

@ARTICLE{Lovell2021,
       author = {{Lovell}, J.~B. and {Marino}, S. and {Wyatt}, M.~C. and {Kennedy}, G.~M. and {MacGregor}, M.~A. and {Stapelfeldt}, K. and {Dent}, B. and {Krist}, J. and {Matr{\`a}}, L. and {Kral}, Q. and {Pani{\'c}}, O. and {Pearce}, T.~D. and {Wilner}, D.},
        title = "{High-resolution ALMA and HST images of q$^{1}$ Eri: an asymmetric debris disc with an eccentric Jupiter}",
      journal = {\mnras},
         year = 2021,
        month = sep,
       volume = {506},
       number = {2},
        pages = {1978-2001},
          doi = {10.1093/mnras/stab1678},
archivePrefix = {arXiv},
       eprint = {2106.05975},
 primaryClass = {astro-ph.EP},
       adsurl = {https://ui.adsabs.harvard.edu/abs/2021MNRAS.506.1978L}
}

@ARTICLE{Lovell2023,
       author = {{Lovell}, Joshua B. and {Lynch}, Elliot M.},
        title = "{Eccentric debris disc morphologies - II. Surface brightness variations from overlapping orbits in narrow eccentric discs}",
      journal = {\mnras},
         year = 2023,
        month = oct,
       volume = {525},
       number = {1},
        pages = {L36-L42},
          doi = {10.1093/mnrasl/slad083},
archivePrefix = {arXiv},
       eprint = {2307.01262},
 primaryClass = {astro-ph.EP},
       adsurl = {https://ui.adsabs.harvard.edu/abs/2023MNRAS.525L..36L}
}

@ARTICLE{Sommer2025,
       author = {{Sommer}, Max and {Wyatt}, Mark and {Han}, Yinuo},
        title = "{A PR drag origin for the Fomalhaut disc's pervasive inner dust: constraints on collisional strengths, icy composition, and embedded planets}",
      journal = {\mnras},
         year = 2025,
        month = may,
       volume = {539},
       number = {1},
        pages = {439-456},
          doi = {10.1093/mnras/staf494},
archivePrefix = {arXiv},
       eprint = {2503.18127},
 primaryClass = {astro-ph.EP},
       adsurl = {https://ui.adsabs.harvard.edu/abs/2025MNRAS.539..439S}
}

@ARTICLE{Dohnanyi1969,
       author = {{Dohnanyi}, J.~S.},
        title = "{Collisional Model of Asteroids and Their Debris}",
      journal = {\jgr},
         year = 1969,
        month = may,
       volume = {74},
        pages = {2531-2554},
          doi = {10.1029/JB074i010p02531},
       adsurl = {https://ui.adsabs.harvard.edu/abs/1969JGR....74.2531D}
}

@ARTICLE{Gaspar2023,
       author = {{G{\'a}sp{\'a}r}, Andr{\'a}s and {Wolff}, Schuyler Grace and {Rieke}, George H. and {Leisenring}, Jarron M. and {Morrison}, Jane and {Su}, Kate Y.~L. and {Ward-Duong}, Kimberly and {Aguilar}, Jonathan and {Ygouf}, Marie and {Beichman}, Charles and {Llop-Sayson}, Jorge and {Bryden}, Geoffrey},
        title = "{Spatially resolved imaging of the inner Fomalhaut disk using JWST/MIRI}",
      journal = {Nature Astronomy},
         year = 2023,
        month = jul,
       volume = {7},
        pages = {790-798},
          doi = {10.1038/s41550-023-01962-6},
archivePrefix = {arXiv},
       eprint = {2305.03789},
 primaryClass = {astro-ph.EP},
       adsurl = {https://ui.adsabs.harvard.edu/abs/2023NatAs...7..790G}
}

@ARTICLE{Su2024,
       author = {{Su}, Kate Y.~L. and {G{\'a}sp{\'a}r}, Andr{\'a}s and {Rieke}, George H. and {Malhotra}, Renu and {Matr{\'a}}, Luca and {Wolff}, Schuyler Grace and {Leisenring}, Jarron M. and {Beichman}, Charles and {Ygouf}, Marie},
        title = "{Imaging of the Vega Debris System Using JWST/MIRI}",
      journal = {\apj},
         year = 2024,
        month = dec,
       volume = {977},
       number = {2},
          eid = {277},
        pages = {277},
          doi = {10.3847/1538-4357/ad8cde},
archivePrefix = {arXiv},
       eprint = {2410.23636},
 primaryClass = {astro-ph.EP},
       adsurl = {https://ui.adsabs.harvard.edu/abs/2024ApJ...977..277S}
}

@ARTICLE{Sefilian2025,
       author = {{Sefilian}, Antranik A. and {Kratter}, Kaitlin M. and {Wyatt}, Mark C. and {Petrovich}, Cristobal and {Th{\'e}bault}, Philippe and {Malhotra}, Renu and {Faramaz-Gorka}, Virginie},
        title = "{The vertical structure of debris discs and the role of disc gravity: a primer using a simplified model}",
      journal = {\mnras},
         year = 2025,
        month = nov,
       volume = {543},
       number = {4},
        pages = {3123-3151},
          doi = {10.1093/mnras/staf1555},
archivePrefix = {arXiv},
       eprint = {2505.09578},
 primaryClass = {astro-ph.EP},
       adsurl = {https://ui.adsabs.harvard.edu/abs/2025MNRAS.543.3123S}
}

@ARTICLE{Kervella2022,
       author = {{Kervella}, Pierre and {Arenou}, Fr{\'e}d{\'e}ric and {Th{\'e}venin}, Fr{\'e}d{\'e}ric},
        title = "{Stellar and substellar companions from Gaia EDR3. Proper-motion anomaly and resolved common proper-motion pairs}",
      journal = {\aap},
         year = 2022,
        month = jan,
       volume = {657},
          eid = {A7},
        pages = {A7},
          doi = {10.1051/0004-6361/202142146},
archivePrefix = {arXiv},
       eprint = {2109.10912},
 primaryClass = {astro-ph.SR},
       adsurl = {https://ui.adsabs.harvard.edu/abs/2022A&A...657A...7K}
}

@ARTICLE{Husser2013,
       author = {{Husser}, T. -O. and {Wende-von Berg}, S. and {Dreizler}, S. and {Homeier}, D. and {Reiners}, A. and {Barman}, T. and {Hauschildt}, P.~H.},
        title = "{A new extensive library of PHOENIX stellar atmospheres and synthetic spectra}",
      journal = {\aap},
         year = 2013,
        month = may,
       volume = {553},
          eid = {A6},
        pages = {A6},
          doi = {10.1051/0004-6361/201219058},
archivePrefix = {arXiv},
       eprint = {1303.5632},
 primaryClass = {astro-ph.SR},
       adsurl = {https://ui.adsabs.harvard.edu/abs/2013A&A...553A...6H}
}

@ARTICLE{Nilsson2010,
       author = {{Nilsson}, R. and {Liseau}, R. and {Brandeker}, A. and {Olofsson}, G. and {Pilbratt}, G.~L. and {Risacher}, C. and {Rodmann}, J. and {Augereau}, J. -C. and {Bergman}, P. and {Eiroa}, C. and {Fridlund}, M. and {Th{\'e}bault}, P. and {White}, G.~J.},
        title = "{Kuiper belts around nearby stars}",
      journal = {\aap},
         year = 2010,
        month = jul,
       volume = {518},
          eid = {A40},
        pages = {A40},
          doi = {10.1051/0004-6361/201014444},
archivePrefix = {arXiv},
       eprint = {1005.3215},
 primaryClass = {astro-ph.EP},
       adsurl = {https://ui.adsabs.harvard.edu/abs/2010A&A...518A..40N}
}

@ARTICLE{Chen2014,
       author = {{Chen}, Christine H. and {Mittal}, Tushar and {Kuchner}, Marc and {Forrest}, William J. and {Lisse}, Carey M. and {Manoj}, P. and {Sargent}, Benjamin A. and {Watson}, Dan M.},
        title = "{The Spitzer Infrared Spectrograph Debris Disk Catalog. I. Continuum Analysis of Unresolved Targets}",
      journal = {\apjs},
         year = 2014,
        month = apr,
       volume = {211},
       number = {2},
          eid = {25},
        pages = {25},
          doi = {10.1088/0067-0049/211/2/25},
       adsurl = {https://ui.adsabs.harvard.edu/abs/2014ApJS..211...25C}
}

@ARTICLE{Su2006,
       author = {{Su}, K.~Y.~L. and {Rieke}, G.~H. and {Stansberry}, J.~A. and {Bryden}, G. and {Stapelfeldt}, K.~R. and {Trilling}, D.~E. and {Muzerolle}, J. and {Beichman}, C.~A. and {Moro-Martin}, A. and {Hines}, D.~C. and {Werner}, M.~W.},
        title = "{Debris Disk Evolution around A Stars}",
      journal = {\apj},
         year = 2006,
        month = dec,
       volume = {653},
       number = {1},
        pages = {675-689},
          doi = {10.1086/508649},
archivePrefix = {arXiv},
       eprint = {astro-ph/0608563},
 primaryClass = {astro-ph},
       adsurl = {https://ui.adsabs.harvard.edu/abs/2006ApJ...653..675S}
}

@ARTICLE{Wright2010,
       author = {{Wright}, Edward L. and {Eisenhardt}, Peter R.~M. and {Mainzer}, Amy K. and {Ressler}, Michael E. and {Cutri}, Roc M. and {Jarrett}, Thomas and {Kirkpatrick}, J. Davy and {Padgett}, Deborah and {McMillan}, Robert S. and {Skrutskie}, Michael and {Stanford}, S.~A. and {Cohen}, Martin and {Walker}, Russell G. and {Mather}, John C. and {Leisawitz}, David and {Gautier}, III, Thomas N. and {McLean}, Ian and {Benford}, Dominic and {Lonsdale}, Carol J. and {Blain}, Andrew and {Mendez}, Bryan and {Irace}, William R. and {Duval}, Valerie and {Liu}, Fengchuan and {Royer}, Don and {Heinrichsen}, Ingolf and {Howard}, Joan and {Shannon}, Mark and {Kendall}, Martha and {Walsh}, Amy L. and {Larsen}, Mark and {Cardon}, Joel G. and {Schick}, Scott and {Schwalm}, Mark and {Abid}, Mohamed and {Fabinsky}, Beth and {Naes}, Larry and {Tsai}, Chao-Wei},
        title = "{The Wide-field Infrared Survey Explorer (WISE): Mission Description and Initial On-orbit Performance}",
      journal = {\aj},
         year = 2010,
        month = dec,
       volume = {140},
       number = {6},
        pages = {1868-1881},
          doi = {10.1088/0004-6256/140/6/1868},
archivePrefix = {arXiv},
       eprint = {1008.0031},
 primaryClass = {astro-ph.IM},
       adsurl = {https://ui.adsabs.harvard.edu/abs/2010AJ....140.1868W}
}

@ARTICLE{Ishihara2010,
       author = {{Ishihara}, D. and {Onaka}, T. and {Kataza}, H. and {Salama}, A. and {Alfageme}, C. and {Cassatella}, A. and {Cox}, N. and {Garc{\'\i}a-Lario}, P. and {Stephenson}, C. and {Cohen}, M. and {Fujishiro}, N. and {Fujiwara}, H. and {Hasegawa}, S. and {Ita}, Y. and {Kim}, W. and {Matsuhara}, H. and {Murakami}, H. and {M{\"u}ller}, T.~G. and {Nakagawa}, T. and {Ohyama}, Y. and {Oyabu}, S. and {Pyo}, J. and {Sakon}, I. and {Shibai}, H. and {Takita}, S. and {Tanab{\'e}}, T. and {Uemizu}, K. and {Ueno}, M. and {Usui}, F. and {Wada}, T. and {Watarai}, H. and {Yamamura}, I. and {Yamauchi}, C.},
        title = "{The AKARI/IRC mid-infrared all-sky survey}",
      journal = {\aap},
         year = 2010,
        month = may,
       volume = {514},
          eid = {A1},
        pages = {A1},
          doi = {10.1051/0004-6361/200913811},
archivePrefix = {arXiv},
       eprint = {1003.0270},
 primaryClass = {astro-ph.IM},
       adsurl = {https://ui.adsabs.harvard.edu/abs/2010A&A...514A...1I}
}

@dataset{Cutri2003,
       author = {{Cutri}, R.~M. and {Skrutskie}, M.~F. and {van Dyk}, S. and {Beichman}, C.~A. and {Carpenter}, J.~M. and {Chester}, T. and {Cambresy}, L. and {Evans}, T. and {Fowler}, J. and {Gizis}, J. and {Howard}, E. and {Huchra}, J. and {Jarrett}, T. and {Kopan}, E.~L. and {Kirkpatrick}, J.~D. and {Light}, R.~M. and {Marsh}, K.~A. and {McCallon}, H. and {Schneider}, S. and {Stiening}, R. and {Sykes}, M. and {Weinberg}, M. and {Wheaton}, W.~A. and {Wheelock}, S. and {Zacarias}, N.},
        title = "{2MASS All Sky Catalog of point sources.}",
         year = 2003,
       adsurl = {https://ui.adsabs.harvard.edu/abs/2003tmc..book.....C}
}

@dataset{Mermilliod2006,
       author = {{Mermilliod}, J.~C.},
        title = "{VizieR Online Data Catalog: Homogeneous Means in the UBV System (Mermilliod 1991)}",
 howpublished = {VizieR On-line Data Catalog: II/168.  Originally published in: Institut d'Astronomie, Universite de Lausanne (1991)},
         year = 2006,
        month = may,
          eid = {II/168},
       adsurl = {https://ui.adsabs.harvard.edu/abs/2006yCat.2168....0M}
}

@PROCEEDINGS{Hipparcos1997,
        title = "{The HIPPARCOS and TYCHO catalogues. Astrometric and photometric star catalogues derived from the ESA HIPPARCOS Space Astrometry Mission}",
    booktitle = {ESA Special Publication},
         year = 1997,
       editor = {{ESA}},
       series = {ESA Special Publication},
       volume = {1200},
        month = jan,
       adsurl = {https://ui.adsabs.harvard.edu/abs/1997ESASP1200.....E}
}

@ARTICLE{Hog2000,
       author = {{H{\o}g}, E. and {Fabricius}, C. and {Makarov}, V.~V. and {Urban}, S. and {Corbin}, T. and {Wycoff}, G. and {Bastian}, U. and {Schwekendiek}, P. and {Wicenec}, A.},
        title = "{The Tycho-2 catalogue of the 2.5 million brightest stars}",
      journal = {\aap},
         year = 2000,
        month = mar,
       volume = {355},
        pages = {L27-L30},
       adsurl = {https://ui.adsabs.harvard.edu/abs/2000A&A...355L..27H}
}

@ARTICLE{Olofsson2022b,
       author = {{Olofsson}, J. and {Th{\'e}bault}, P. and {Kennedy}, G.~M. and {Bayo}, A.},
        title = "{The halo around HD 32297: {\ensuremath{\mu}}m-sized cometary dust}",
      journal = {\aap},
         year = 2022,
        month = aug,
       volume = {664},
          eid = {A122},
        pages = {A122},
          doi = {10.1051/0004-6361/202243794},
archivePrefix = {arXiv},
       eprint = {2206.07068},
 primaryClass = {astro-ph.EP},
       adsurl = {https://ui.adsabs.harvard.edu/abs/2022A&A...664A.122O}
}

@ARTICLE{Ballering2016,
       author = {{Ballering}, Nicholas P. and {Su}, Kate Y.~L. and {Rieke}, George H. and {G{\'a}sp{\'a}r}, Andr{\'a}s},
        title = "{A Comprehensive Dust Model Applied to the Resolved Beta Pictoris Debris Disk from Optical to Radio Wavelengths}",
      journal = {\apj},
         year = 2016,
        month = jun,
       volume = {823},
       number = {2},
          eid = {108},
        pages = {108},
          doi = {10.3847/0004-637X/823/2/108},
archivePrefix = {arXiv},
       eprint = {1605.01731},
 primaryClass = {astro-ph.EP},
       adsurl = {https://ui.adsabs.harvard.edu/abs/2016ApJ...823..108B}
}

@ARTICLE{Su2009,
       author = {{Su}, K.~Y.~L. and {Rieke}, G.~H. and {Stapelfeldt}, K.~R. and {Malhotra}, R. and {Bryden}, G. and {Smith}, P.~S. and {Misselt}, K.~A. and {Moro-Martin}, A. and {Williams}, J.~P.},
        title = "{The Debris Disk Around HR 8799}",
      journal = {\apj},
         year = 2009,
        month = nov,
       volume = {705},
       number = {1},
        pages = {314-327},
          doi = {10.1088/0004-637X/705/1/314},
archivePrefix = {arXiv},
       eprint = {0909.2687},
 primaryClass = {astro-ph.EP},
       adsurl = {https://ui.adsabs.harvard.edu/abs/2009ApJ...705..314S}
}

@dataset{SEIP2020,
       author = {{IRSA} and {SSC}},
        title = "{Spitzer Enhanced Imaging Products}",
 howpublished = {NASA IPAC DataSet, IRSA433},
         year = 2020,
        month = jan,
          doi = {10.26131/IRSA433},
       adsurl = {https://ui.adsabs.harvard.edu/abs/2020ipac.data.I433I}
}

@ARTICLE{Pilbratt2010,
       author = {{Pilbratt}, G.~L. and {Riedinger}, J.~R. and {Passvogel}, T. and {Crone}, G. and {Doyle}, D. and {Gageur}, U. and {Heras}, A.~M. and {Jewell}, C. and {Metcalfe}, L. and {Ott}, S. and {Schmidt}, M.},
        title = "{Herschel Space Observatory. An ESA facility for far-infrared and submillimetre astronomy}",
      journal = {\aap},
         year = 2010,
        month = jul,
       volume = {518},
          eid = {L1},
        pages = {L1},
          doi = {10.1051/0004-6361/201014759},
archivePrefix = {arXiv},
       eprint = {1005.5331},
 primaryClass = {astro-ph.IM},
       adsurl = {https://ui.adsabs.harvard.edu/abs/2010A&A...518L...1P}
}

@ARTICLE{Su2005,
       author = {{Su}, K.~Y.~L. and {Rieke}, G.~H. and {Misselt}, K.~A. and {Stansberry}, J.~A. and {Moro-Martin}, A. and {Stapelfeldt}, K.~R. and {Werner}, M.~W. and {Trilling}, D.~E. and {Bendo}, G.~J. and {Gordon}, K.~D. and {Hines}, D.~C. and {Wyatt}, M.~C. and {Holland}, W.~S. and {Marengo}, M. and {Megeath}, S.~T. and {Fazio}, G.~G.},
        title = "{The Vega Debris Disk: A Surprise from Spitzer}",
      journal = {\apj},
         year = 2005,
        month = jul,
       volume = {628},
       number = {1},
        pages = {487-500},
          doi = {10.1086/430819},
archivePrefix = {arXiv},
       eprint = {astro-ph/0504086},
 primaryClass = {astro-ph},
       adsurl = {https://ui.adsabs.harvard.edu/abs/2005ApJ...628..487S}
}

@ARTICLE{Strubbe2006,
       author = {{Strubbe}, Linda E. and {Chiang}, Eugene I.},
        title = "{Dust Dynamics, Surface Brightness Profiles, and Thermal Spectra of Debris Disks: The Case of AU Microscopii}",
      journal = {\apj},
         year = 2006,
        month = sep,
       volume = {648},
       number = {1},
        pages = {652-665},
          doi = {10.1086/505736},
archivePrefix = {arXiv},
       eprint = {astro-ph/0510527},
 primaryClass = {astro-ph},
       adsurl = {https://ui.adsabs.harvard.edu/abs/2006ApJ...648..652S}
}

@ARTICLE{Stapelfeldt2004,
       author = {{Stapelfeldt}, K.~R. and {Holmes}, E.~K. and {Chen}, C. and {Rieke}, G.~H. and {Su}, K.~Y.~L. and {Hines}, D.~C. and {Werner}, M.~W. and {Beichman}, C.~A. and {Jura}, M. and {Padgett}, D.~L. and {Stansberry}, J.~A. and {Bendo}, G. and {Cadien}, J. and {Marengo}, M. and {Thompson}, T. and {Velusamy}, T. and {Backus}, C. and {Blaylock}, M. and {Egami}, E. and {Engelbracht}, C.~W. and {Frayer}, D.~T. and {Gordon}, K.~D. and {Keene}, J. and {Latter}, W.~B. and {Megeath}, T. and {Misselt}, K. and {Morrison}, J.~E. and {Muzerolle}, J. and {Noriega-Crespo}, A. and {Van Cleve}, J. and {Young}, E.~T.},
        title = "{First Look at the Fomalhaut Debris Disk with the Spitzer Space Telescope}",
      journal = {\apjs},
         year = 2004,
        month = sep,
       volume = {154},
       number = {1},
        pages = {458-462},
          doi = {10.1086/423135},
       adsurl = {https://ui.adsabs.harvard.edu/abs/2004ApJS..154..458S}
}

@ARTICLE{Malhotra1993,
       author = {{Malhotra}, Renu},
        title = "{The origin of Pluto's peculiar orbit}",
      journal = {\nat},
         year = 1993,
        month = oct,
       volume = {365},
       number = {6449},
        pages = {819-821},
          doi = {10.1038/365819a0},
       adsurl = {https://ui.adsabs.harvard.edu/abs/1993Natur.365..819M}
}

@ARTICLE{Lee2016,
       author = {{Lee}, Eve J. and {Chiang}, Eugene},
        title = "{A Primer on Unifying Debris Disk Morphologies}",
      journal = {\apj},
         year = 2016,
        month = aug,
       volume = {827},
       number = {2},
          eid = {125},
        pages = {125},
          doi = {10.3847/0004-637X/827/2/125},
archivePrefix = {arXiv},
       eprint = {1605.06118},
 primaryClass = {astro-ph.EP},
       adsurl = {https://ui.adsabs.harvard.edu/abs/2016ApJ...827..125L}
}

@ARTICLE{Burns1979,
       author = {{Burns}, J.~A. and {Lamy}, P.~L. and {Soter}, S.},
        title = "{Radiation forces on small particles in the solar system}",
      journal = {\icarus},
         year = 1979,
        month = oct,
       volume = {40},
       number = {1},
        pages = {1-48},
          doi = {10.1016/0019-1035(79)90050-2},
       adsurl = {https://ui.adsabs.harvard.edu/abs/1979Icar...40....1B}
}

@ARTICLE{Wright2023,
       author = {{Wright}, Gillian S. and {Rieke}, George H. and {Glasse}, Alistair and {Ressler}, Michael and {Garc{\'\i}a Mar{\'\i}n}, Macarena and {Aguilar}, Jonathan and {Alberts}, Stacey and {{\'A}lvarez-M{\'a}rquez}, Javier and {Argyriou}, Ioannis and {Banks}, Kimberly and {Baudoz}, Pierre and {Boccaletti}, Anthony and {Bouchet}, Patrice and {Bouwman}, Jeroen and {Brandl}, Bernard R. and {Breda}, David and {Bright}, Stacey and {Cale}, Steven and {Colina}, Luis and {Cossou}, Christophe and {Coulais}, Alain and {Cracraft}, Misty and {De Meester}, Wim and {Dicken}, Daniel and {Engesser}, Michael and {Etxaluze}, Mireya and {Fox}, Ori D. and {Friedman}, Scott and {Fu}, Henry and {Gasman}, Danny and {G{\'a}sp{\'a}r}, Andr{\'a}s and {Gastaud}, Ren{\'e} and {Geers}, Vincent and {Glauser}, Adrian Michael and {Gordon}, Karl D. and {Greene}, Thomas and {Greve}, Thomas R. and {Grundy}, Timothy and {G{\"u}del}, Manuel and {Guillard}, Pierre and {Haderlein}, Peter and {Hashimoto}, Ryan and {Henning}, Thomas and {Hines}, Dean and {Holler}, Bryan and {Detre}, {\"O}rs Hunor and {Jahromi}, Amir and {James}, Bryan and {Jones}, Olivia C. and {Justtanont}, Kay and {Kavanagh}, Patrick and {Kendrew}, Sarah and {Klaassen}, Pamela and {Krause}, Oliver and {Labiano}, Alvaro and {Lagage}, Pierre-Olivier and {Lambros}, Scott and {Larson}, Kirsten and {Law}, David and {Lee}, David and {Libralato}, Mattia and {Lorenzo Alverez}, Jose and {Meixner}, Margaret and {Morrison}, Jane and {Mueller}, Migo and {Murray}, Katherine and {Mycroft}, Matthew and {Myers}, Richard and {Nayak}, Omnarayani and {Naylor}, Bret and {Nickson}, Bryony and {Noriega-Crespo}, Alberto and {{\"O}stlin}, G{\"o}ran and {O'Sullivan}, Brian and {Ottens}, Richard and {Patapis}, Polychronis and {Penanen}, Konstantin and {Pietraszkiewicz}, Martin and {Ray}, Tom and {Regan}, Michael and {Roteliuk}, Anthony and {Royer}, Pierre and {Samara-Ratna}, Piyal and {Samuelson}, Bridget and {Sargent}, Beth A. and {Scheithauer}, Silvia and {Schneider}, Analyn and {Schreiber}, J{\"u}rgen and {Shaughnessy}, Bryan and {Sheehan}, Evan and {Shivaei}, Irene and {Sloan}, G.~C. and {Tamas}, Laszlo and {Teague}, Kelly and {Temim}, Tea and {Tikkanen}, Tuomo and {Tustain}, Samuel and {van Dishoeck}, Ewine F. and {Vandenbussche}, Bart and {Weilert}, Mark and {Whitehouse}, Paul and {Wolff}, Schuyler},
        title = "{The Mid-infrared Instrument for JWST and Its In-flight Performance}",
      journal = {\pasp},
         year = 2023,
        month = apr,
       volume = {135},
       number = {1046},
          eid = {048003},
        pages = {048003},
          doi = {10.1088/1538-3873/acbe66},
       adsurl = {https://ui.adsabs.harvard.edu/abs/2023PASP..135d8003W}
}

@ARTICLE{Plavchan2005,
       author = {{Plavchan}, Peter and {Jura}, M. and {Lipscy}, S.~J.},
        title = "{Where Are the M Dwarf Disks Older Than 10 Million Years?}",
      journal = {\apj},
         year = 2005,
        month = oct,
       volume = {631},
       number = {2},
        pages = {1161-1169},
          doi = {10.1086/432568},
archivePrefix = {arXiv},
       eprint = {astro-ph/0506132},
 primaryClass = {astro-ph},
       adsurl = {https://ui.adsabs.harvard.edu/abs/2005ApJ...631.1161P}
}

@ARTICLE{Mamajek2012,
       author = {{Mamajek}, Eric E.},
        title = "{On the Age and Binarity of Fomalhaut}",
      journal = {\apjl},
         year = 2012,
        month = aug,
       volume = {754},
       number = {2},
          eid = {L20},
        pages = {L20},
          doi = {10.1088/2041-8205/754/2/L20},
archivePrefix = {arXiv},
       eprint = {1206.6353},
 primaryClass = {astro-ph.SR},
       adsurl = {https://ui.adsabs.harvard.edu/abs/2012ApJ...754L..20M}
}

@ARTICLE{Yoon2010,
       author = {{Yoon}, Jinmi and {Peterson}, Deane M. and {Kurucz}, Robert L. and {Zagarello}, Robert J.},
        title = "{A New View of Vega's Composition, Mass, and Age}",
      journal = {\apj},
         year = 2010,
        month = jan,
       volume = {708},
       number = {1},
        pages = {71-79},
          doi = {10.1088/0004-637X/708/1/71},
       adsurl = {https://ui.adsabs.harvard.edu/abs/2010ApJ...708...71Y}
}

@ARTICLE{Vican2012,
       author = {{Vican}, Laura},
        title = "{Age Determination for 346 Nearby Stars in the Herschel DEBRIS Survey}",
      journal = {\aj},
         year = 2012,
        month = jun,
       volume = {143},
       number = {6},
          eid = {135},
        pages = {135},
          doi = {10.1088/0004-6256/143/6/135},
archivePrefix = {arXiv},
       eprint = {1203.1966},
 primaryClass = {astro-ph.SR},
       adsurl = {https://ui.adsabs.harvard.edu/abs/2012AJ....143..135V}
}

@ARTICLE{Gaspar2016,
       author = {{G{\'a}sp{\'a}r}, Andr{\'a}s and {Rieke}, George H. and {Ballering}, Nicholas},
        title = "{The Correlation between Metallicity and Debris Disk Mass}",
      journal = {\apj},
         year = 2016,
        month = aug,
       volume = {826},
       number = {2},
          eid = {171},
        pages = {171},
          doi = {10.3847/0004-637X/826/2/171},
archivePrefix = {arXiv},
       eprint = {1604.07403},
 primaryClass = {astro-ph.SR},
       adsurl = {https://ui.adsabs.harvard.edu/abs/2016ApJ...826..171G}
}

@ARTICLE{Marshall2023,
       author = {{Marshall}, Jonathan P. and {Milli}, J. and {Choquet}, E. and {del Burgo}, C. and {Kennedy}, G.~M. and {Kemper}, F. and {Wyatt}, M.~C. and {Kral}, Q. and {Soummer}, R.},
        title = "{Stirred but not shaken: a multiwavelength view of HD 16743's debris disc}",
      journal = {\mnras},
         year = 2023,
        month = jun,
       volume = {521},
       number = {4},
        pages = {5940-5951},
          doi = {10.1093/mnras/stad913},
archivePrefix = {arXiv},
       eprint = {2303.17128},
 primaryClass = {astro-ph.EP},
       adsurl = {https://ui.adsabs.harvard.edu/abs/2023MNRAS.521.5940M}
}

@ARTICLE{SearchCal,
       author = {{Chelli}, Alain and {Duvert}, Gilles and {Bourg{\`e}s}, Laurent and {Mella}, Guillaume and {Lafrasse}, Sylvain and {Bonneau}, Daniel and {Chesneau}, Olivier},
        title = "{Pseudomagnitudes and differential surface brightness: Application to the apparent diameter of stars}",
      journal = {\aap},
         year = 2016,
        month = may,
       volume = {589},
          eid = {A112},
        pages = {A112},
          doi = {10.1051/0004-6361/201527484},
archivePrefix = {arXiv},
       eprint = {1604.07700},
 primaryClass = {astro-ph.SR},
       adsurl = {https://ui.adsabs.harvard.edu/abs/2016A&A...589A.112C}
}

@ARTICLE{Argyriou2023,
       author = {{Argyriou}, Ioannis and {Lage}, Craig and {Rieke}, George H. and {Gasman}, Danny and {Bouwman}, Jeroen and {Morrison}, Jane and {Libralato}, Mattia and {Dicken}, Daniel and {Brandl}, Bernhard R. and {{\'A}lvarez-M{\'a}rquez}, Javier and {Labiano}, Alvaro and {Regan}, Michael and {Ressler}, Michael E.},
        title = "{The brighter-fatter effect in the JWST MIRI Si:As IBC detectors. I. Observations, impact on science, and modeling}",
      journal = {\aap},
         year = 2023,
        month = dec,
       volume = {680},
          eid = {A96},
        pages = {A96},
          doi = {10.1051/0004-6361/202346490},
archivePrefix = {arXiv},
       eprint = {2303.13517},
 primaryClass = {astro-ph.IM},
       adsurl = {https://ui.adsabs.harvard.edu/abs/2023A&A...680A..96A}
}

@ARTICLE{Wolff2025,
       author = {{Wolff}, Schuyler G. and {G{\'a}sp{\'a}r}, Andr{\'a}s and {Rieke}, George and {Leisenring}, Jarron M. and {Sefilian}, Antranik A. and {Ygouf}, Marie and {Llop-Sayson}, Jorge},
        title = "{JWST/MIRI Imaging of the Warm Dust Component of the ϵ Eridani Debris Disk}",
      journal = {\aj},
         year = 2025,
        month = oct,
       volume = {170},
       number = {4},
          eid = {244},
        pages = {244},
          doi = {10.3847/1538-3881/adfcd6},
archivePrefix = {arXiv},
       eprint = {2509.24976},
 primaryClass = {astro-ph.EP},
       adsurl = {https://ui.adsabs.harvard.edu/abs/2025AJ....170..244W}
}

@ARTICLE{Marshall2025,
       author = {{Marshall}, Jonathan P. and {Mu{\~n}oz-Guti{\'e}rrez}, Marco A. and {Sefilian}, Antranik A. and {Peimbert}, A.},
        title = "{Testing the Impact of Planet-stirring, Self-stirring, and Mixed-stirring on Debris Disc Architecture: A Case Study of HD 16743}",
      journal = {\mnras},
         year = 2025,
        month = nov,
          doi = {10.1093/mnras/staf1990},
archivePrefix = {arXiv},
       eprint = {2509.22822},
 primaryClass = {astro-ph.EP},
       adsurl = {https://ui.adsabs.harvard.edu/abs/2025MNRAS.tmp.1878M}
}

@ARTICLE{Gaspar2008,
       author = {{G{\'a}sp{\'a}r}, A. and {Su}, K.~Y.~L. and {Rieke}, G.~H. and {Balog}, Z. and {Kamp}, I. and {Mart{\'\i}nez-Galarza}, J.~R. and {Stapelfeldt}, K.},
        title = "{Modeling the Infrared Bow Shock at {\ensuremath{\delta}} Velorum: Implications for Studies of Debris Disks and {\ensuremath{\lambda}} Bo{\"o}tis Stars}",
      journal = {\apj},
         year = 2008,
        month = jan,
       volume = {672},
       number = {2},
        pages = {974-983},
          doi = {10.1086/523299},
archivePrefix = {arXiv},
       eprint = {0709.4247},
 primaryClass = {astro-ph},
       adsurl = {https://ui.adsabs.harvard.edu/abs/2008ApJ...672..974G}
}

@ARTICLE{Jankovic2024,
       author = {{Jankovic}, Marija R. and {Wyatt}, Mark C. and {L{\"o}hne}, Torsten},
        title = "{Collisional damping in debris discs: Only significant if collision velocities are low}",
      journal = {\aap},
         year = 2024,
        month = nov,
       volume = {691},
          eid = {A302},
        pages = {A302},
          doi = {10.1051/0004-6361/202451080},
archivePrefix = {arXiv},
       eprint = {2411.13991},
 primaryClass = {astro-ph.EP},
       adsurl = {https://ui.adsabs.harvard.edu/abs/2024A&A...691A.302J}
}

@ARTICLE{Thebault2007,
       author = {{Th{\'e}bault}, P. and {Augereau}, J.-C.},
        title = "{Collisional processes and size distribution in spatially extended debris discs}",
      journal = {\aap},
         year = 2007,
        month = sep,
       volume = {472},
       number = {1},
        pages = {169-185},
          doi = {10.1051/0004-6361:20077709},
archivePrefix = {arXiv},
       eprint = {0706.0344},
 primaryClass = {astro-ph},
       adsurl = {https://ui.adsabs.harvard.edu/abs/2007A&A...472..169T}
}

@BOOK{Murray1999,
       author = {{Murray}, Carl D. and {Dermott}, Stanley F.},
        title = "{Solar System Dynamics}",
         year = 1999,
          doi = {10.1017/CBO9781139174817},
       adsurl = {https://ui.adsabs.harvard.edu/abs/1999ssd..book.....M}
}

@ARTICLE{Thebault2019,
       author = {{Thebault}, P. and {Kral}, Q.},
        title = "{Is there more than meets the eye? Presence and role of sub-micron grains in debris discs}",
      journal = {\aap},
         year = 2019,
        month = jun,
       volume = {626},
          eid = {A24},
        pages = {A24},
          doi = {10.1051/0004-6361/201935341},
archivePrefix = {arXiv},
       eprint = {1904.05395},
 primaryClass = {astro-ph.EP},
       adsurl = {https://ui.adsabs.harvard.edu/abs/2019A&A...626A..24T}
}
\bibliographystyle{aasjournalv7}



\end{document}